\documentclass[useAMS,usenatbib,onecolumn]{mn2e}
\usepackage{epsfig}

\newcommand{\mHt}{{\rm H_{2}}}
\newcommand{\me}{{\rm e^{-}}}
\newcommand{\mH}{{\rm H}}
\newcommand{\Hp}{\rm{H}^{+}}
\newcommand{\Hm}{{\rm H^{-}}}
\newcommand{\mHtp}{\rm{H}_{2}^{+}}
\newcommand{\htp}{\rm{H}_{3}^{+}}
\newcommand{\He}{\rm{He}}
\newcommand{\Hep}{\rm{He}^{+}}
\newcommand{\Cp}{\rm{C^{+}}}
\newcommand{\Op}{\rm{O}^{+}}
\newcommand{\mC}{\rm{C}}
\newcommand{\ch}{{\rm CH}}
\newcommand{\Cm}{{\rm C^{-}}}
\newcommand{\mO}{\rm{O}}
\newcommand{\Om}{\rm{O^{-}}}
\newcommand{\oh}{{\rm OH}}
\newcommand{\hto}{\rm{H_{2}O}}
\newcommand{\co}{{\rm CO}}
\newcommand{\expf}[3]{\exp \left(#1\frac{#2}{#3}\right)}
\newcommand{\tmpt}[1]{\left(\frac{T}{300}\right)^{#1}}
\newcommand{\tmptscl}[2]{\left(\frac{T}{#1}\right)^{#2}}
\newcommand{\mwfrac}[1]{{\langle x_{#1}}\rangle_{\rm M}}

\title[CO formation in the turbulent ISM]{Modelling CO formation in
  the turbulent interstellar medium}
\author[Glover et al.]{S.~C.~O.~Glover$^{1,2}$, C.~Federrath$^{1,3}$,
M.-M.~{Mac Low}$^4$, \& R.~S.~Klessen$^{1}$ \\
$^1$Zentrum f\"ur Astronomie der Universit\"at Heidelberg, Institut f\"ur Theoretische  Astrophysik, Albert-Ueberle-Str.\ 2, 69120 Heidelberg, Germany \\
$^2$Astrophysikalisches Institut Potsdam, An der Sternwarte 16, D-14482 Potsdam, Germany \\
$^3$Max-Planck-Institut f\"ur Astronomie, K\"onigstuhl 17, D-69117 Heidelberg, Germany \\
$^4$Department of Astrophysics, American Museum of Natural History, Central Park West at 79th Street, New York, NY 10024
}

\begin{document}

\maketitle

\begin{abstract}
We present results from high-resolution three-dimensional simulations of turbulent interstellar gas
that self-consistently follow its coupled thermal, chemical and dynamical evolution, with a particular
focus on the formation and destruction of H$_{2}$ and CO.  We quantify the formation timescales for 
H$_{2}$ and CO in physical conditions corresponding to those found in nearby giant molecular clouds, 
and show that both species form rapidly, with chemical timescales that are comparable to the dynamical 
timescale of the gas. 

We also investigate the spatial distributions of H$_{2}$ and CO, and how they relate to the underlying gas 
distribution. We show that H$_{2}$ is a good tracer of the gas distribution, but that the relationship between
CO abundance and gas density is more complex. 
The CO abundance is not well-correlated with either the gas number density $n$ or the visual 
extinction $A_{\rm V}$: both have a large influence on the CO abundance, but the inhomogeneous nature 
of the density field produced by the turbulence means that $n$ and $A_{\rm V}$ are only poorly correlated. 
There is a large scatter in $A_{\rm V}$, and hence CO abundance, for gas with any particular density, 
and similarly a large scatter in density and CO abundance for gas with any particular visual extinction. This will 
have important consequences for the interpretation of the CO emission observed from real molecular 
clouds.

Finally, we also examine the temperature structure of the simulated gas. We show that the molecular gas
is not isothermal. Most of it has a temperature in the range of 10--20~K, but there is also a significant fraction
of warmer gas, located in low-extinction regions where photoelectric heating remains effective.
\end{abstract}


\section{Introduction}
As essentially all star formation within the Milky Way occurs within cold, dense 
clouds of molecular gas \citep{ll03}, understanding how these clouds form, 
and how and why they then form stars, is crucial if we are to develop any real 
understanding of the process of star formation \citep{mk04,bp07,mo07}.
Observations of molecular emission 
lines provide us with a great deal of information on the internal structure and 
dynamics of molecular clouds, and may also hold important clues to their 
past dynamical history \citep{ferr01}.

Traditionally, molecular clouds have been viewed as quasi-static objects that
form stars slowly over a long lifetime \citep{ze74,sal87,kmm06}. In this picture, the dynamical 
evolution of a cloud and the chemical evolution of the gas within it are only loosely 
coupled, and can be modelled separately. In the past few years, however, this traditional 
picture has begun to give way to a new picture of clouds as inherently dynamical entities, 
whose formation and evolution are dominated by the effects of
turbulence \citep{bhv99, khm00, kl01, es04,se04}.
The dynamical evolution of the cloud is rapid, with a timescale comparable to 
those of key chemical processes such as the conversion of atomic to molecular hydrogen
(H$_{2}$) or the freeze-out of molecules onto the surfaces of interstellar dust grains in dark cores. 
If this picture is correct, then the dynamics and chemistry of the gas are strongly coupled, 
with one directly influencing the evolution of the other, and to model them correctly we must
model them together.

An additional impetus towards the development of coupled models of 
molecular cloud dynamics and chemistry comes from the realization that 
if the internal motions of the clouds are dominated by supersonic 
turbulence, then it is far more difficult than is often appreciated to infer
details of their three-dimensional structure from emission or absorption
line observations. Because we see clouds in projection, we have no direct
information about the line-of-sight positions of emission or absorption features,
only about their radial {\em velocities}. In a supersonically turbulent cloud,
random velocity variations along the line of sight can create coherent 
features in position-position-velocity (PPV) space that actually correspond to 
multiple, physically-separated regions in real space, or, conversely, can
break up a real physical feature into multiple distinct features in PPV 
space (\citealt{bpm02}; see also \citealt{khm00}, \citealt{hmk01}, \citealt{fed09}). 
To properly interpret these 
observations, we need to be able to compare them with {\em simulated 
observations} produced from numerical cloud models that capture the full 
dynamical and chemical complexity of the clouds and the dense cores 
within them.

Previous attempts to model cloud formation have not properly addressed the
coupling between the cloud dynamics and the cloud chemistry. Most attempts
have focussed either on a detailed treatment of cloud chemistry within the framework 
of a highly simplified dynamical model \citep[e.g.][]{hp99,hp00,ki00,ki02,berg04}
or on a detailed treatment of  the dynamics, but without any treatment of the chemistry 
\citep[e.g.][]{bal04,kn04,slyz05,ha07,henn08,ban09}. Recently, 
some studies have begun to treat the formation of molecular hydrogen ($\mHt$)
in high-resolution, three-dimensional simulations of cloud formation 
\citep{dbp06,db07,gm07a,gm07b,dgck08}. 
This is a useful step forward, but although $\mHt$ makes up most of the mass of 
a giant molecular cloud (GMC), it is very difficult to observe, owing to the weakness 
of its rotational lines and their large energy separations. Much of what we know about 
molecular clouds comes instead from observations of carbon monoxide (CO), but until 
now there has  been no attempt to model the far more complex chemistry of CO in a 
high-resolution three-dimensional simulation.

In this paper, we present a lightweight treatment of gas-phase chemistry and radiative
cooling that we have developed to tackle this problem. Our treatment is outlined in
Section~\ref{method} and includes a simplified model for CO formation and destruction 
that tracks the abundances of thirty-two distinct chemical species (\S\ref{basic}),
an approximate treatment of $\mHt$ self-shielding and dust extinction that uses the ``six-ray'' 
approximation of \citet{gm07a,gm07b} (\S\ref{photochem}) and a detailed cooling function
(\S\ref{cool_func}). We show the results of some tests of the model in Section~\ref{tests}, 
and discuss the initial conditions used for our simulations in Section~\ref{initial_conditions}. 
In Sections~\ref{chem_time_evol}--\ref{spatial}, we present a few preliminary results of these 
simulations, concerning the time evolution of the mean chemical abundances 
(\S\ref{chem_time_evol}), the density and temperature probability distribution functions 
(\S\ref{pdfs}), and the spatial distribution of molecular hydrogen and CO (\S\ref{spatial}). 
We conclude with a brief summary in Section~\ref{summ}.


\section{Method}
\label{method}
\subsection{Basic framework}
\label{basic}
We solve the equations of fluid flow for a magnetised interstellar gas using a 
modified version of
the ZEUS-MP hydrodynamical code \citep{norman00,hayes06}. An earlier version
of this modified code was presented in \citet{gm07a}. In the present version, we have
updated and extended the cooling function, as described in \S\ref{cool_func} 
below. We have also improved and significantly extended our treatment of gas phase
chemistry. We now track the abundances of thirty-two species. Thirteen of these species -- 
$\Hm$, $\mHtp$, $\htp$, $\ch^{+}$, $\ch_{2}^{+}$, $\oh^{+}$, $\hto^{+}$, 
${\rm H_{3}O^{+}}$, $\co^{+}$, ${\rm HOC^{+}}$, $\Om$, $\Cm$ and ${\rm O_{2}^{+}}$
-- are assumed to be instantaneously in chemical equilibrium. For the remaining nineteen
species -- $\me$, $\Hp$, $\mH$, $\mHt$, $\He$, $\Hep$, $\mC$, $\Cp$, $\mO$, 
$\Op$, $\oh$, $\hto$, $\co$, ${\rm C_{2}}$, $\mO_{2}$, ${\rm HCO^{+}}$, $\ch$, 
$\ch_{2}$ and $\ch_{3}^{+}$ -- we follow the full non-equilibrium evolution.

As in \citet{gm07a}, we represent the abundance of  each of the non-equilibrium 
species with a tracer field that advects as a density. To ensure consistent advection
of the chemical species, such that elemental abundances are conserved locally,
as well as globally, we use a modified version of the Consistent Multi-fluid Advection
(CMA) algorithm of \citet{pm99},
described in Appendix~\ref{cma}. The chemical rate equations governing the 
creation and destruction of these species are solved in an operator-split fashion, 
using the implicit integrator DVODE \citep{bbh89}. The gas energy equation is also 
operator-split: the effects of any radiative heating or cooling of the gas are combined 
with those of the pressure-work term into a rate equation that is coupled to the chemical
rate equations and so is solved implicitly by DVODE, while the advection of the
gas energy density is handled as in the unmodified ZEUS-MP code.  For reasons 
of computational efficiency, we use conservation laws for charge and elemental 
abundance to determine the abundances of $\me$, H, He, C and O,  reducing the 
number of rate equations that DVODE must solve to fifteen (fourteen chemical rate
equations plus the energy equation). 

To model the chemistry of the gas, we use a chemical network consisting of 218 reactions
between 32 species. Details of the reactions included in the network, along with the rate
coefficients adopted in each case, are given in Tables~\ref{chem_gas}--\ref{tab:cosmic}
in Appendix B.  Note that in our present study, we include no grain surface reactions other 
than the formation of $\mHt$ (reaction 165). The chemical effects of grain surface 
recombination and the freeze-out of CO, ${\rm H_{2}O}$, etc., in dense gas will be 
treated in future work. 

\subsection{Photochemistry}
\label{photochem}
To model the photochemistry of optically thin gas, we assume an incident radiation
field corresponding to the standard interstellar radiation field, as determined by
\citet{d78}. This field has a strength $G_{0} = 1.7$ in \citet{habing68} units, corresponding
to an integrated flux of $2.7 \times 10^{-3} \: {\rm erg} \: {\rm cm^{-2}} \: {\rm s^{-1}}$.
Photodissociation and photoionization rates appropriate for this field strength are
listed in Table~\ref{chem_photo}. 

In optically thick gas, it is necessary to account for absorption by dust, by the
Lyman-Werner lines of $\mHt$ (important for $\mHt$ self-shielding and CO
shielding), and by the ultraviolet absorption lines of CO (important for CO
self-shielding). To treat these processes with full accuracy in a three-dimensional
hydrodynamic simulation is beyond our current capabilities. The problem is
one of computational cost: in a hydrodynamical simulation with $N$ fluid elements,
the cost of resolving a monochromatic radiation field with the same spatial resolution,
and with a comparable angular resolution, is of order $O(N^{5/3})$, as we require 
$O(N^{2/3})$ rays to fully sample the angular distribution of the radiation field for each
of the $N$ fluid elements. To model line
absorption, the radiation field must also be discretised in frequency-space, increasing
the cost to $O(N_{\nu} \times N^{5/3})$, where $N_{\nu}$ is the number of frequency
bins required. In comparison, the cost of modelling the hydrodynamics is only of
order $O(N)$. Thus, the cost of properly treating the radiation field in an 
optically thick $128^{3}$ zone hydrodynamical simulation is 
$128^{2} N_{\nu}$ times greater than the cost of solving for the hydrodynamical 
evolution, which is far out of the reach of our current computational resources.

To overcome this problem, we are forced to approximate. The approach that
we have chosen to adopt is the ``six-ray'' method used in \citet{gm07a,gm07b},
which is based on an original idea by \citet{nl97}. In this approximation, we 
compute photochemical rates in each zone in our simulation volume by 
averaging over a small number of lines of sight. Specifically, we compute
the column densities of $\mHt$, CO, and H nuclei in all forms (i.e. $N_{\rm H, tot}
\simeq N_{\rm H} + 2N_{\mHt} + N_{\Hp}$, where $N_{\rm H}$, $N_{\mHt}$
and $N_{\Hp}$ are the column densities of atomic H, molecular $\mHt$, and
protons, respectively) in both the positive and negative directions along each 
of the three coordinate axes of the simulation. For each zone we therefore
know the column densities in six different directions (or, alternatively, along
six rays).

The rates of most of the reactions listed in Table~\ref{chem_photo} are sensitive
only to the amount of dust extinction, and for a plane-parallel slab geometry,
the rates in optically thick gas are related to those in optically thin gas by
the expression
\begin{equation}
R_{\rm thick} = R_{\rm thin} f_{\rm dust} \equiv  R_{\rm thin} \exp(-\gamma A_{\rm V}),
\label{thick}
\end{equation}
where $R_{\rm thin}$ is the optically thin rate,  $f_{\rm dust} \equiv \exp(-\gamma A_{\rm V})$
is the dust shielding factor, $A_{\rm V}$ is the visual extinction 
in magnitudes, and where the appropriate value of $\gamma$ for each reaction 
is listed in Table~\ref{chem_photo}. Using the relationship 
\begin{equation}
A_{\rm V} = \frac{N_{\rm H, tot}}{1.87 \times 10^{21} \: {\rm cm^{-2}}}
\end{equation}
between $A_{\rm V}$ and $N_{\rm H, tot}$, as is appropriate for gas in the diffuse
ISM \citep{db96}, we can associate a value of $A_{\rm V}$ with each of our six rays for any 
given zone in our simulation volume. We can then calculate the total rate in that zone,
$R_{\rm thick, i}$, as
\begin{equation}
R_{\rm thick, i} = \frac{1}{6} R_{\rm thin} \sum_{j=1}^{6}  \exp(-\gamma A_{{\rm V}, j}),
\end{equation}
where $A_{{\rm V}, j}$ is the visual extinction along ray $j$. 

To treat the photodissociation of ${\rm H_{2}O^{+}}$ and ${\rm H_{3}O^{+}}$
(reactions 188--195), we use a very similar approach. However, for these reactions
the optically thick and optically thin rates are related by the more complicated
expression \citep{sd95},
\begin{equation}
R_{\rm thick}  = R_{\rm thin} \exp(-2.55 A_{\rm V} + 0.0165 A_{\rm V}^{2}),
\end{equation}
for $A_{\rm V} \leq 15$, and by Equation~\ref{thick} with $\gamma = 2.8$
for $A_{\rm V} > 15$. Hence, the total rate in zone $i$ is given in this case by
\begin{equation}
R_{\rm thick, i}  = \frac{1}{6} R_{\rm thin} \sum_{j=1}^{6} \left[ f_{\rm A} \exp(-2.55 A_{{\rm V}, j} 
+ 0.0165 A_{{\rm V}, j}^{2}) + (1 - f_{\rm A}) \exp(-2.8 A_{{\rm V}, j}) \right],
\end{equation}
where $f_{\rm A} = 1$ for $A_{{\rm V}, j} \leq 15$, and $f_{\rm A} = 0$ for $A_{{\rm V}, j} > 15$, 
and where  $A_{{\rm V}, j}$ is computed in the same manner as before.

To treat $\mHt$ photodissociation accurately, one must take into account not
only absorption by dust, but also the effects of $\mHt$ self-shielding. If we assume
that the effects of dust absorption and self-shielding can be treated separately, then
for a plane-parallel geometry the effects of dust absorption can be modeled by a 
dust shielding factor given by Equation~\ref{thick}, with $\gamma = 3.74$
\citep{db96}.
Similarly, the effects of $\mHt$ self-shielding can be modeled by a factor
\citep{db96}
\begin{equation}
f_{\rm shield} = \frac{0.965}{(1 + x/b_{5})^{2}} + \frac{0.035}{(1 + x)^{1/2}}
 \exp\left[-8.5 \times 10^{-4} (1+x)^{1/2}\right], \label{fsh}
\end{equation}
where $x = N_{\mHt} / (5 \times 10^{14} \: {\rm cm}^{-2})$, and 
$b_{5} = b / (10^{5} \: {\rm cm} \:{\rm s}^{-1})$, where $b$ is the Doppler 
broadening parameter. The  fully shielded $\mHt$ photodissociation rate
then follows as
\begin{equation}
R_{\rm H_{2},  thick} = f_{\rm dust} f_{\rm shield} R_{\rm H_{2}, thin}.
\end{equation}
Using our six-ray approximation, we can compute $f_{\rm dust}$
and $f_{\rm shield}$ for each ray, and hence can compute the total rate as
\begin{equation}
R_{\rm H_{2}, thick, i} = \frac{1}{6} R_{\rm H_{2}, thin}
\sum_{j=1}^{6} f_{{\rm dust}, j} f_{{\rm shield}, j} .
\end{equation}
Finally, to treat CO photodissociation, it is necessary to take account of three
separate contributions to the shielding: CO self-shielding, shielding of CO by
the $\mHt$ Lyman-Werner lines, and dust shielding. For a plane-parallel geometry, 
the  shielded CO photodissociation rate is related to the optically thin rate by 
\begin{equation}
R_{\rm CO, thick} = f_{\rm CO} f_{\rm H_{2}} f_{\rm dust} R_{\rm CO, thin},
\end{equation}
where $f_{\rm CO}$ and $f_{\rm H_{2}}$ are functions of the CO and $\mHt$
column densities, respectively, and have been tabulated by \citet{lee96},
and where $f_{\rm dust}$ is given by Equation~\ref{thick}, using the value
of $\gamma$ from Table~\ref{chem_photo}. Using our six-ray approximation, we can 
compute values of 
$f_{\rm CO}$, $f_{\rm H_{2}}$ and $A_{\rm V}$ for each ray, from which the partial
contribution from that ray to the total rate follows as
\begin{equation}
R_{\rm CO, thick, i} = \frac{1}{6} R_{\rm CO, thin} \sum_{j=1}^{6}
f_{{\rm CO}, j} f_{{\rm H_{2}}, j} f_{{\rm dust}, j}.
\end{equation}

\subsection{Thermal model}
\label{cool_func}
We model the radiative and chemical heating and cooling of the gas with a cooling 
function that contains contributions from 18 different processes, listed in 
Table~\ref{cool_model}. Our treatment of most of these processes largely
follows that in \citet{gm07a}, to which we refer readers desiring further details;
the few exceptions are noted below.

\begin{table}
\centering
\begin{tabular}{ll}
\hline
Process & Reference(s) \\
\hline
{\bf Cooling:} & \\
C fine structure lines & Atomic data -- \citet{sv02} \\
& Collisional rates ($\mH$) -- \citet{akd07} \\
& Collisional rates ($\mHt$) -- \citet{sch91} \\
& Collisional rates ($\me$) -- \citet{joh87} \\
& Collisional rates ($\Hp$) -- \citet{rlb90} \\
$\Cp$ fine structure lines &  Atomic data -- \citet{sv02} \\
&  Collisional rates ($\mHt$) -- \citet{fl77}  \\
&  Collisional rates ($\mH$, $T < 2000 \: {\rm K}$) -- \citet{hm89} \\
&  Collisional rates ($\mH$, $T > 2000 \: {\rm K}$) -- \citet{k86} \\
&  Collisional rates (${\rm e^{-}}$) -- \citet{wb02} \\
O fine structure lines &  Atomic data -- \citet{sv02} \\
& Collisional rates ($\mH$) -- \citet{akd07} \\
& Collisional rates ($\mHt$) -- see \citet{gj07} \\ 
& Collisional rates (${\rm e^{-}}$) -- \citet{bbt98} \\
& Collisional rates ($\Hp$) -- \citet{p90,p96} \\
$\mHt$ rovibrational lines & \citet{lpf99} \\
CO and ${\rm H_{2}O}$ rovibrational lines & \citet{nk93,nlm95} \\
OH rotational lines & \citet{pav02} \\
Gas-grain energy transfer & \citet{hm89} \\
Recombination on grains & \citet{w03} \\
Atomic resonance lines & \citet{sd93} \\
$\mH$ collisional ionisation& \citet{a97} \\
$\mHt$ collisional dissociation & See Table~\ref{chem_gas} \\
Compton cooling & \citet{cen92} \\
\hline
{\bf Heating:} & \\
Photoelectric effect & \citet{bt94,w03} \\
$\mHt$ photodissociation & \citet{bd77} \\ 
UV pumping of $\mHt$ & \citet{bht90}  \\
$\mHt$ formation on dust grains & \citet{hm89} \\
Cosmic ray ionisation & \citet{gl78}  \\
\hline
\end{tabular}
\caption{Processes included in our thermal model. \label{cool_model}}
\end{table}

\subsubsection{Fine structure cooling}
The extremely simplified model of ISM chemistry presented in \citet{gm07a} 
assumed that all of the carbon in the gas would be kept in singly ionized form 
by the interstellar radiation field, and so in that paper there was no need to 
treat cooling from the fine structure lines of neutral carbon.
However, the significantly improved treatment of ISM chemistry
presented in this paper removes this assumption and so it is necessary
to include them in our thermal model. To do this, we largely follow the
same prescription as in \citet{gj07}. The one exception is in the rates for
the collisional excitation of the fine structure lines of atomic carbon by collisions with
atomic hydrogen. \citet{gj07} use rates for this process that were taken
from \citet{hm89}, but in this work we use instead the more accurate rates recently
calculated by \citet{akd07}. We also adopt their new rates
for the excitation of atomic oxygen by atomic hydrogen in place of the rates
used in our previous work.

\subsubsection{${\rm CO}$ and ${\rm H_{2}O}$ cooling}
\label{co-cooling-section}
To treat rotational cooling from CO and ${\rm H_{2}O}$, we use the tabulated cooling 
functions of \citet{nk93} and \citet{nlm95}. They use a large velocity gradient (LVG) 
approach to compute the cooling rates from CO and ${\rm H_{2}O}$ as a function of 
temperature, density and effective optical depth. Based on the results of their 
calculations, they present fits to the CO and ${\rm H_{2}O}$ rotational cooling 
functions of the form
\begin{equation}
\frac{1}{L_{\rm m}} = \frac{1}{L_{0}} + \frac{n_{\mHt}}{L_{\rm LTE}} + 
\frac{1}{L_{0}}\left[\frac{n_{\mHt}}{n_{1/2}}\right]^{\alpha} 
\left(1 - \frac{n_{1/2}L_{0}}{L_{\rm LTE}} \right), \label{nlm_cool}
\end{equation}
where $L_{\rm m}$ is a cooling rate coefficient defined such that the cooling 
rate per unit volume from species m (where m = CO or ${\rm H_{2}O}$) is 
given by $\Lambda_{\rm m} = L_{\rm m} n_{\rm m} n_{\mHt}$, and where
$L_{\rm 0}$ is the cooling rate coefficient in the low density limit, 
$L_{\rm LTE}$ is the cooling rate per molecule when the rotational level
populations are in local thermodynamic equilibrium (LTE), and $n_{1/2}$
is the $\mHt$ number density at which $L_{\rm m} = 0.5 L_{\rm 0}$. 
$L_{0}$ is purely a function of temperature, but the other three fit 
parameters ($L_{\rm LTE}$, $n_{1/2}$ and $\alpha$) depend on both the
temperature and the effective optical depth of the gas. \citet{nlm95}
parameterise the latter in terms of an effective column density per unit velocity,
$\tilde{N}({\rm m})$, for each coolant ${\rm m}$. For a flow without any special 
symmetry in which the LVG approximation applies, \citet{nk93} give this parameter 
as  
\begin{equation}
\tilde{N}({\rm m}) = \frac{n({\rm m})}{|{\bf{\nabla \cdot v}}|}.
\end{equation}
For CO rotational cooling, \citet{nk93} and \citet{nlm95}  tabulate values of 
the fitting parameters for temperatures in the range $10 \: {\rm K} < T < 2000 \: {\rm K}$  
and effective column densities in the range $14.5 < \log{\tilde{N}({\rm CO})} < 19.0$,
where $\tilde{N}({\rm CO})$ has units of cm$^{-2}$ per km s$^{-1}$.
For ${\rm H_{2}O}$ rotational cooling, they tabulate values for temperatures
$10 \: {\rm K} < T < 4000 \: {\rm K}$  and optical depth parameters 
$10.0 < \log{\tilde{N}({\rm H_{2}O})} < 19.0$. At low temperatures, \citet{nlm95}
list values of the cooling rate for both the ortho and the para variants of
${\rm H_{2}O}$. To compute the total water cooling rate, we assume that 
the ortho:para ratio is fixed at 3:1.

To properly treat gas below $10 \: {\rm K}$, we would not only have to extend the
\citet{nlm95} cooling functions to lower temperatures, but would also have to
take into account several other physical processes that are not included in our
current model, such as the freeze-out of CO and water onto dust grains, or the
fact that  the dust temperature in the cloud will decrease as the extinction 
increases \citep{gold01}. To avoid this, in the simulations presented here we have 
introduced an artificial temperature floor at $10 \: {\rm K}$ and switch off radiative
cooling in gas colder than this.

To account for cooling from CO and water at very high temperatures ($T > 2000 
\: {\rm K}$ for CO, $T > 4000 \: {\rm K}$ for ${\rm H_{2}O}$), we adopt cooling 
rates that are the same as the rates at the largest tabulated temperature. As
we expect CO and ${\rm H_{2}O}$ to be rapidly destroyed by collisional 
dissociation and chemical reactions with atomic hydrogen at these high 
temperatures, any uncertainty in the cooling rates per CO or ${\rm H_{2}O}$ 
molecule in this regime will not have a large effect on the total cooling rate,
owing to the small molecular abundances. This approximation should 
therefore be reasonable for most uses.
Nevertheless, a treatment of CO and ${\rm H_{2}O}$ 
cooling along the same lines as \citet{nk93} and \citet{nlm95}, but which 
extended to higher temperatures, would clearly be desirable.

To treat gas with $\tilde{N}$ below the tabulated range, we simply adopt 
the same fitting parameters as are given for the smallest value of $\tilde{N}$
that is tabulated. As the latter generally corresponds to gas that is
already very close to the optically thin limit, this assumption should give
accurate results. To handle the case where $\tilde{N}$ in the simulation
exceeds the largest tabulated value, we again use rates corresponding
to the largest value that is tabulated. Consequently, we will overestimate
the cooling rate in very dense, highly shielded gas, particularly when the
velocity divergence of this gas is small. However, as this gas is unlikely
to be well-resolved in our simulations in any case, this simplification is
again unlikely to introduce large uncertainties into our results.

The \citet{nk93} and \citet{nlm95} treatments assume that only collisions with
$\mHt$ are important in determining the CO or ${\rm H_{2}O}$ rotational
cooling rates, as is appropriate in a fully molecular gas. However, at early
times or at low gas densities, the $\mHt$ abundance is small, and collisions
with atomic hydrogen or with electrons may also become important. We therefore
follow \citet{yan97} and \citet{ms05} and replace $n_{\mHt}$ in 
Equation~\ref{nlm_cool} with an effective number density $n_{\rm eff}$. For
CO rotational cooling, $n_{\rm eff}$ is given by
\begin{equation}
n_{\rm eff, CO, rot} = n_{\mHt} + \sqrt{2}
\left(\frac{\sigma_{\rm H}}{\sigma_{\mHt}}\right)  n_{\mH} 
+ \left(\frac{1.3 \times 10^{-8} \: {\rm cm^{-1}} \: {\rm s^{-1}}}{\sigma_{\mHt}v_e}\right) n_{\rm e},
\end{equation}
where $\sigma_{\mH} = 2.3 \times 10^{-15} \: {\rm cm^{-2}}$, 
$\sigma_{\mHt} = 3.3 \times 10^{-16} (T / 1000 \: {\rm K})^{-1/4}
\: {\rm cm^{-2}}$, and $v_e = 1.03 \times 10^{4} \, (T / 1 \: {\rm K})^{1/2} \: {\rm cm} \: {\rm s^{-1}}$. 
For ${\rm H_{2}O}$ rotational cooling, $n_{\rm eff}$ is given by
\begin{equation}
n_{\rm eff, H_{2}O, rot} = n_{\mHt} + 10 \, n_{\mH} + \left( \frac{k_{\rm e}}{k_{\mHt}} \right) n_{\rm e},
\end{equation}
where $k_{\mHt} = 7.4 \times 10^{-12} T^{1/2} \: {\rm cm^{3}} \: {\rm s^{-1}}$,
and $k_{\rm e}$ is given by
\begin{equation}
k_{\rm e} = {\rm dex} \left[-8.020 + 15.749 / T^{1/6} - 47.137 / T^{1/3} + 76.648 / T^{1/2} - 60.191 / T^{2/3} \right].
\end{equation}
These formulae for $n_{\rm eff}$ are adapted from those in \citet{ms05}, with one exception: the expression
for $k_{\rm e}$ is taken from \citet{fgt04}, because the expression given by \citet{ms05} for $k_{\rm e}$ blows 
up at low temperatures.

To treat CO and ${\rm H_{2}O}$ vibrational cooling, we again use the
results of \citet{nk93}. They present fitting functions of the form
\begin{equation}
\frac{1}{L_{\rm M}} = \frac{1}{L_{0}} + \frac{n_{\mHt}}{L_{\rm LTE}}
\label{nlm_cool_2}
\end{equation}
for both CO and ${\rm H_{2}O}$, and give analytical fits to $L_{0}(T)$
for both coolants, as well as numerical values for $L_{\rm LTE}$ for
temperatures $100 \: {\rm K} < T < 4000 \: {\rm K}$ and effective column 
densities $13.0 < \log{\tilde{N}} < 20.0$. Our treatment of cooling 
at temperatures and effective column densities that lie outside 
these bounds is the same as that used for CO and ${\rm H_{2}O}$ 
rotational cooling, as described above. As before, we account for
CO and ${\rm H_{2}O}$ cooling in gas that is not fully molecular
by replacing $n_{\mHt}$ in Equation~\ref{nlm_cool_2} with an
effective number density $n_{\rm eff}$, taken from \citet{ms05}.
For CO vibrational cooling, this is given by
\begin{equation}
n_{\rm eff, CO, vib} = n_{\mHt} + 50 \, n_{\mH} + \left( \frac{L_{\rm CO, e}}{L_{\rm CO, 0}}
\right) n_{\rm e},  
\end{equation}
where
\begin{eqnarray}
L_{\rm CO, e} & = & 1.03 \times 10^{-10} \left(\frac{T}{300}\right)^{0.938} \exp \left(\frac{-3080}{T}\right), \\
L_{\rm CO, 0} & = & 1.14 \times 10^{-14} \exp \left(\frac{-68.0}{T^{1/3}} \right)  \exp \left(\frac{-3080}{T}\right),
\end{eqnarray}
while for $\hto$ vibrational cooling we have
\begin{equation}
 n_{\rm eff, \hto, vib} = n_{\mHt} + 10 n_{\mH} + \left( \frac{L_{\rm \hto, e}}{L_{\rm \hto, 0}} \right) n_{\rm e},  
\end{equation}
where
\begin{eqnarray}
L_{\rm \hto, e} & = & 2.6 \times 10^{-6} T^{-1/2} \exp \left(\frac{-2325}{T}\right), \\
L_{\rm \hto, 0} & = & 0.64 \times 10^{-14} \exp \left(\frac{-47.5}{T^{1/3}} \right)  
\exp \left(\frac{-2325}{T}\right),
\end{eqnarray}
and where in all of these expressions, $T$ is the gas temperature in K.

Finally, we note that when $\tilde{N}$ is large, isotopic variants of
CO and ${\rm H_{2}O}$ can contribute significantly to the total
cooling rate, as they will have much smaller effective column 
densities and so will be less affected by self-absorption. To 
model cooling from the isotopic species $^{13}{\rm C}^{16}{\rm O}$, 
$^{12}{\rm C}^{18}{\rm O}$ and ${\rm H_{2}}^{18}{\rm O}$, we
follow \citet{nlm95} and assume that the cooling rate coefficients
for these isotopic species are the same as for the standard variants
$^{12}{\rm C}^{16}{\rm O}$ and ${\rm H_{2}}^{16}{\rm O}$, and 
that they have abundance ratios of 1\%, 0.2\% and 0.2\%, 
respectively, compared to the standard variants.
We do not include the effects of other isotopic variants (e.g.\
deuterated water, ${\rm HDO}$ or ${\rm D_{2}O}$) as we 
expect their abundances to be too small for them to contribute
significantly.

\subsubsection{${\rm OH}$ cooling}
To model cooling from OH, we use a rate taken from \citet{pav02},
based on \citet{hm79}. This rate assumes that the OH molecules
are not in LTE, which is a reasonable assumption provided that the
gas density $n < 10^{10} \: {\rm cm^{-3}}$ \citep{hm79}. 

\section{Code tests}
\label{tests}
In testing our modified version of ZEUS-MP, our main focus was on verifying the 
additional physics that we have added to the code, as the unmodified code has 
already undergone significant testing \citep[see e.g.][]{sn92a,sn92b,hayes06}.

To verify that our simplified model of CO formation and destruction performs as
expected, we have used our chemical network and cooling function to model
the chemical and thermal evolution of static gas for a range of different number
densities and obscuring average column densities (or visual extinctions). 
We then compared the results of these 
single zone models with the results of similar simulations performed using a
detailed chemical network derived from the UMIST99 compilation of reaction rates
\citep{TEU00}, and using the same cooling function. By modelling the thermal
as well as the chemical evolution of the gas, we are able to quantify the effect
of errors in the chemical abundances of the major coolants on the thermal
state of the gas. In all of these tests, we use the same incident radiation field,
cosmic ray ionization rate, elemental abundance of carbon and oxygen, etc.\
as in our three-dimensional simulations described below. We ran each of our
test models for a total time of $3.1 \times 10^{14} \: {\rm s}$, or about 10 million
years.

Our initial comparisons showed significant discrepancies between the results
obtained using our simplified chemical model and the UMIST99-derived model.
The abundances of the dominant carbon and oxygen carrying species were 
generally reproduced accurately in both models, but we found large differences
in the abundances of some of the trace species.
However, further investigation showed that these discrepancies were caused 
primarily by differences in the reaction rate coefficients adopted for several key 
reactions in our model compared to those used in the UMIST model. Specifically,
we found that to get good agreement between the models, it was necessary to
ensure that the same values were used for the rates of the following reactions:
H$^{+}$ recombination (reaction 12); He$^{+}$ recombination (reaction 17); 
charge transfer from He$^{+}$ to H (reaction 18);  the destruction of CH by atomic 
hydrogen (reaction 35); the formation of CO from C and OH (reaction 46);
the formation of O$_{2}$ from O and OH (reaction 47);
dissociative charge transfer from He$^{+}$ to CO (reaction 104);
H$_{3}^{+}$ dissociative recombination (reactions 110--111); 
CH$^{+}$ dissociative recombination (reaction 112);
H$_{2}$O$^{+}$ dissociative recombination (reactions 120--122);
H$_{3}$O$^{+}$ dissociative recombination (reactions 123--126);
H$_{2}$ formation on dust grains (reaction 165),
and, lastly, H$_{2}$ photodissociation (reaction 168).
In a couple of cases, the differences in rate coefficients are
a result of the use of a different low temperature extrapolation from the same 
experimental data, but in most cases, the differences are due to our use of
recent experimental or theoretical values that post-date the construction of
the UMIST99 model. We therefore expect the values used in our model (and
listed in Table~\ref{chem_gas}) to be the more accurate ones. We also note
that we are not the first group to remark on the sensitivity of astrochemical
reaction networks to the large uncertainties that exist in some reaction rate
coefficients \citep[see e.g.][]{vas04,wake05}, and that uncertainties of this kind in the input
physics are an important, but currently unavoidable, limitation on the accuracy of
simulations of molecular cloud chemistry.

\begin{figure}
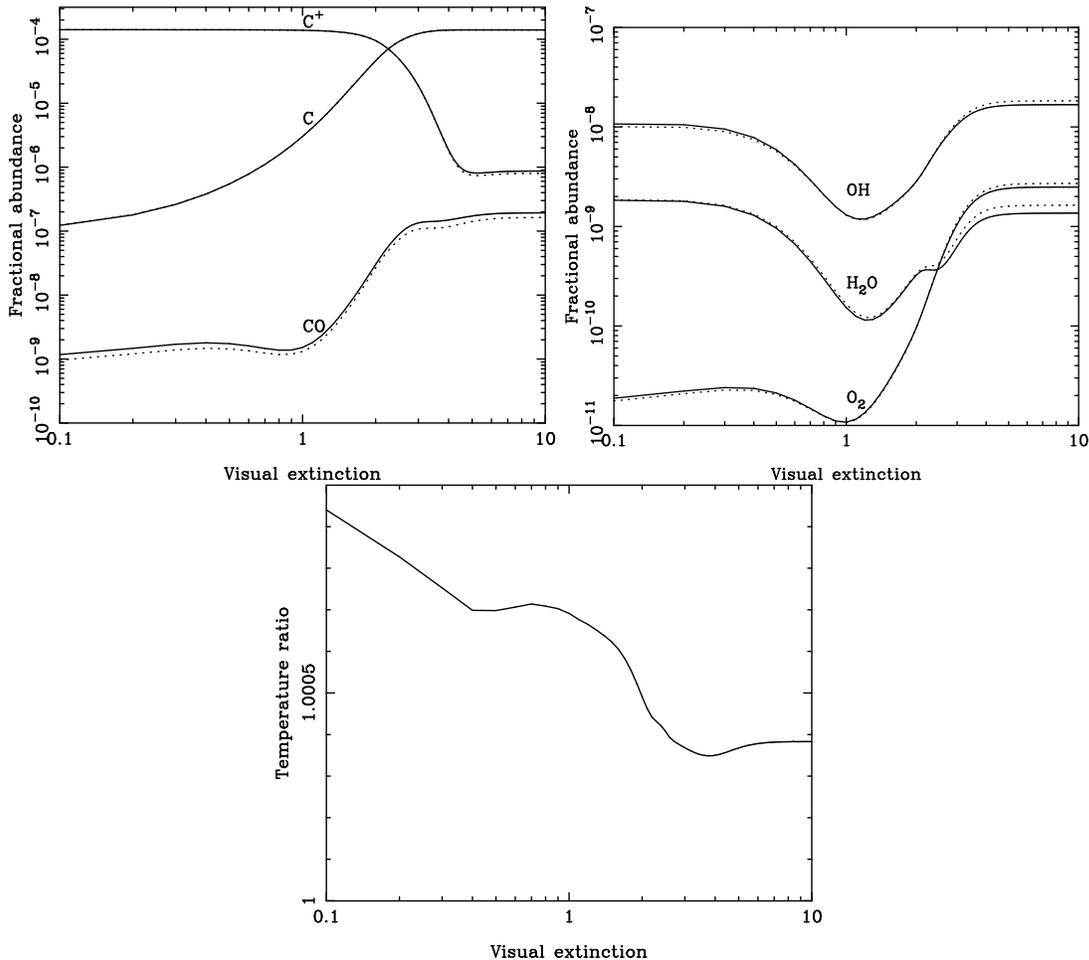

\centering
\epsfig{figure=f1a.eps,width=15pc,angle=270,clip=}
\epsfig{figure=f1b.eps,width=15pc,angle=270,clip=}
\epsfig{figure=f1c.eps,width=15pc,angle=270,clip=}
\caption{
(a)  Abundances of C$^{+}$, C and CO, plotted as a 
function of  $A_{\rm V}$, at the end of our static, single-zone
simulations with $n_{0} = 100 \: {\rm cm^{-3}}$. The results 
produced by our simplified chemical model are given as dotted
lines, while the results of the UMIST model are shown by 
solid lines.
(b) As (a), but for the OH, H$_{2}$O and O$_{2}$ abundances.
(c) As (a), but showing the ratio of gas temperature produced by the simplified
model to that produced by the full UMIST model. Note that we plot the ratio 
rather than the individual temperatures because the difference between the
models is very small, of the order of 0.05\%.
\label{test-n100}}
\end{figure}

\begin{figure}
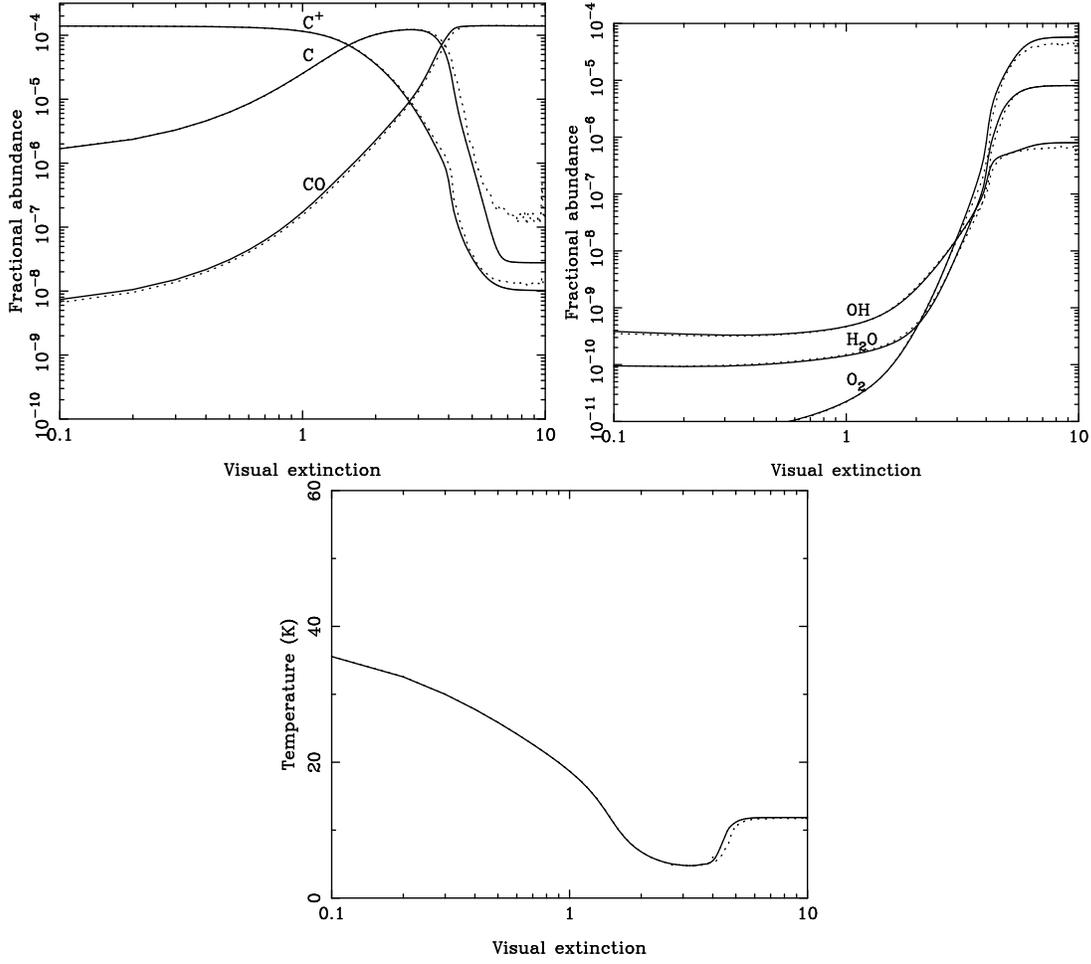

\centering
\epsfig{figure=f2a.eps,width=15pc,angle=270,clip=}
\epsfig{figure=f2b.eps,width=15pc,angle=270,clip=}
\epsfig{figure=f2c.eps,width=15pc,angle=270,clip=}
\caption{(a) As Figure~\ref{test-n100}a, but for 
$n_{0} = 1000 \: {\rm cm^{-3}}$.
(b) As (a), but for the OH, H$_{2}$O and O$_{2}$ abundances.
(c) As (a), but for the gas temperature $T$. 
 \label{test-n1000}}
\end{figure}

\begin{figure}
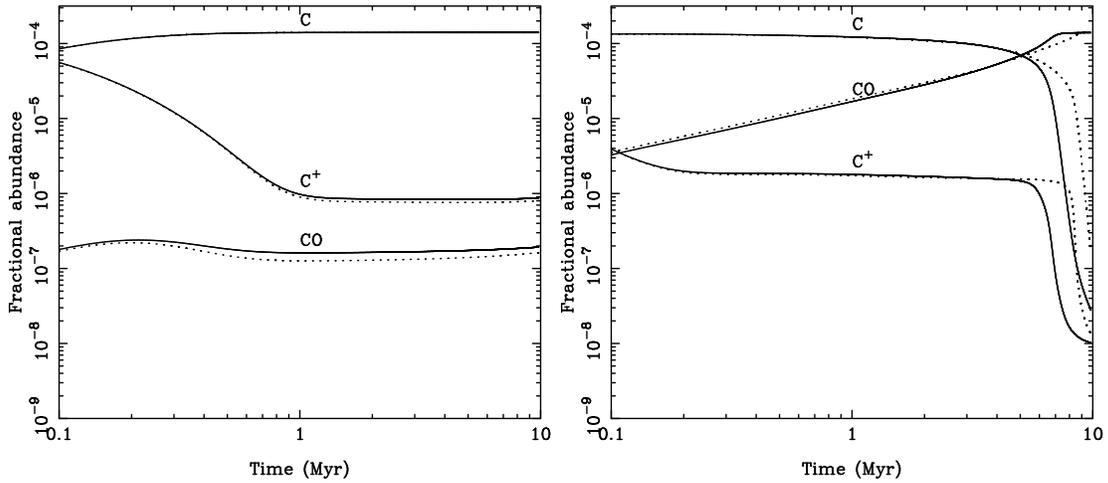

\centering
\epsfig{figure=f3a.eps,width=15pc,angle=270,clip=}
\epsfig{figure=f3b.eps,width=15pc,angle=270,clip=}
\caption{(a) Time evolution of the C$^{+}$, C and CO abundances in a test run
with $n_{0} = 100 \: {\rm cm^{-3}}$ and $A_{\rm V} = 10$.
(b) As (a), but for $n_{0} = 1000 \: {\rm cm^{-3}}$
\label{test-time}}
\end{figure}

Having made these adjustments, so that we are comparing like with like,
we find good agreement between the results produced using our simplified
chemical network and the results produced using the full UMIST network.
This is illustrated in Figures~\ref{test-n100}--\ref{test-time}. In Figure~\ref{test-n100}a, 
we show how the C$^{+}$, C and CO abundances vary as a function of $A_{\rm V}$ 
at the end of our test runs for an initial density of $n_{0} = 100 \: {\rm cm^{-3}}$. 
The results from our simplified model are plotted as dotted lines, while those from the UMIST 
model are plotted as solid lines. Figure~\ref{test-n100}b gives a similar
comparison of the OH, H$_{2}$O and O$_{2}$ abundances, while in Figure~\ref{test-n100}c 
we plot the ratio of the gas temperature in the simplified model to that in the full model.
We find very good agreement for all of the plotted abundances at all $A_{\rm V}$, and
only very small differences in the temperature. 

In Figure~\ref{test-n1000}, we give a similar comparison for the case of gas with
$n_{0} = 1000 \: {\rm cm^{-3}}$. Again we find good agreement for most species,
although in this case the simplified model predicts too large an abundance of 
neutral carbon at $A_{\rm V} > 6$. However, this disagreement is possibly a little
misleading. As Figure~\ref{test-time} demonstrates, the evolution of the C abundance with time
at high $A_{\rm V}$ is very similar in both the simplified and the full models, but the
final conversion of the residual C and C$^{+}$ to CO occurs slightly later in the 
simplified model. It should also be noted that we expect our neglect of the effects
of freeze-out to have a larger impact on the gas-phase chemistry of high $A_{\rm V}$
gas than the small discrepancy noted here.

Finally, regarding the cooling function, we note that most of its features
are already well-tested, as described in \citet{gm07a}. The main
addition that requires testing is our implementation of CO and
H$_{2}$O cooling. We have verified that this is implemented 
correctly by ensuring that we can reproduce all of the tabulated
values for the CO and H$_{2}$O cooling rates given in \citet{nk93}
and \citet{nlm95}, and by ensuring that the rates vary smoothly
in between the tabulated values. 

\section{Initial conditions}
\label{initial_conditions}
In this paper, we present the results of a small set of simulations of supersonic
turbulence designed to address the issue of numerical convergence, and to 
highlight the capabilities of the code. We performed three simulations with 
numerical resolutions of $64^{3}$, $128^{3}$ and $256^{3}$ zones, respectively, 
that we will hereafter denote as runs R1, R2, and R3. All three simulations shared 
the same set of initial conditions: a periodic box of side length $L = 20 \: {\rm pc}$, 
filled with initially uniform atomic gas with a density $n_{0} = 300 \: {\rm cm^{-3}}$ and
a temperature $T_{0} = 60 \: {\rm K}$, permeated by a magnetic field with an
initial field strength $B_{0} = 2 \: \mu {\rm G}$ oriented parallel to the $z$-axis 
of the box. The initial turbulent velocity field had an RMS velocity of 
$v_{\rm rms} = 5 \: {\rm km} \: {\rm s^{-1}}$ and was constructed in the same fashion as
in \citet{gm07b}. The turbulence was driven as outlined in \citet{mkbs98} and \citet{ml99} with
a driving power $\dot{E} = 2.805 \times 10^{35} \: {\rm erg} \: {\rm s^{-1}}$ so that the RMS 
velocity of the turbulence remained approximately $5 \: {\rm km} \: {\rm s^{-1}}$ throughout 
each of the runs. To allow us to make meaningful comparisons between the
different resolution runs, we ensured that the same turbulence driving pattern 
was used in each case.

As in \citet{gm07a,gm07b}, we adopted standard solar abundances 
of hydrogen and helium, and abundances of carbon and oxygen taken from
\citet{sem00}, i.e.\ $x_{\rm C} = 1.41 \times 10^{-4}$ and $x_{\rm O} = 3.16 \times 10^{-4}$,
where $x_{\rm C}$ and $x_{\rm O}$ are the fractional abundances by number
of carbon and oxygen relative to hydrogen. 

We ran each simulation until a time $t_{\rm end} = 1.8 \times 10^{14} \: {\rm s}
\simeq 5.7 \: {\rm Myr}$. This corresponds to almost three turbulent crossing 
times, $t_{\rm cross} = L / (2 v_{\rm rms}) \simeq 2 \: {\rm Myr}$. As we shall see 
later, this is long enough to allow most of the fluid quantities to reach a state of 
statistical equilibrium.

It should be noted that our simulations are somewhat inconsistent in that they adopt
periodic boundary conditions for the gas, but do not do the same for the radiation.
This is an unfortunate but necessary compromise. Our simulations  do not have
sufficient dynamical range to follow both the process of cloud assembly and the
evolution of gas within the cloud, and so we use periodic boundary conditions for
the gas in order to be able to focus on a small volume of already assembled 
material, while continuing to use a more physically appropriate boundary condition
for the radiation that has it simply penetrating inwards from the edges of the box.


\section{Time evolution of mean chemical abundances}
\label{chem_time_evol}
We begin our discussion of the results of our simulations
by examining the time evolution of the spatially-averaged
mass-weighted mean abundances of
several key chemical species in our three simulations. We define the mass-weighted 
mean abundance of  a species m as
\begin{equation}
\mwfrac{\rm m} = \frac{\sum_{i,j,k} x_{\rm m}(i,j,k) \rho \Delta V(i,j,k)}{M_{\rm tot}},
\end{equation}
where  $x_{\rm m}(i,j,k)$ is the fractional abundance of species m, $\rho(i,j,k)$ is the mass density in 
zone $(i,j,k)$, $\Delta V(i,j,k)$ is the volume of zone $(i,j,k)$, $M_{\rm tot}$ is the total mass of gas
present in the simulation, and where we sum over all grid zones. In all but one case, we define 
the fractional abundance $x_{\rm m}$ of species m as the abundance by number 
of the species relative to the abundance of hydrogen nuclei, i.e.\
\begin{equation}
x_{\rm m} = \frac{n_{\rm m}}{n},
\end{equation}
where $n_{\rm m}$ is the number density of species m and $n$ is the number density of hydrogen
nuclei. The exceptional case is that of $\mHt$, for which this convention would give a value of 
$x_{\mHt} = 0.5$ for gas in which the hydrogen is fully molecular. This has the potential to be
confusing for readers who are unfamiliar with this convention, and so for clarity we define 
the fractional abundance of $\mHt$ to be
\begin{equation}
x_{\mHt} = \frac{2 n_{\mHt}}{n},
\end{equation}
so that $x_{\mHt} = 1$ for fully molecular hydrogen. Note also that this is the same convention
as is used in \citet{gm07a,gm07b}, and so the values of $x_{\mHt}$ and $\mwfrac{\mHt}$ 
discussed here can be directly compared to those in our previous papers. 

\subsection{Molecular hydrogen}
\begin{figure}
\centering
\epsfig{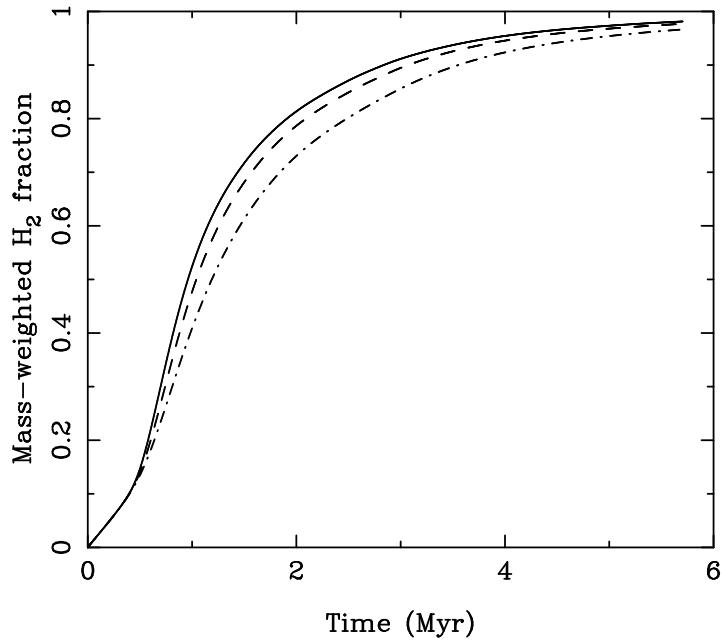}
\caption{Time evolution of the mass-weighted $\mHt$ abundance in 
simulations R1, R2 and R3, which have numerical resolutions of $64^{3}$ 
zones (dot-dashed), $128^{3}$ zones (dashed) and $256^{3}$ zones (solid),
respectively.
\label{h2-abundance}}
\end{figure}

In Figure~\ref{h2-abundance}, we show how the mass-weighted mean abundance of 
$\mHt$, $\mwfrac{\mHt}$, evolves in runs R1, R2, and R3. At early times, we see some 
dependence on the numerical resolution of the simulation. This is probably a 
consequence of the fact that denser structures are formed as we 
increase the numerical resolution, on account of the 
better resolution of turbulent compression \citep[see e.g.][Figure 5]{fed09}. 
As the $\mHt$ formation rate depends on the density, the formation rate thus also increases.

Similar results were found previously by \citet{gm07b}. At late times,
most of the resolution dependence disappears. The value of
$\mwfrac{\mHt}$ in the $64^{3}$ zone run remains slightly smaller than
that in the higher resolution runs, but there is almost no difference
between the results of the $128^{3}$ and $256^{3}$ zone runs,
suggesting that in this case, a numerical resolution of $128^3$
zones is enough to reproduce the final H$_{2}$ abundance accurately.

Regarding the time evolution of the $\mHt$ fraction, we first note that we see the
same rapid growth in the $\mHt$ abundance as we found in \citet{gm07b}. Within 
only 1~Myr, the hydrogen has already become 50\% molecular. Nevertheless, it
is also clear that  $\mwfrac{\mHt}$ has yet to settle into a steady state by the end
of the simulations at $t = 5.7 \: {\rm Myr}$. Although the $\mHt$ chemistry in the 
denser gas has largely reached a steady state by this point, there remain some
low-density regions in which the $\mHt$ formation timescale is longer than the
time elapsed in the simulation (despite the acceleration of $\mHt$ formation
caused by the turbulence).

\subsection{Carbon chemistry: ${\mathbf C^{+}}$, C and CO}
\begin{figure}
\centering
\epsfig{figure=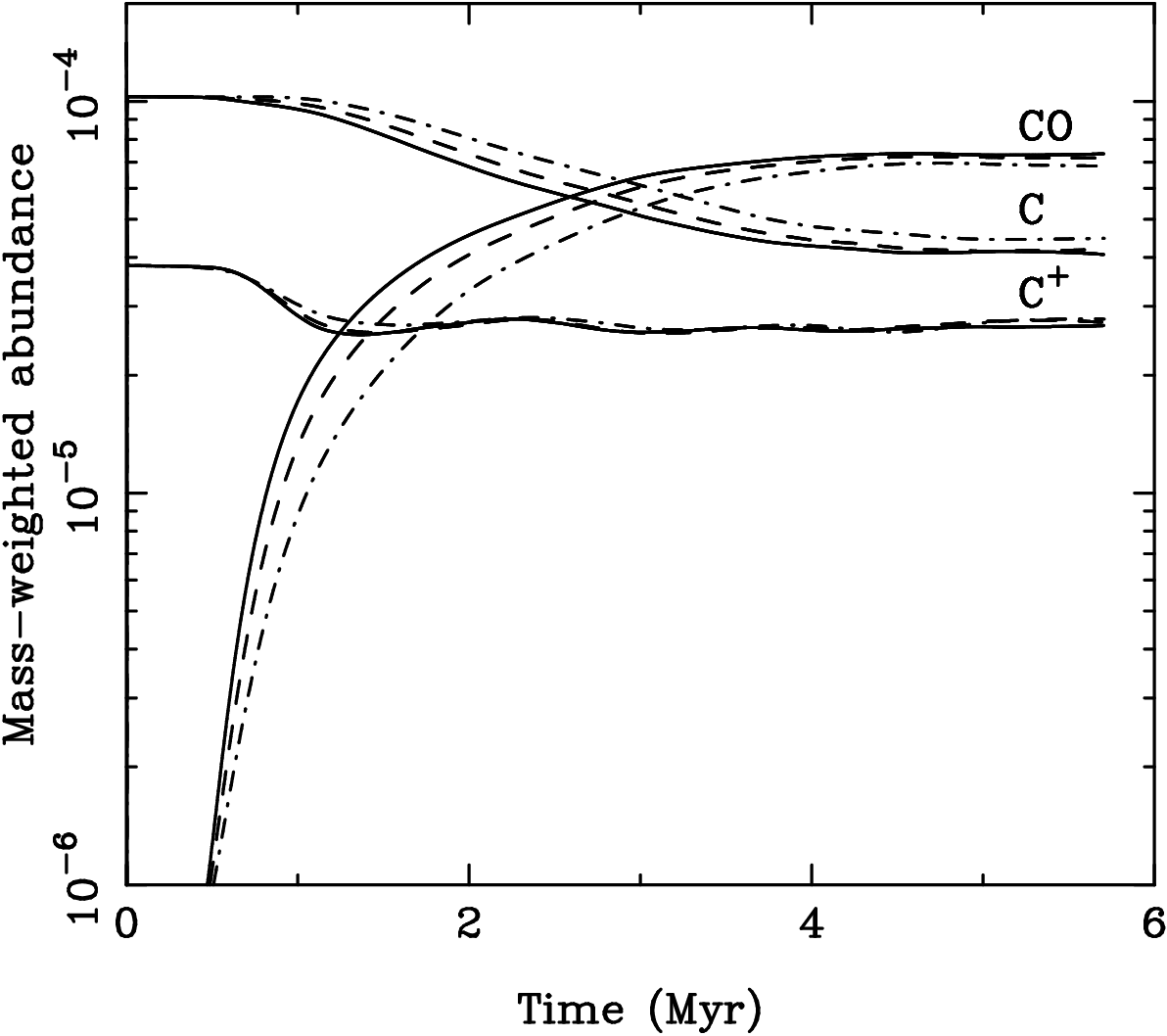,width=20pc,angle=270,clip=}
\caption{Time evolution of the mass-weighted abundances of atomic carbon, CO, and $\Cp$
in simulations with numerical resolutions  of $64^{3}$ zones (dot-dashed), $128^{3}$ zones 
(dashed) and $256^{3}$ zones (solid).
\label{carbon-abundances}}
\end{figure}

In Figure~\ref{carbon-abundances}, we examine the time evolution of the
mass-weighted mean abundances of $\Cp$, ${\rm C}$, and CO. The first point to note
here is the very rapid change in $\mwfrac{\Cp}$ and $\mwfrac{\mC}$ at the beginning
of the simulation.  This is a consequence of our choice of initial conditions.
We began with all of our carbon in the form of $\Cp$, which has a short recombination 
timescale $t_{\rm rec, C^{+}} < 0.1 \: {\rm Myr}$ for our initial temperature of 
$60 \: {\rm K}$ and initial density of $300 \: {\rm cm^{-3}}$. In the interior of the 
simulation volume, the value of $x_{\Cp}$ in photoionisation equilibrium is
typically much smaller than this starting value. Thus, there is a rapid
conversion of $\Cp$ to $\mC$, resulting in the rapid changes in  $\mwfrac{\Cp}$ and 
$\mwfrac{\mC}$ visible in the Figure. Once we are past this initial transient, the subsequent 
evolution of $\mwfrac{\Cp}$ and $\mwfrac{\mC}$ occurs on a much slower timescale. 
It is driven by a combination of two main factors: the changes that are occuring in the 
density structure of the gas, in response to the turbulence, and the conversion of $\Cp$ 
and $\mC$ into CO.

The CO abundance evolves rapidly within the first few million years of the simulations.
During the first 0.5~Myr, there is almost no CO present, but by $t = 1 \: {\rm Myr}$,
12\% of the total carbon has been incorporated into CO. This fraction increases to
32\% by $t = 2 \: {\rm Myr}$, and reaches a steady-state value of around 50\% at 
$t > 4$~Myr. Atomic carbon dominates for $t < 3$~Myr, while CO dominates at
$t > 3$~Myr. The steady-state value of the C/CO ratio is approximately 55\%. This
is consistent with the values computed by \citet{ptv04} for gas with a density of
order $10^{3} \: {\rm cm^{-3}}$ and a visual extinction $A_{\rm V} \sim 1$--10
that is illuminated by the standard interstellar radiation field, and is compatible
with the ratio observed in a number of nearby molecular clouds 
\citep[see e.g.][]{plume94}.

At early times, the C and CO abundances clearly depend on the numerical resolution
of the simulation: in higher resolution simulations, we find more CO and less C than
in lower resolution simulations. However, the difference between the runs lessens
with time and with increasing resolution. For times $t > 4 \: {\rm Myr}$, there is very 
little difference between the results of the $128^{3}$  and $256^{3}$ zone runs. The 
$\Cp$ abundance shows very little sensitivity to the numerical resolution throughout 
the simulation, as it is located primarily in large, low-density regions that are well
resolved in all of our simulations.

\subsection{Oxygen chemistry: O, OH, $\mathbf{H_{2}O}$ and $\mathbf{O_{2}}$}
\begin{figure}
\centering
\epsfig{figure=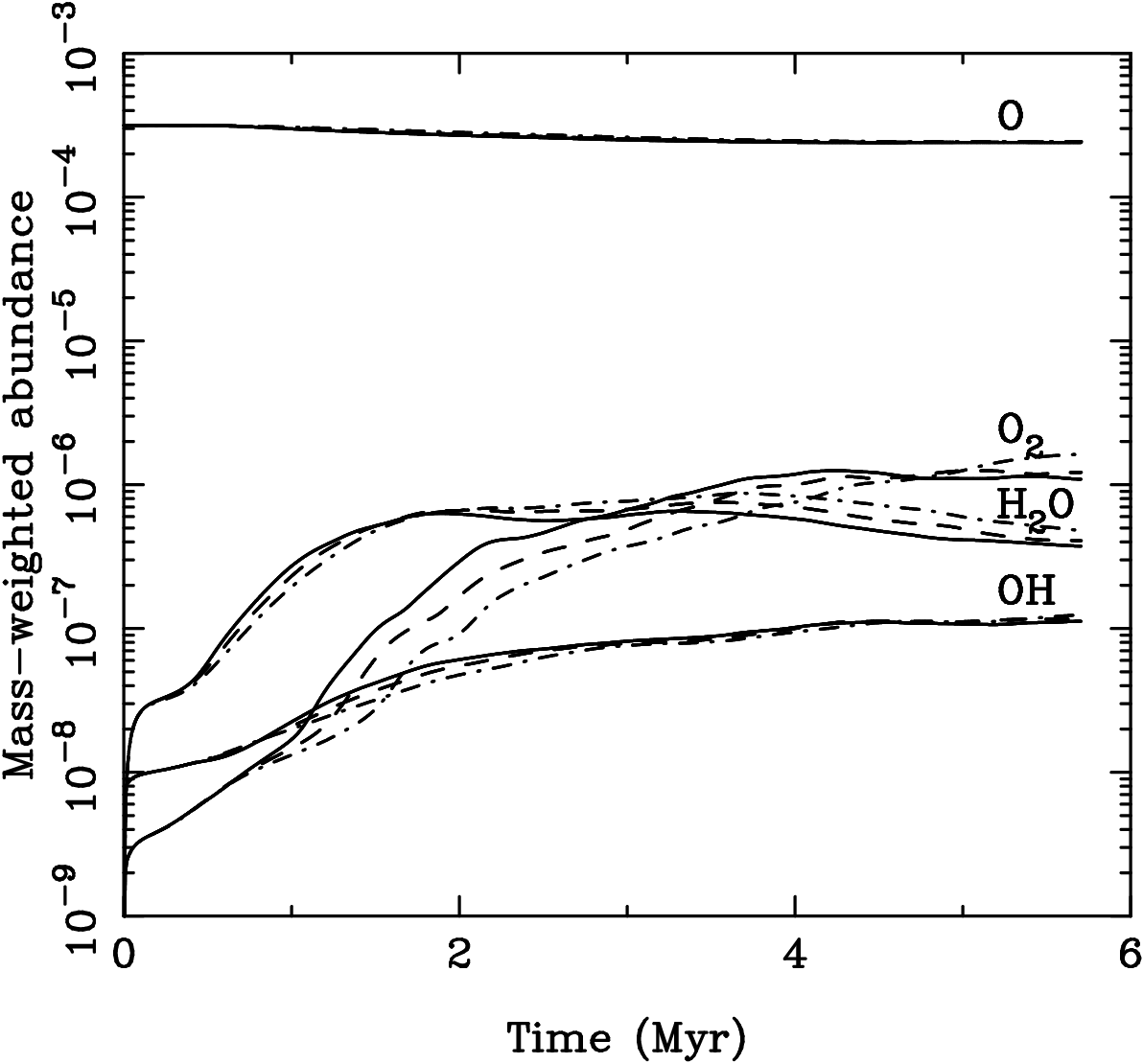,width=20pc,angle=270,clip=}
\caption{Time evolution of the mass-weighted abundances of atomic oxygen, OH, 
$\hto$ and ${\rm O_{2}}$  in simulations
with numerical resolutions of $64^{3}$ zones (dot-dashed), $128^{3}$ zones (dashed) and
$256^{3}$ zones (solid).
\label{oxygen-abundances}}
\end{figure}

In Figure~\ref{oxygen-abundances}, we examine the time evolution of the
mass-weighted mean abundances of O, OH, $\hto$, and ${\rm O_{2}}$. It is clear
from the Figure that most of the oxygen remains in atomic form at the end of the 
simulation, with roughly 23\% having been incorporated into CO (not plotted here),
and less than 1\% into other molecules (primarily $\hto$ at early times and
${\rm O_{2}}$ at late times). Our values for the O and OH abundances appear to
be numerically well converged during most of the period simulated, with the exception
of a short period around $t = 2 \: {\rm Myr}$; note that this corresponds to a turbulent
crossing time, and hence to the time at which the first major shock-shock interactions
are occurring. This is also the time at which most
of the CO is forming. Our values for the $\hto$ and  ${\rm O_{2}}$ abundances are
less well converged, although there is some indication that the results of the $128^{3}$
and $256^{3}$ runs have converged by the end of the simulation. The O and OH 
abundances appear to have reached a steady state by $t = t_{\rm end}$, but there is
no indication that the  $\hto$ and ${\rm O_{2}}$ abundances have done so.

If we compare the mean mass-weighted abundances of $\hto$ and ${\rm O_{2}}$ that we
find in these simulations with the values (or upper limits) measured in local star-forming
regions, then it becomes immediately apparent that there is a significant discrepancy.
In our simulations, we find that at times $t > 2 \: {\rm Myr}$ (corresponding to roughly
a single turbulent crossing time), $\mwfrac{\hto} \sim 3$--$5 \times 10^{-7}$ and 
$\mwfrac{{\rm O_{2}}} 
\sim 1$--$10 \times 10^{-7}$. However, observations of a number of local
star-forming regions with the {\em Submillimeter Wave Astronomy Satellite} (SWAS) and 
the {\rm Odin} satellite find $\hto$ and ${\rm O_{2}}$ abundances that are more than a 
factor of ten smaller \citep[see e.g.][]{berg00,gold00,pag03,lars07}. 
This discrepancy is probably due to the neglect of freeze-out processes in our
current study. Static gas-phase chemical models of molecular clouds overproduce 
$\hto$ and ${\rm O_{2}}$ in a similar fashion to our dynamical models \citep{berg00,gold00}
and the inclusion of grain-surface processes in these models is widely seen as the
most promising way to restore agreement with the observations. A few studies have
considered the effects of turbulent mixing, using an approach based in mixing-length 
theory, and have suggested that this could also suppress the gas-phase $\hto$ 
and ${\rm O_{2}}$ abundances \citep{cp89,xie95}. However, our current results would 
appear to rule this out as a solution to the `water problem'. Another possibility is
that there is some as-yet unidentified problem with the reaction rate coefficients used
for the oxygen gas-phase chemistry, but this is not an issue that our dynamical models
can address.

Since our results for the $\hto$ and  ${\rm O_{2}}$ abundances are unrealistic
compared to those measured in real molecular clouds, for one or more of the reasons
noted above,  will not discuss these molecules any further in this paper, and will focus our 
attention in the following sections on $\mHt$ and CO.

\section{Density and temperature probability distribution functions}
\label{pdfs}
\subsection{Probability density functions (PDFs) at $t = t_{\rm end}$}
\begin{figure}
\centering
\epsfig{figure=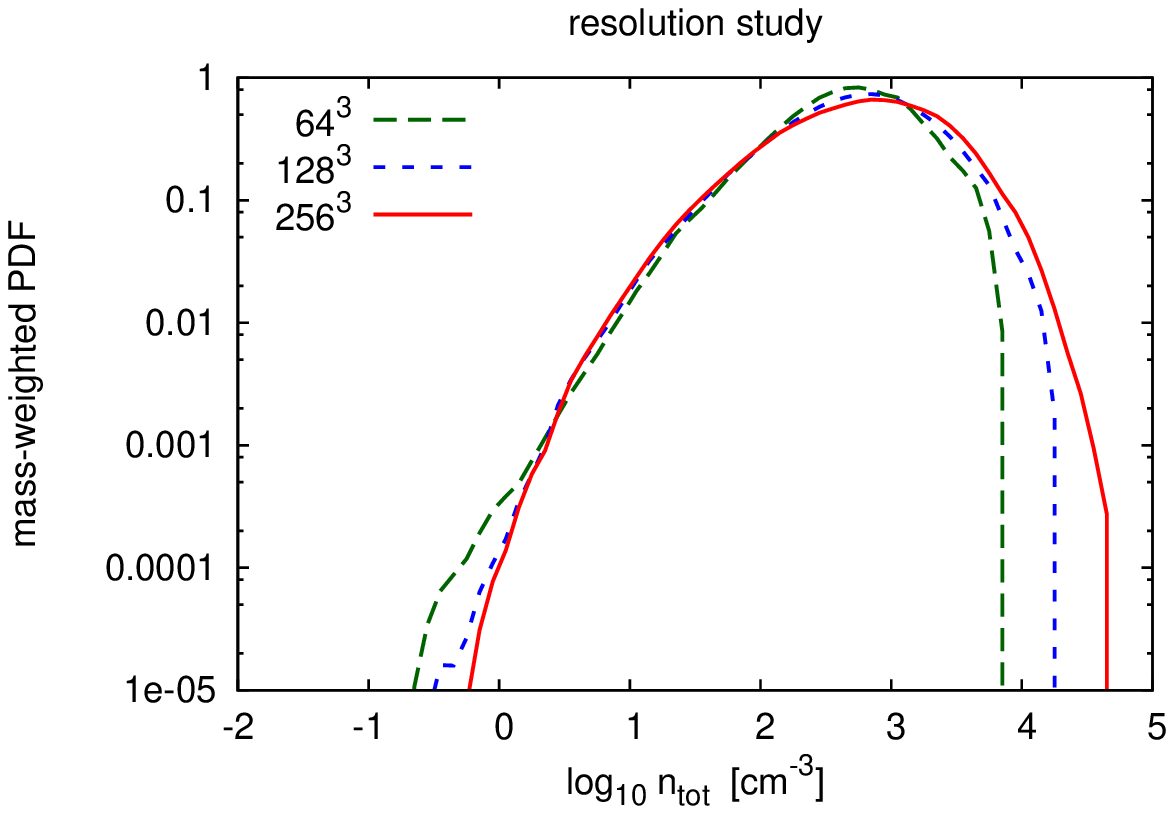,width=20pc,angle=0,clip=}
\epsfig{figure=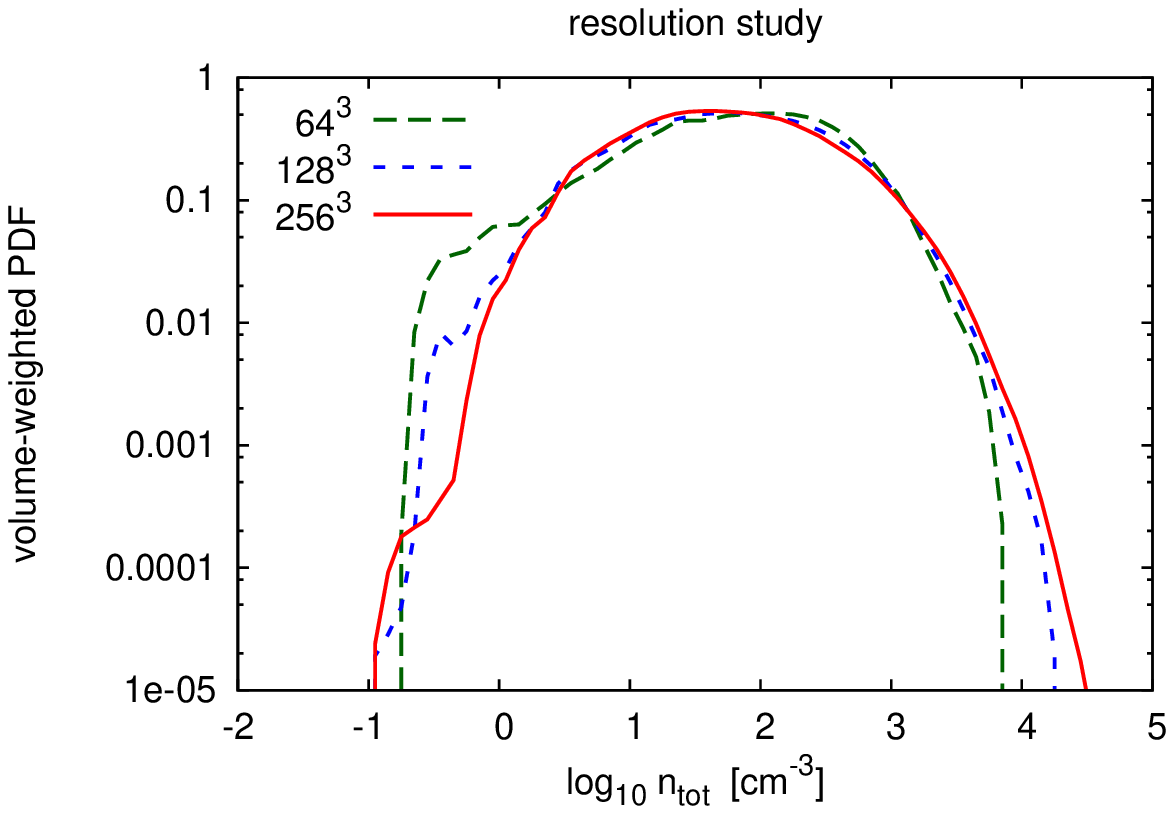,width=20pc,angle=0,clip=}
\caption{(a) Mass-weighted PDF of total number density $n_{\rm tot}$ at a time
$t = t_{\rm end}$ in runs R1 (green line), R2 (blue line) and R3 (red line).
(b) As (a), but for the volume-weighted PDF.
\label{ntot-pdfs}}
\end{figure}

In Figure~\ref{ntot-pdfs}, we plot the mass-weighted and volume-weighted
probability density functions of the total gas number density $n_{\rm tot}$ at the
end of simulations R1, R2 and R3, i.e.\ at $t_{\rm end} = 5.7 \: {\rm Myr}$.
We note first  that both PDFs have a log-normal shape around their peak,
although they deviate from this shape in the far wings of the distribution. This
form for the density distribution is not unexpected. Previous studies of the
density PDF produced by fully-developed supersonic turbulence in isothermal
gas find that it has a log-normal form \citep[e.g.][]{pnj97,pvs98,np99,k00,osg01,li04,ls08,fks08}:
\begin{equation}
p_{s} \, {\rm d}s = \frac{1}{\sqrt{2 \pi \sigma^{2}_{s}}} \exp \left[- \frac{\left(s - \left<s\right>\right)^{2}}{2
\sigma_{s}^{2}} \right] \, {\rm d}s,
\end{equation}
where $s = \ln(\rho / \left<\rho\right>)$, $\left<\rho\right>$ is the mean density of the gas, and where
the mean $\left<s\right>$ is related to the dispersion $\sigma_{s}$ by $\left<s\right> = - \sigma_{s}^{2} / 2$
due to the constraint of mass conservation. However, the tails of the PDFs may significantly deviate from this
log-normal distribution due to intermittent fluctuations  \citep{krit07,schm09,fed09}.

\citet{pnj97} argue that the logarithmic density dispersion $\sigma_{s}$ is related to the
RMS Mach number of the flow, ${\cal M}$, by
\begin{equation}  \label{eq:b}
 \sigma_{s}^{2} = \ln \left(1 + b^{2} {\cal M}^{2} \right),
\end{equation}
where $b \approx 0.5$. More recently, \citet{fks08} have shown that the proportionality
parameter $b$ depends on the relative strength of solenoidal compared to
compressive modes in the forcing field used to initialize and drive the turbulence.
For purely solenoidal forcing, they find that $b \simeq 1/3$, while for purely
compressive forcing, $b \simeq 1$.

We have measured the proportionality constant $b$ using equation~(\ref{eq:b}) in the regime of
fully developed turbulence ($3.8\,\rm Myr < t < t_{\rm end}$) and find a mean value of $b=0.32$. However,
$b$ decreases systematically in time from $b=0.36$ at $t=3.8\,{\rm Myr}$ to $b=0.27$ at $t=t_{\rm end}$.

Instead of using the total number density to estimate $b$, we additionally used the PDF of CO number density
to compute $b$. Since the PDF of CO number density significantly departs from a log-normal distribution, we
make use of the expression for the linear density dispersion \citep{pnj97,pvs98,fed09},
\begin{equation} \label{eq:b_lin}
 \sigma_\rho = b {\cal M}\,.
\end{equation}
This equation for the dispersion does not assume a log-normal distribution \citep{fks08}. Again, we find a systematic
decrease of $b$ in time with $b=0.39$ at $t=3.8\,{\rm Myr}$ and $b=0.33$ at $t=t_{\rm end}$. The mean value in
the regime of fully developed turbulence is $b=0.35$, slightly larger than our estimate using the total number density.
This is most likely due to the broad plateau of small CO number density seen in its PDF (see Fig.~\ref{nco-pdfs}).


Although the gas in our simulations is non-isothermal, the deviations from isothermality
do not appear to cause major changes in the density PDF compared to the isothermal
case. This is consistent with the previous findings of \citet{gm07b} for gas at a slightly
lower mean density ($n_{0} = 100 \: {\rm cm^{-3}}$ in the majority of their runs, compared
with $n_{0} = 300 \: {\rm cm^{-3}}$ here). However, simulations such as that of 
\citet{ha07} that consider much lower mean densities and so probe the regime in which the 
gas is thermally unstable produce a broad, bimodal PDF that is not log-normal. 

Regarding the numerical convergence of the density PDF, we note that we find good
convergence over much of the density range probed by our simulations, but that there
is not yet convergence in the wings of the distribution (see also \citealt{ha07}; \citealt{fed09}). 
In particular, there appears to
be a systematic shift to higher densities with increasing numerical resolution that is 
particularly apparent in the mass-weighted version of the plot, and that causes a slight
shift in the position of the peak. This behaviour is to be expected, since we are unable
to fully resolve the shocks in our simulations, even at our highest numerical resolution.
As \citet{gm07a} demonstrate, the characteristic cooling length of shock-heated gas, $L_{\rm cool}$, 
at our mean density is of the order of 0.01~pc, and so to resolve these cooling lengths with 
four grid cells in our 20~pc box, we would need to use a numerical resolution of 8000$^{3}$,
far larger than is currently possible (although \citealt{ha07} have successfully resolved the
post-shock cooling regions in two-dimensional simulations of interstellar turbulence). 
To properly resolve shocks occuring in gas denser
than the mean, where the cooling length is smaller, we would require an even higher
numerical resolution. 
Since our shocks are under-resolved, the effect of increasing the numerical resolution
is to decrease the size of the cooling regions behind the shocks (which cannot be 
smaller than the grid spacing $\Delta x$, even if $L_{\rm cool} \ll \Delta x$), which 
also allows for greater compression of the cold, post-shock gas. Fortunately, the 
effect of this on the density PDF appears to be relatively small, and hence we should
be able to trust the results of our simulations, provided that the quantities of interest
are not dominated by the behaviour in the wings of the density PDF.
We also note that quantities that are dependent on the behaviour of the wings of the
PDF are in any case difficult to characterise based on a single simulation, since the
wings are strongly affected by turbulent intermittency \citep{krit07,fks09,fed09}, and so vary from
realisation to realisation of the same turbulence simulation, even if the overall shape of the PDF 
remains the same when averaged over long enough times.

\begin{figure}
\centering
\epsfig{figure=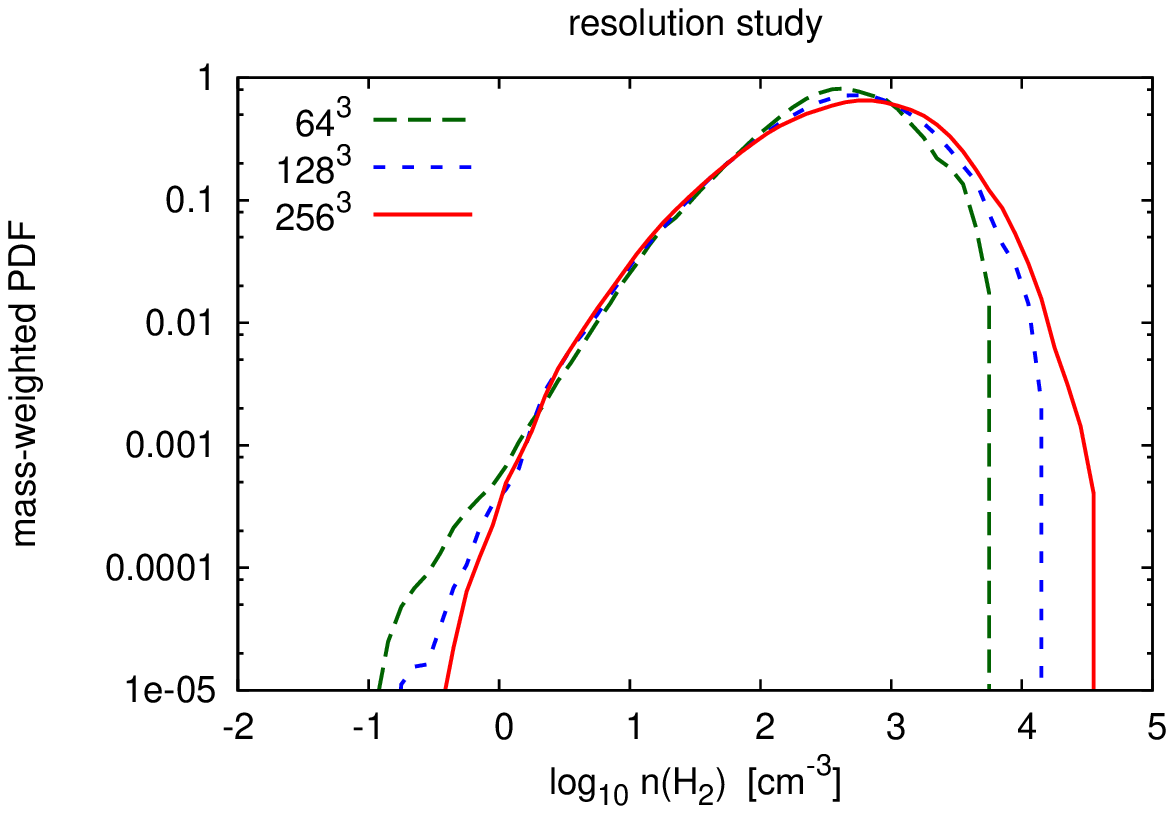,width=20pc,angle=0,clip=}
\epsfig{figure=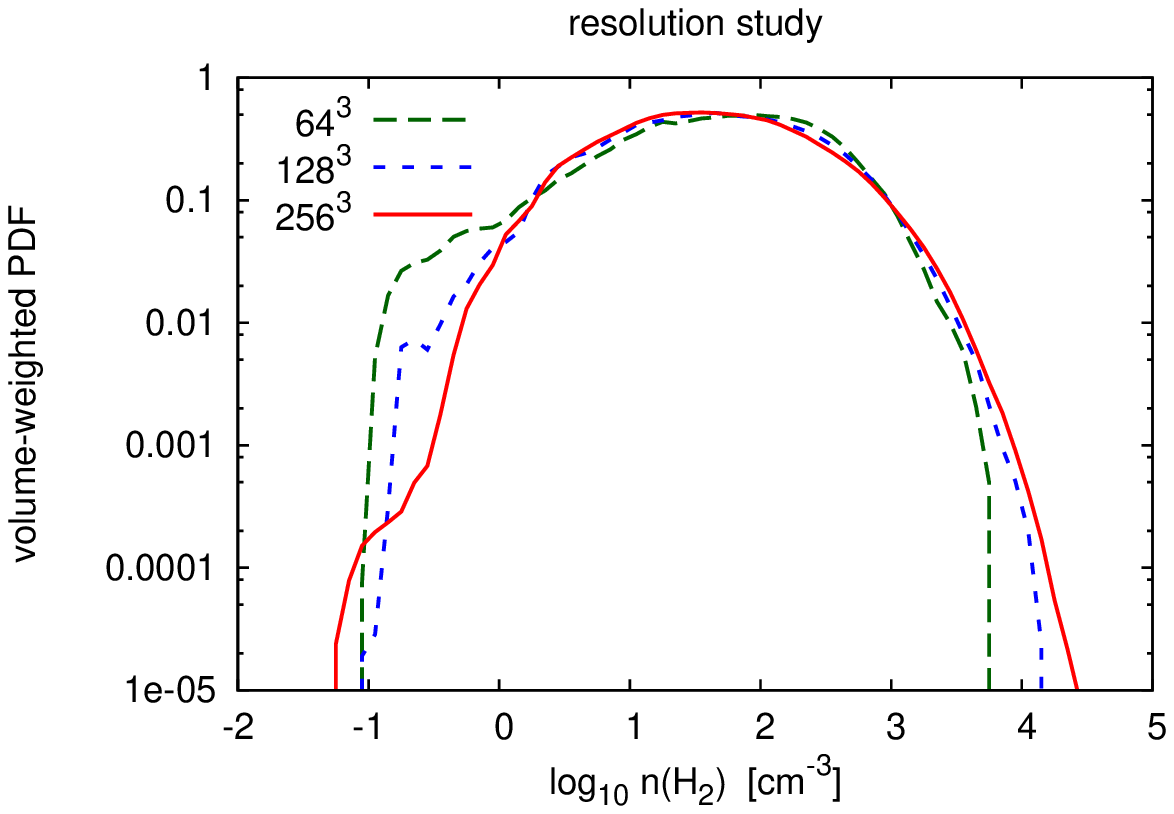,width=20pc,angle=0,clip=}
\caption{(a) Mass-weighted PDF of $\mHt$ number density $n_{\mHt}$ at a time
$t = t_{\rm end}$ in runs R1 (green line), R2 (blue line) and R3 (red line).
(b) As (a), but for the volume-weighted PDF.
\label{nh2-pdfs}}
\end{figure}

In Figure~\ref{nh2-pdfs}, we plot the mass-weighted and volume-weighted PDFs
of the H$_{2}$ number density $n_{\rm H_{2}}$ at $t = t_{\rm end}$. Comparing
these with the PDFs of the total number density shown in the previous Figure, we
see that there is very good agreement. This is to be expected: given that only a
few percent of the gas remains in atomic form at this point in the simulations, it
is unsurprising that the PDF of the H$_{2}$ number density closely follows the
PDF of total number density.

In Figure~\ref{nco-pdfs}, we plot the mass-weighted and volume-weighted PDFs
of the CO number density $n_{\rm CO}$ at $t = t_{\rm end}$. Unlike the PDFs of
total number density and H$_{2}$ number density, this is not lognormal. The
mass-weighted PDF has a clear peak at $n_{\rm CO} \sim 10^{-1} \: {\rm cm^{-3}}$,
and falls off sharply at higher CO number densities in a fashion similar to a 
lognormal, but at lower CO  number densities there is a clear feature in the
distribution function, which decreases only slightly
from $n_{\rm CO} \sim 10^{-2} \: {\rm cm^{-3}}$
down to  $n_{\rm CO} \sim 10^{-8} \: {\rm cm^{-3}}$. The volume-weighted PDF
also shows a peak at $n_{\rm CO} \sim 10^{-1} \: {\rm cm^{-3}}$, but is actually
bimodal, with a second peak at  $n_{\rm CO} \sim 10^{-7} \: {\rm cm^{-3}}$, 
although it is also clear that a large number of zones have CO number densities in between
these two values.

The curious shape of the CO number density PDFs can be better understood
once we realise that we are dealing with a PDF made up of two separate
contributions, one coming from grid zones with $x_{\rm CO} \simeq x_{\rm C, tot}$ 
and a second from grid zones with $x_{\rm CO} \ll x_{\rm C, tot}$, and that the
dependence of $n_{\rm CO}$ on $n_{\rm tot}$ is very different in these two
sets of zones.

If $x_{\rm CO} \simeq x_{\rm C, tot}$, or in other words if almost all of the carbon 
in a zone has been converted to CO, then clearly $n_{\rm CO}$ is directly
proportional to $n_{\rm tot}$.
The contribution that these zones make to the PDF therefore simply mirrors the
lognormal shape of the underlying mass density PDF. A significant fraction of the 
carbon in our simulation is located in such fully molecular zones, and it is
the contribution of the gas in these regions that gives us our high density peak
in the CO number density PDF.

On the other hand, if $x_{\rm CO} \ll x_{\rm C, tot}$ and, crucially, if the CO fraction
is not tightly correlated with the total gas density, then we would expect to find
only a weak correlation between $n_{\rm CO}$ and $n_{\rm tot}$. In other words,
if the scatter in the values of $x_{\rm CO}$ in gas with a given $n_{\rm tot}$ is large,
then this scatter will wash out the effects of any correlation between $n_{\rm CO}$ 
and $n_{\rm tot}$. As we will see later, in Section~\ref{spatial}, a considerable fraction of the
gas in our simulations behaves in this fashion, and it is this that is responsible for
the extended low-density plateau that we see in the CO number density PDF. We
will return to this point in Section~\ref{spatial}. 

As far as the numerical convergence of the PDFs is concerned, we see very good
agreement between the results of our three runs over a very wide range of CO
number densities. Differences between the three runs are only apparent in the tails
of the distribution. Increasing the resolution increases the mass fraction in regions
with very high CO number density, which simply reflects the fact that we better 
resolve the dense, post-shock gas in the highest resolution simulation, as already
noted above. We also find fewer regions with very low CO number density in our
higher resolution simulations, which again seems to be a consequence of the
resolution-dependence of the wings of the underlying mass-density PDF.

\begin{figure}
\centering
\epsfig{figure=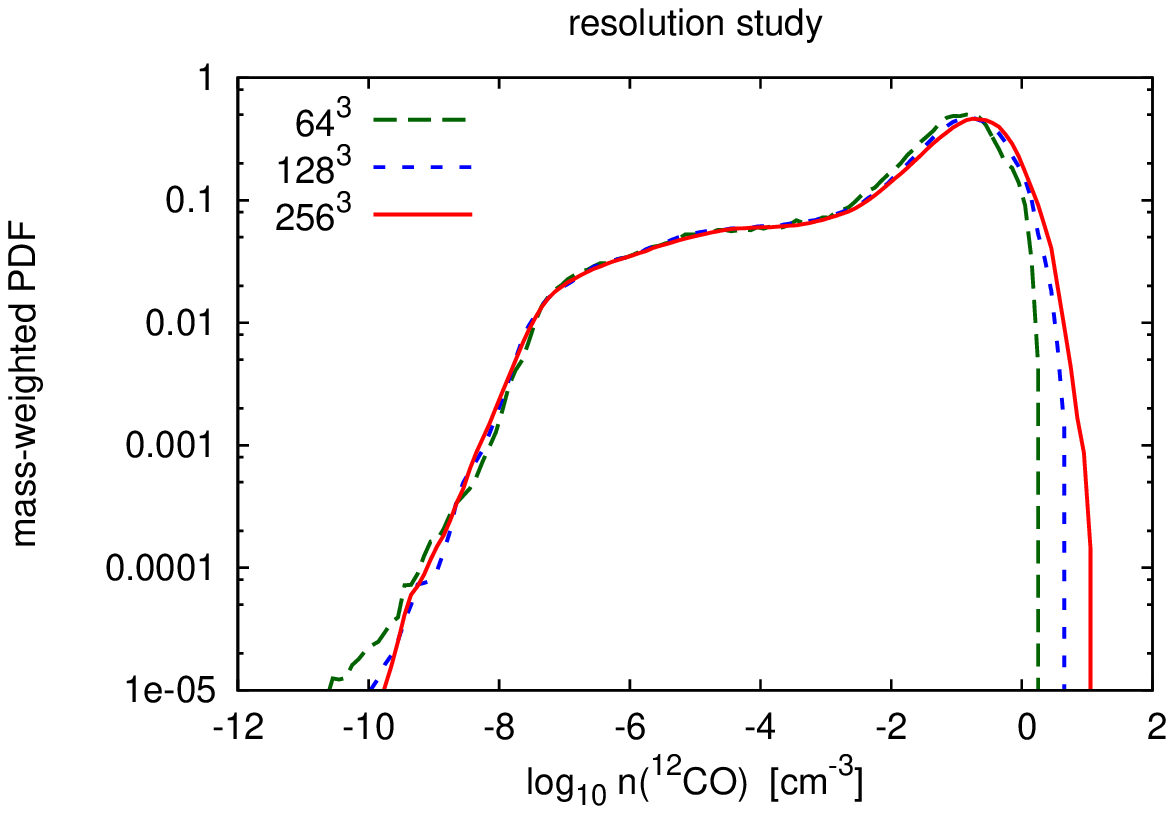,width=20pc,angle=0,clip=}
\epsfig{figure=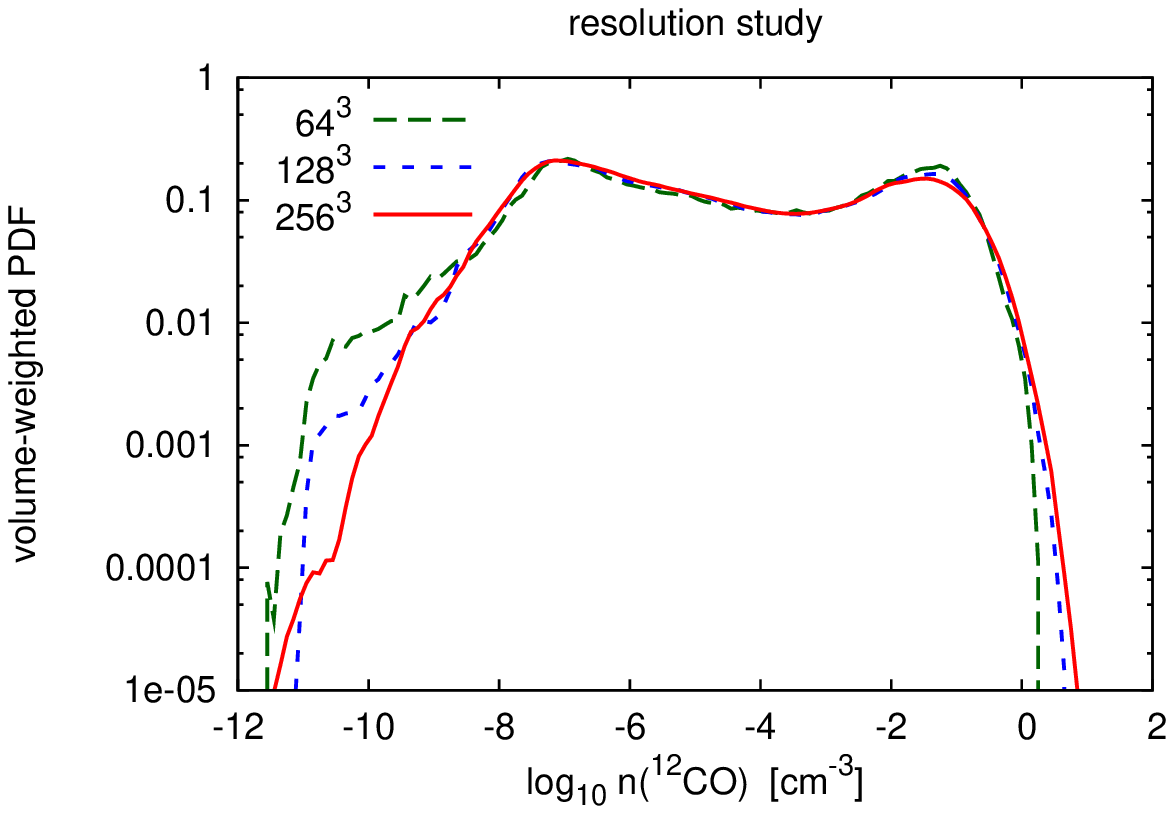,width=20pc,angle=0,clip=}
\caption{(a) Mass-weighted PDF of CO number density $n_{\rm CO}$ at a time
$t = t_{\rm end}$ in runs R1 (green line), R2 (blue line) and R3 (red line).
(b) As (a), but for the volume-weighted PDF.
\label{nco-pdfs}}
\end{figure}

\begin{figure}
\centering
\epsfig{figure=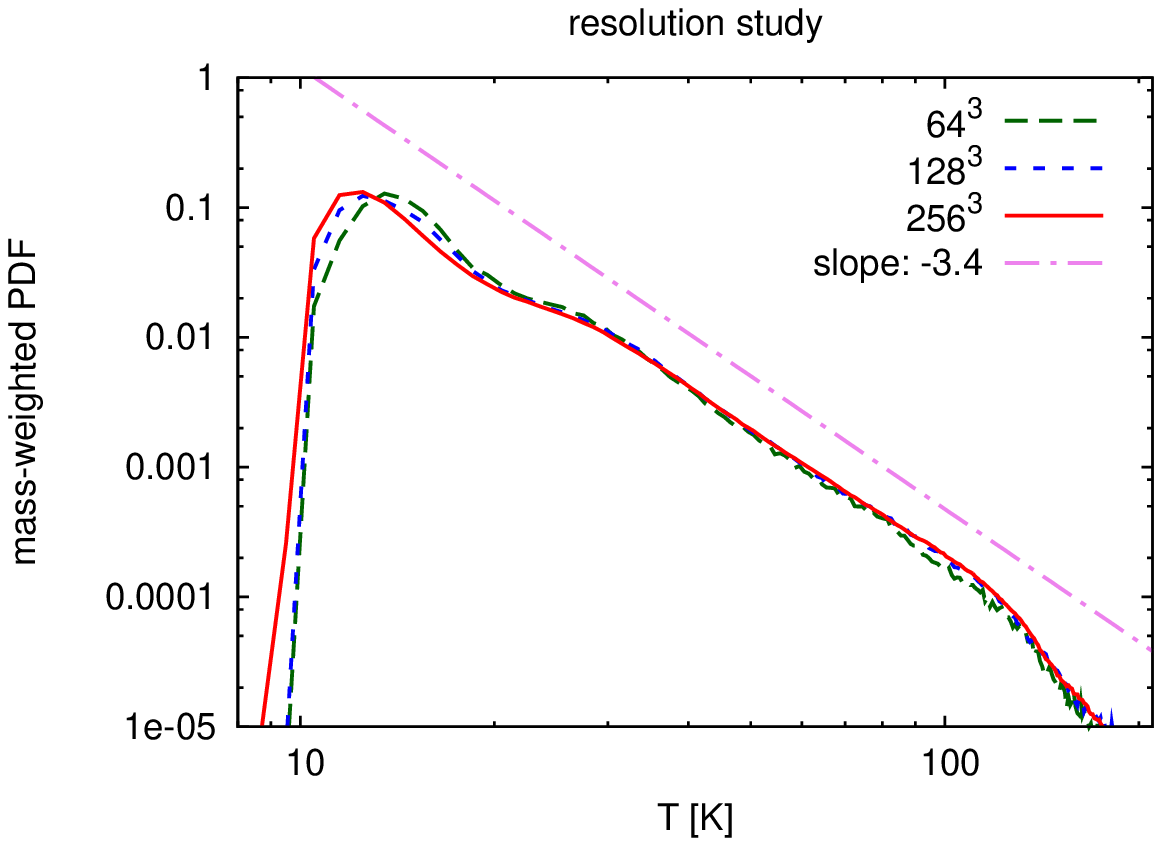,width=20pc,angle=0,clip=}
\epsfig{figure=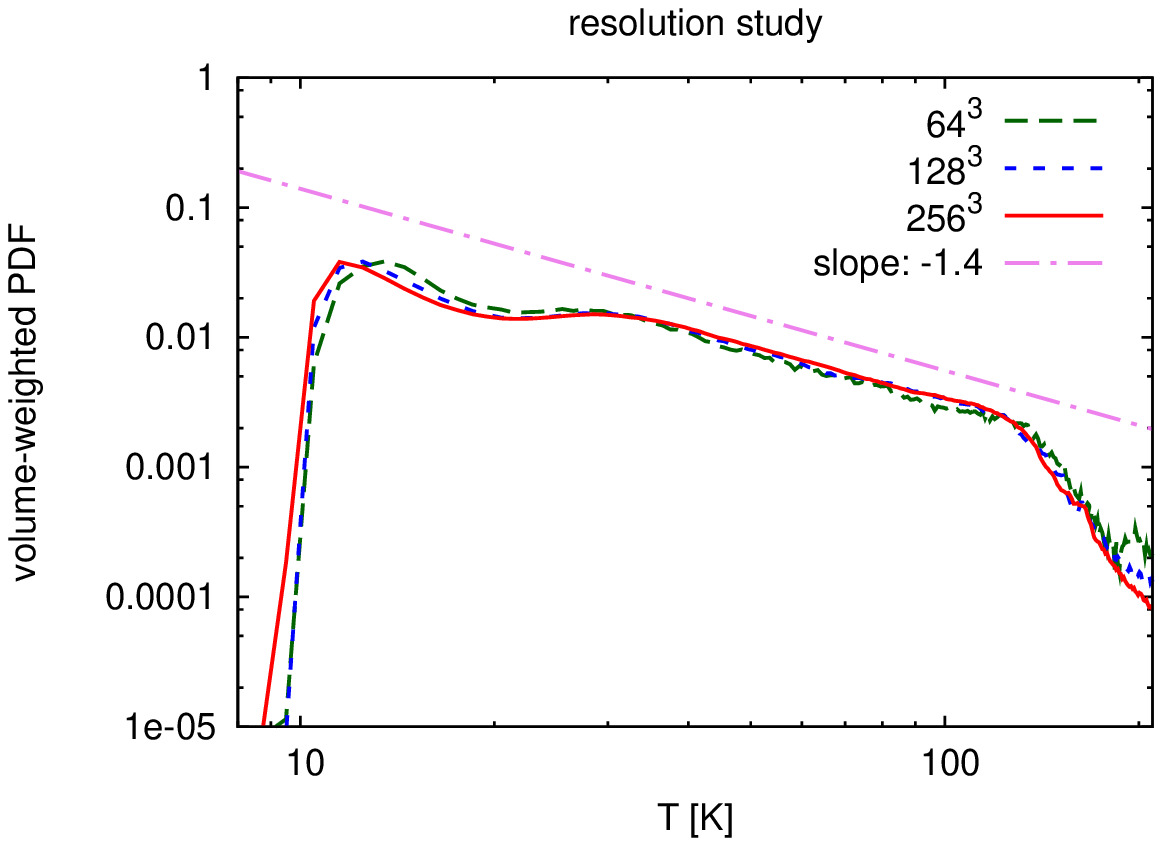,width=20pc,angle=0,clip=}
\caption{(a) Mass-weighted PDF of gas temperature $T$ at a time
$t = t_{\rm end}$ in runs R1 (green line), R2 (blue line) and R3 (red line).
(b) As (a), but for the volume-weighted PDF.
\label{temp-pdfs}}
\end{figure}

Finally, in Figure~\ref{temp-pdfs}, we plot the mass-weighted and volume-weighted 
PDFs of the gas temperature $T$ at $t = t_{\rm end}$. Several features of these plots
stand out. First, as the temperature approaches 10~K, the PDF falls 
steeply, and there is almost no gas in the simulation with $T < 10 \: {\rm K}$. This 
feature of the plot is artificial, and is a consequence of our adoption of a temperature
floor at $T = T_{\rm dust} = 10 \: {\rm K}$ in our treatment of the radiative cooling
(see \S\ref{co-cooling-section} above). Thus, the only fluid elements that 
have $T < 10 \: {\rm K}$ are those that had temperatures close to 10~K and 
then underwent a strong rarefaction, leading to significant adiabatic cooling. 
Moreover, this must have happened within the last 0.1~Myr, or else cosmic ray 
heating would have warmed the gas up above 10~K again. As is apparent from 
the PDFs, very few of the fluid elements in our simulations find themselves in this 
situation at any given time.

The second obvious feature of the temperature PDFs is the clear power-law
tail between 30~K and 120~K in both the mass and the volume-weighted PDFs. 
This tail is composed of gas with a low $A_{\rm V, eff}$ that is heated primarily
by photoelectric emission from dust grains and is cooled primarily by C$^{+}$
fine structure emission. The equilibrium temperature of this gas varies approximately 
as a power-law function of density, $T_{\rm eq} \propto n^{0.7}$ \citep{larson05,gm07b}.


Regarding numerical convergence, we again find good convergence for the
majority of the PDF, with significant differences visible only around the low
temperature peak in the distribution. Increasing the numerical resolution
shifts the peak to slightly lower gas temperatures, reflecting the fact that the
coldest gas is also, typically, the densest, and that this dense material is 
better resolved in the higher resolution simulations. 

\subsection{Time evolution of the PDFs}
\begin{figure}
\centering
\epsfig{figure=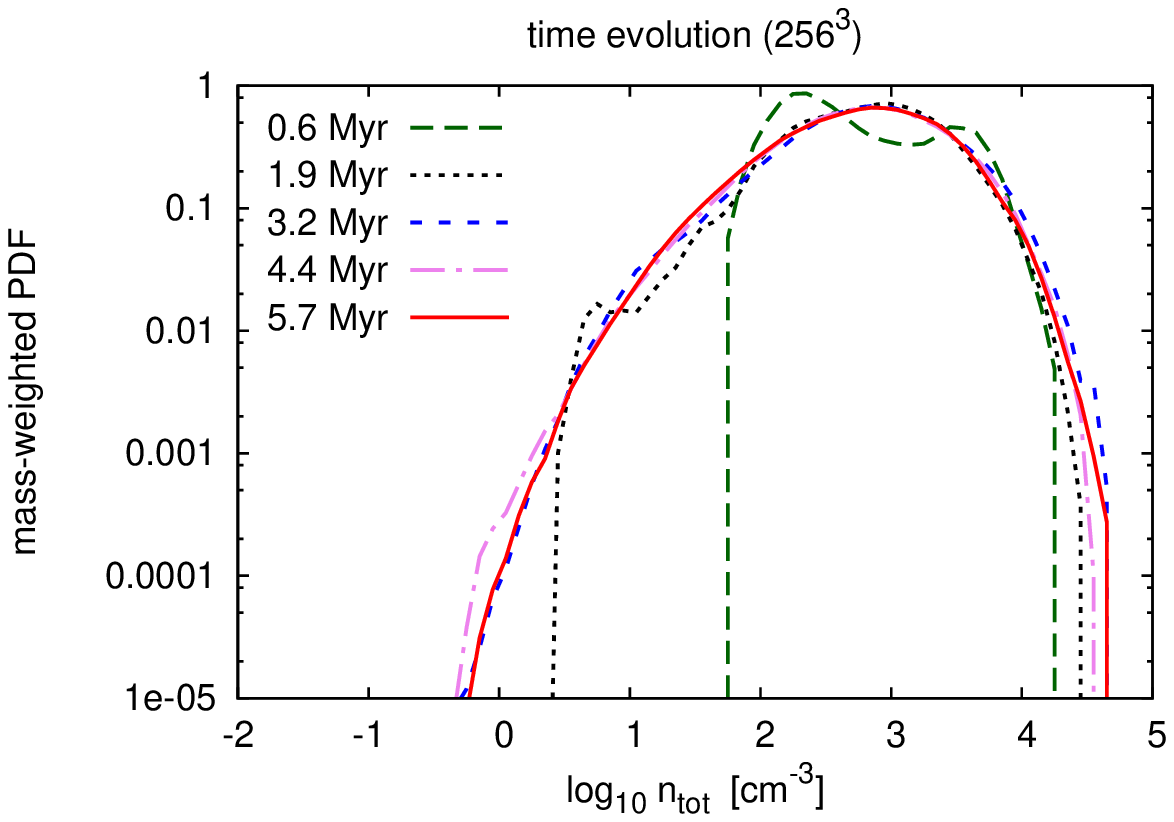,width=20pc,angle=0,clip=}
\epsfig{figure=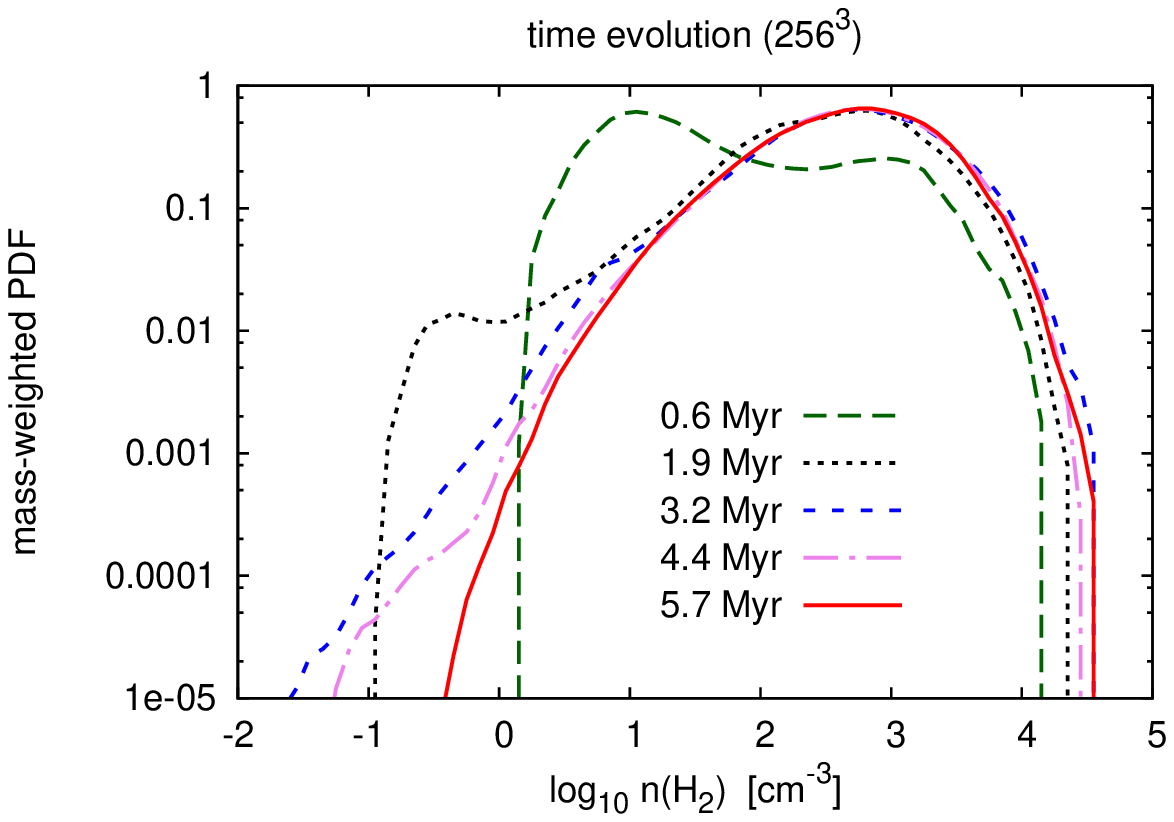,width=20pc,angle=0,clip=}
\epsfig{figure=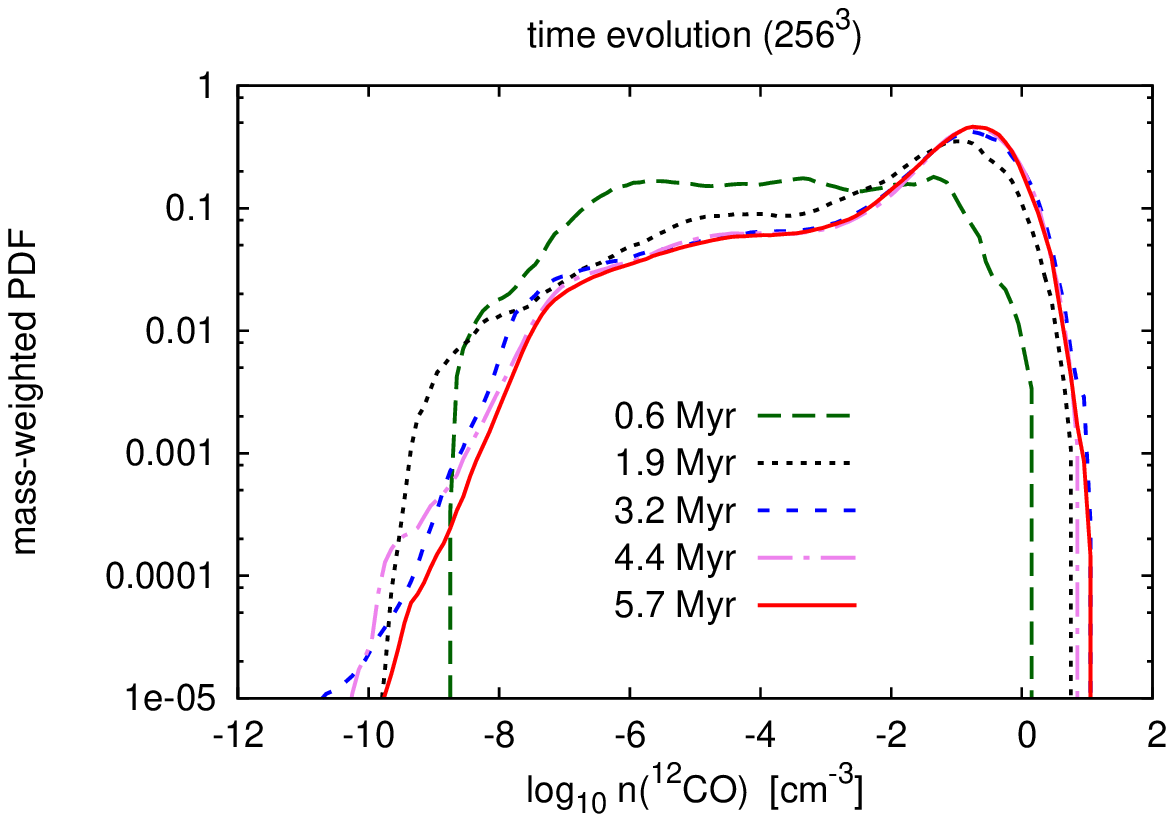,width=20pc,angle=0,clip=}
\epsfig{figure=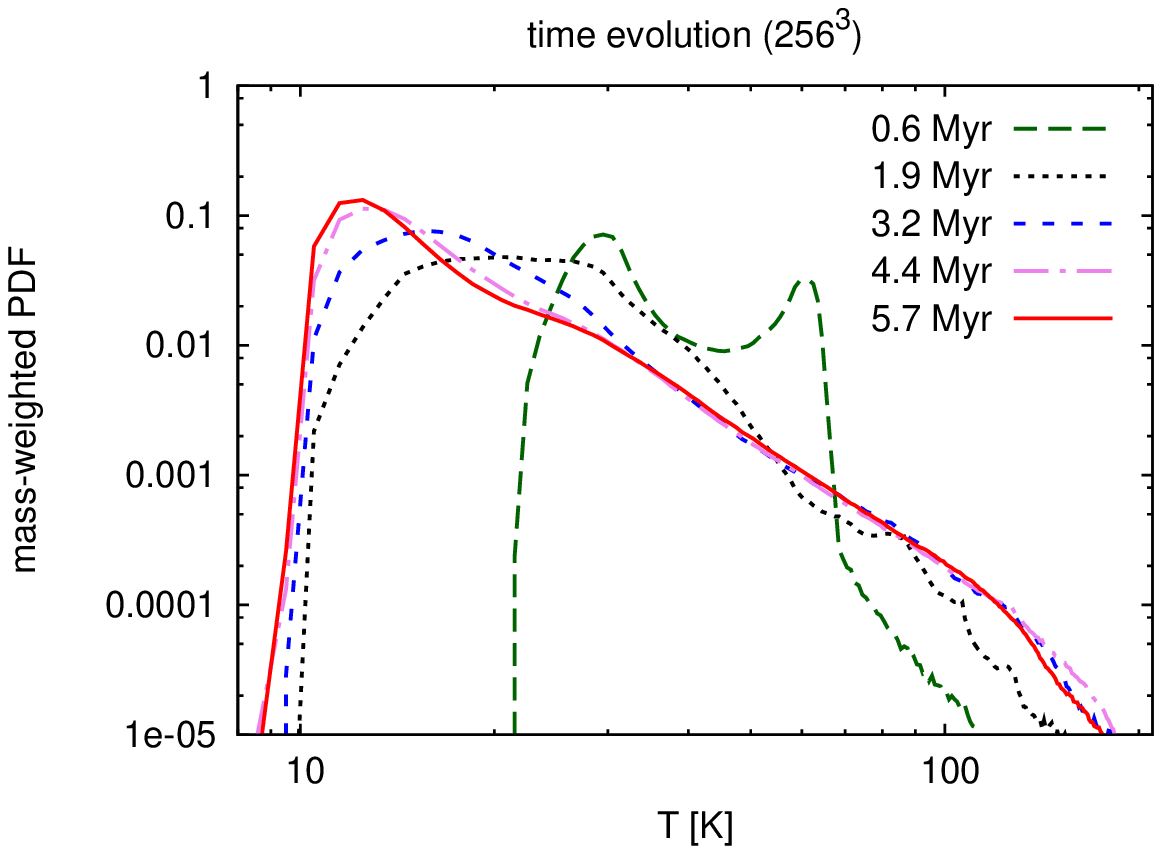,width=20pc,angle=0,clip=}
\caption{(a) Evolution with time of the mass-weighted PDF of  the total number density 
$n_{\rm tot}$ in our $256^{3}$ zone run, R3. Values are plotted for $t  = 0.6 \: {\rm Myr}$
(green long-dashed line),  $t  = 1.9 \: {\rm Myr}$ (black dotted line),  $t  = 3.2 \: {\rm Myr}$
(blue short-dashed line),  $t  = 4.4 \: {\rm Myr}$ (mauve dot-dashed line) and  
$t  = 5.7 \: {\rm Myr}$ (red solid line). 
(b) As (a), but for the $\mHt$ number density.
(c) As (a), but for the CO number density.
(d) As (a), but for the gas temperature.
\label{pdf-time}}
\end{figure}

In Figure~\ref{pdf-time}a, we show how the mass-weighted PDF of  the total number 
density evolves with time in  the $256^3$ run R3. 
The first output time for which data is plotted,
$t = 0.6 \: {\rm Myr}$, corresponds to less than half of a turbulent crossing time, and at 
this point in the simulation, the imprint of the initial conditions is still quite apparent.
The density PDF at this time is clearly not lognormal. Instead, it is bimodal, with one 
peak close to the starting density $n_{0} = 300 \: {\rm cm^{-3}}$, corresponding to gas
which has not yet been significantly affected by the turbulence, and a second peak at 
$n \sim 3000 \: {\rm cm^{-3}}$, corresponding to gas that has already been compressed
by the strong, large-scale shocks present in the initial turbulent velocity field.

By the time of the second output dump at $t = 1.9 \: {\rm Myr}$, corresponding to roughly
one turbulent crossing time, the picture is quite different. The characteristic lognormal
density PDF has now been established, although a few fluctuations in the low density
tail of the PDF are still apparent. These have vanished by the time of our third output
dump, at $t = 3.2 \: {\rm Myr}$, and from this point on we see very little evidence for
any change in the PDF, suggesting that the density distribution has reached a statistical
steady state.

In Figure~\ref{pdf-time}b, we plot the mass-weighted PDF of  the H$_{2}$ number
density in run R3 at various output times. As in Figure~\ref{pdf-time}a, the PDF at
the earliest output time is clearly not lognormal, but after one turbulent crossing time,
it has become lognormal over much of the range of densities plotted.  This is easily
understood when one considers that by $t = 1.9 \: {\rm Myr}$, roughly 80\% of the
hydrogen in the simulation has already become molecular (see Figure~\ref{h2-abundance}) 
and that we expect there to be a clear correlation between gas density and molecular fraction
(see \citealt{gm07b}, or \S\ref{spatial} below). At gas densities where the hydrogen is
almost fully molecular, we expect that the H$_{2}$ number density PDF will simply track
the underlying density PDF. Comparison of Figures~\ref{pdf-time}a and
\ref{pdf-time}b shows that this is the case for H$_{2}$ number densities greater
than $n_{\rm H_{2}} \sim 10 \: {\rm cm^{-3}}$ at  $t = 1.9 \: {\rm Myr}$. Lower H$_{2}$ 
number densities correspond to regions of low density gas where a significant fraction
of atomic hydrogen remains, and in these regions there is no simple mapping between
gas density and H$_{2}$ number density, since here the H$_{2}$ fraction is sensitive
not only to the current density and the degree of shielding, but also to the previous 
dynamical history of the gas \citep{gm07b,fed08}. It is therefore not surprising that at these low
densities the PDF is not lognormal.

At later output times, we see little change in the PDF at densities $n_{\rm H_{2}} > 
10 \: {\rm cm^{-3}}$, which simply reflects the fact that there is very little change in the
underlying density PDF. At lower densities, the PDF continues to evolve with time.
The mass fraction of gas with extremely low H$_{2}$ number densities, $n_{\rm H_{2}} <
0.3 \: {\rm cm^{-3}}$, significantly decreases. This gas physically corresponds to gas
with a low total number density and low H$_{2}$ fraction. As time passes, the H$_{2}$ 
fraction in this gas increases, and so the H$_{2}$ number densities increase, even
though the statistical distribution of the total number densities is unchanged. At the latest
output time, $t  = 5.7 \: {\rm Myr}$, the shape of the full H$_{2}$ number density 
PDF is almost identical to that of the full density PDF, since by this point, almost all
of the hydrogen has become molecular.

In Figure~\ref{pdf-time}c, we show how the mass-weighted PDF of the CO number 
density evolves with time in run R3. Just as with $n_{\rm tot}$ and $n_{\rm H_{2}}$, the
shape of the PDF at our earliest output time bears little resemblance to its later shape,
while by the time of the second output dump, the shape of the final PDF has become 
far better established. Unlike  $n_{\rm tot}$ and $n_{\rm H_{2}}$, however, we continue
to see evolution in both the low and the high density portions of the PDF, and it is only
after $3.2 \: {\rm Myr}$, corresponding to 1.6 turbulent crossing times, that the
CO number density PDF reaches a statistical steady state. This longer evolutionary
timescale is a consequence of the longer time required to form CO, compared to 
H$_{2}$. Between $t = 1.9 \: {\rm Myr}$ and $t = 3.2 \: {\rm Myr}$, the total mass 
of H$_{2}$ in the simulation increases by about 15\%, while the total mass of 
CO increases by about 50\%.

Finally, in Figure~\ref{pdf-time}d, we show how the mass-weighted temperature 
PDF  varies with time. At the earliest output time, this has a two-peaked structure: a
high temperature peak, centered on $T \sim 65 \: {\rm K}$, corresponding to 
unshocked gas that still has a density and temperature
close to its initial value, and a low temperature
peak, centered on  $T \sim 30 \: {\rm K}$, corresponding to shocked, higher-density
gas (compare with the density PDF at $t = 0.6 \: {\rm Myr}$ in Figure~\ref{pdf-time}a). By
$t = 1.9 \: {\rm Myr}$, this double-peaked structure has disappeared, but as with the
CO number density PDF, it is not until $t = 3.2 \: {\rm Myr}$ that large portions of the 
PDF settle into their final form. By this point, the power-law high temperature tail has
become fully established, and the PDF at temperatures $T > 30 \: {\rm K}$ shows
very little evolution at later times. On the other hand, the portion of the PDF around
the low temperature peak continues to evolve. The PDF peaks at 
$T \simeq 16 \: {\rm K}$ at $t = 3.2 \: {\rm Myr}$, but by $t = 4.4 \: {\rm Myr}$ the peak
has shifted to $T \simeq 13 \: {\rm K}$, while by $t = 5.7 \: {\rm Myr}$, it has shifted
further, to $T \simeq 12 \: {\rm K}$. We therefore see a systematic increase with time
of the amount of cold gas present in the simulation. This increase has not yet come to 
an end by the end of our simulation, although it has significantly slowed: there is far
more evolution in the temperature PDF between $t = 3.2 \: {\rm Myr}$ and  
$t = 4.4 \: {\rm Myr}$  than there is between  $t = 4.4 \: {\rm Myr}$  and 
$t = 5.7 \: {\rm Myr}$. This increase in the mass of cold gas is simply driven by
radiative cooling of dense, fully-molecular gas. At temperatures $T \gg 10 \: {\rm K}$,
the cooling time of the gas is very short, but for temperatures close to our temperature 
floor of 10~K, it becomes comparable to the dynamical time \citep[see e.g.\ Figure~2a in][]{nlm95}.



\section{Spatial distribution of $\mathbf{H_{2}}$ and CO}
\label{spatial}
\begin{figure}
\centering
\epsfig{figure=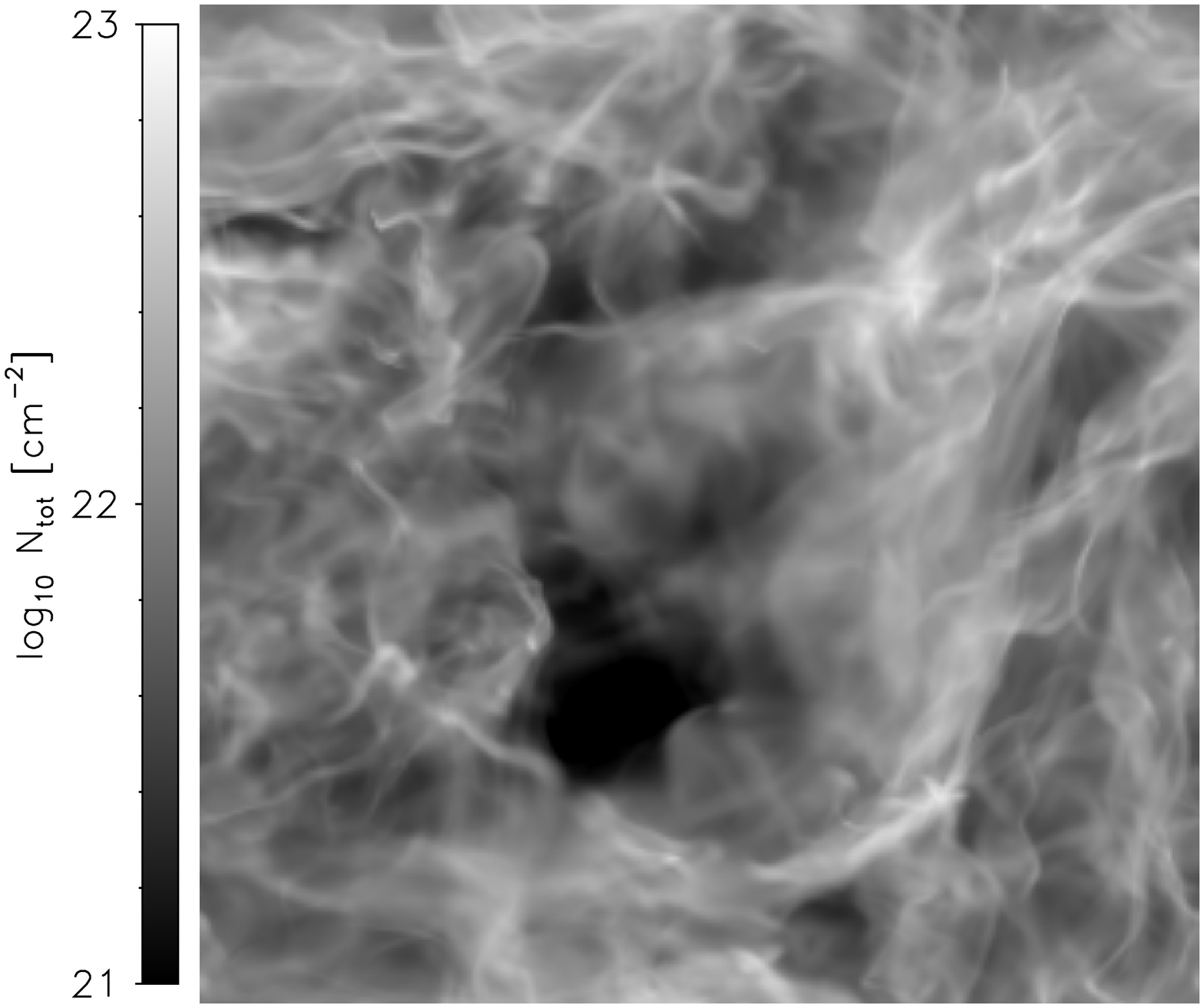,width=20pc,angle=0,clip=}
\epsfig{figure=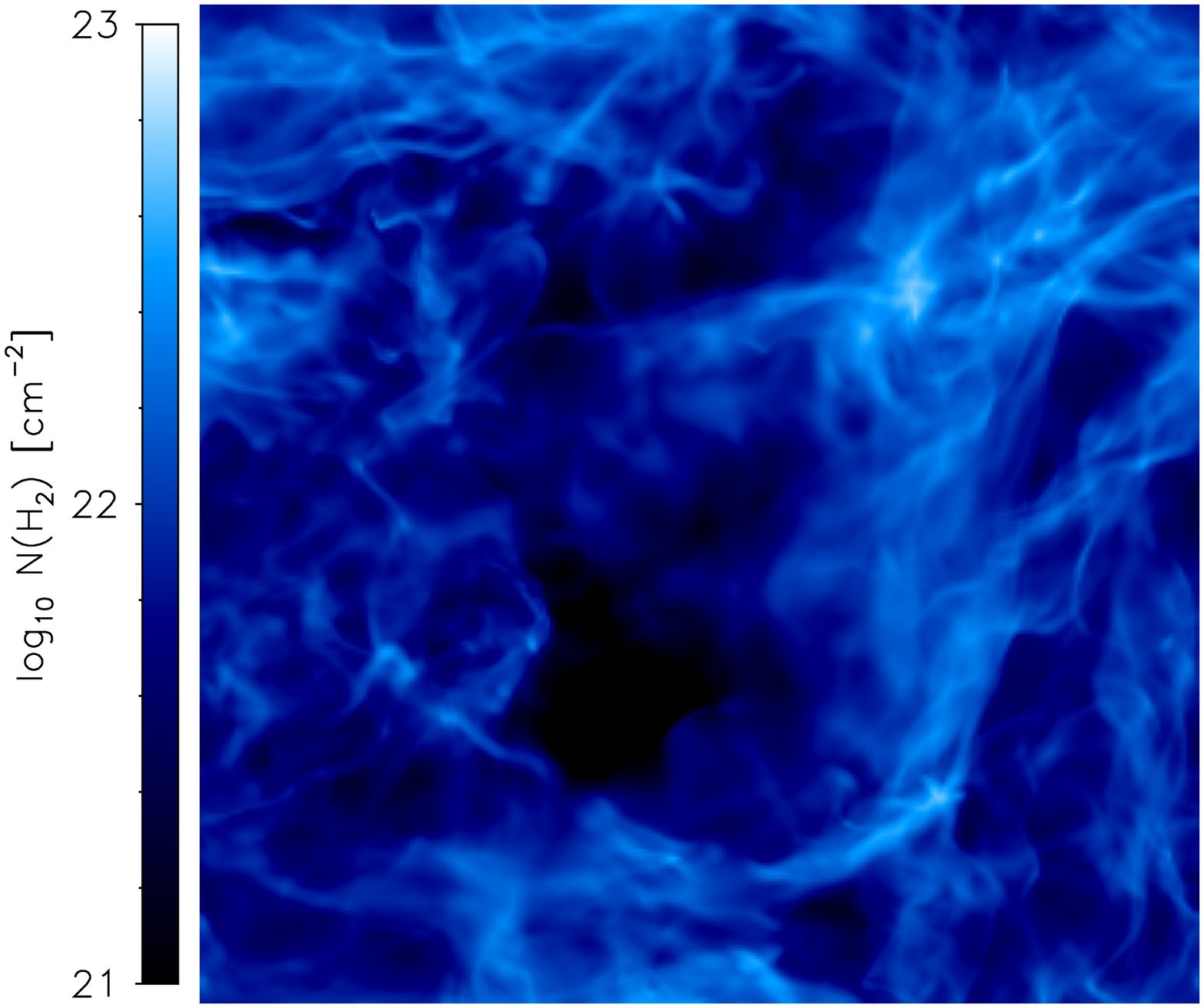,width=20pc,angle=0,clip=}
\epsfig{figure=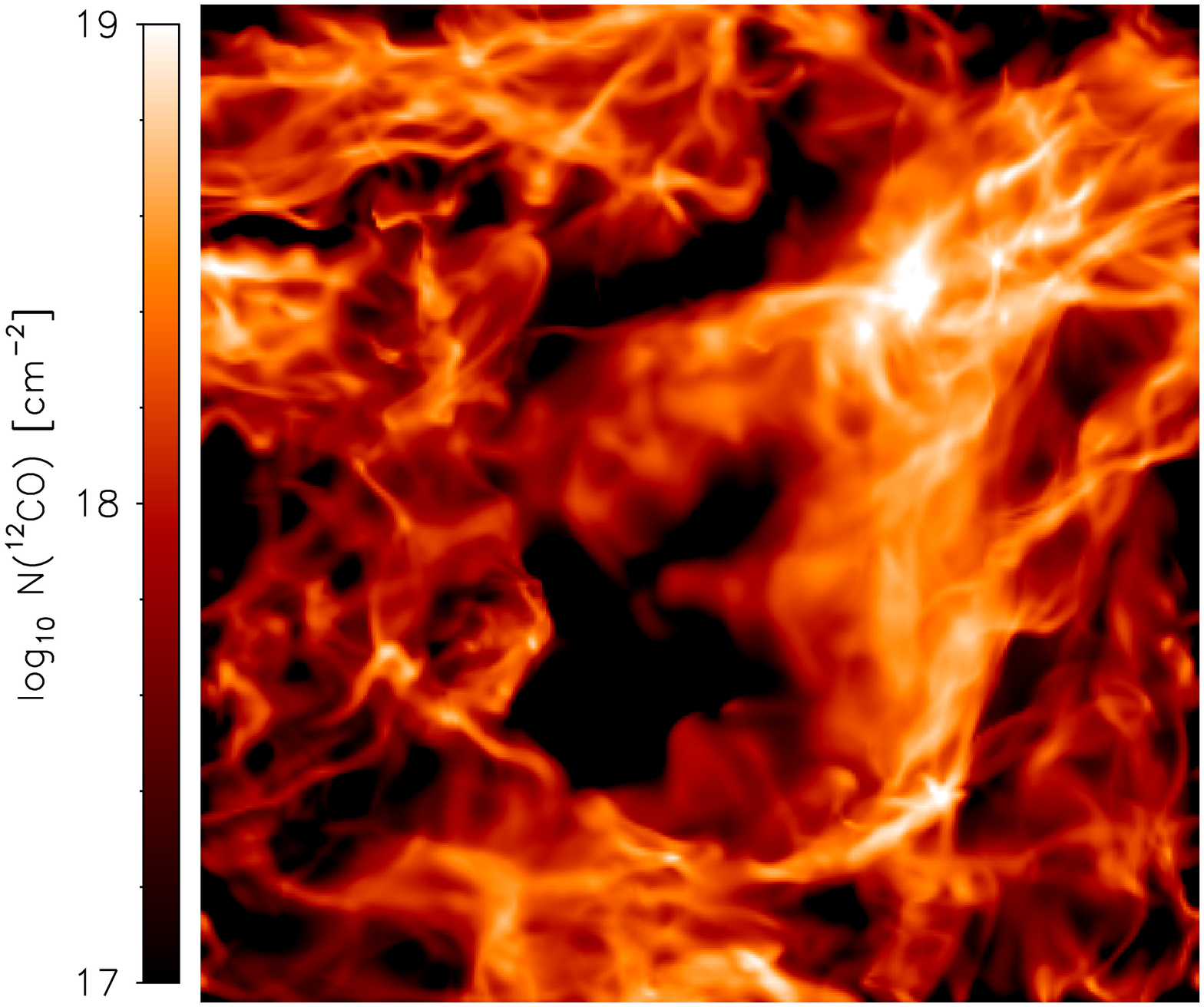,width=20pc,angle=0,clip=}
\epsfig{figure=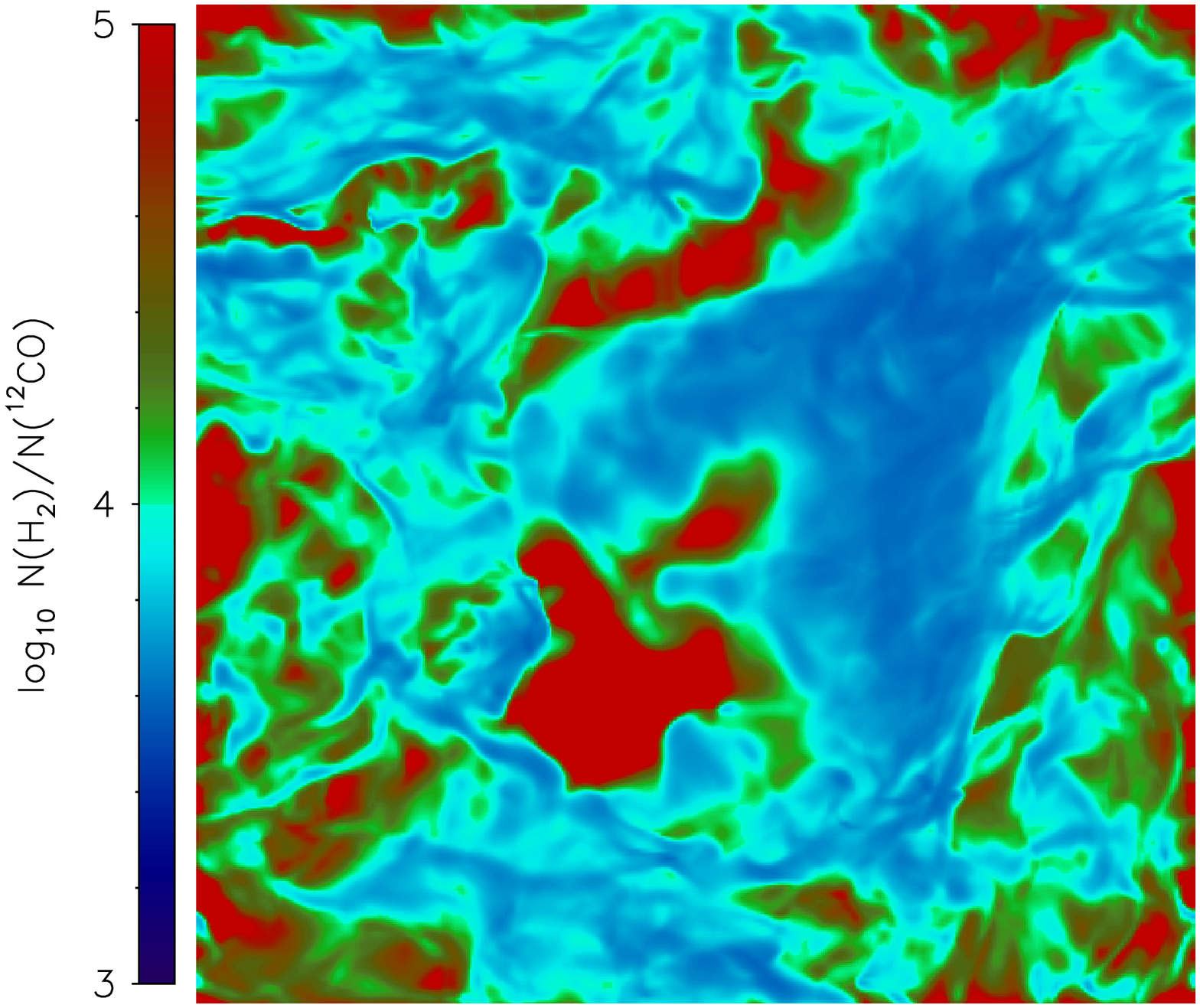,width=20pc,angle=0,clip=}
\caption{(a) Column density of hydrogen nuclei, $N_{\rm H, tot}$,  in run R3 at time 
$t = t_{\rm end}$, viewed along a line of sight parallel to the $z$-axis of the simulation volume.
This direction is also parallel to the initial orientation of the magnetic field. 
(b) As (a), but for the $\mHt$ column density.
(c) As (a), but for the CO column density.
(d) Ratio of $\mHt$ column density to CO column density along the same line of sight 
through run R3 at time  $t = t_{\rm end}$. \label{N-xy}}
\end{figure}

In Figure~\ref{N-xy}a, we plot the column density of hydrogen nuclei, $N_{\rm H, tot}$,
projected along the $z$-axis of the simulation (i.e.\ along the axis parallel to the direction
of the initial magnetic field) in run R3 and at time $t = t_{\rm end}$. In Figures~\ref{N-xy}b
and \ref{N-xy}c, we show similar plots of the H$_{2}$ and CO column densities. The
basic morphology of the gas is the same in all three plots. The gas has a filamentary
distribution, and large spatial variations in the column densities are apparent, including,
coincidentally, a rather prominent under-density visible toward the bottom-left of the
figures. The plots of 
$N_{\rm H, tot}$ and $N_{\rm H_{2}}$ are very difficult to distinguish, which is
unsurprising since most of the hydrogen gas in the simulation is  molecular by this 
point. On the other hand, the plot of CO column density is clearly different from the
other two plots: the underdense regions are larger, and also more numerous, 
particularly towards the edges of the box. Similar results are found if we consider
projections along lines of sight perpendicular to the initial magnetic field.

The difference between the spatial 
distributions of $N_{\rm CO}$ and $N_{\rm H_{2}}$ can be more clearly highlighted
by plotting the ratio $N_{\rm H_{2}} / N_{\rm CO}$, as we do in Figure~\ref{N-xy}d. 
Along lines of sight corresponding to the highest column densities, this ratio is
around $10^{3.5}$, as we would expect for gas in which all of the hydrogen is in the
form of H$_{2}$ and all of the carbon is in the form of CO. However, if we look 
along lines of sight that pass through regions of lower total column density, then 
we find values for this ratio that are up to a factor of thirty larger. Along these 
lines of sight, much of the carbon remains in the form of C or C$^{+}$. It is clear
from Figure~\ref{N-xy}d that the regions with low CO column densities are 
found preferentially towards the edges of the simulation volume. This is only to
be expected, given our treatment of the external ultraviolet radiation. Gas close 
to an edge of the simulation volume  will tend to have a low column density of 
dust between itself and the edge, and so will be more readily affected by UV
photons propagating inwards in that direction, even if in other directions, that
same gas is well-shielded. However, it should be stressed that this is not simply
an edge effect: when the column density is low, UV photons can penetrate in to
the volume  to considerable depths, and are not confined to a narrow region at the surface
of the box.

Similarly, the two large regions with high H$_{2}$:CO column density
ratios that are close to the center of the projections are well-shielded 
by dust along the $x$ and $y$ axes of the simulation, but have a low column density 
of material shielding them in the $z$ direction, and so are strongly affected by UV 
photons propagating inwards in that direction. 

Clearly, if we were to solve for the UV radiation field using much higher
angular resolution, we would expect to obtain slightly different results for
the spatial distributions of the CO column density and the H$_{2}$:CO
column density ratio. Nevertheless, the picture we obtain here is
qualitatively correct. The clumpiness created by the turbulence opens up
channels in the gas distribution, allowing UV photons to propagate far
deeper into the cloud than would be possible if it were a single
homogeneous mass of gas \citep[see also][]{boisse90,pad04,bzl07}.

\subsection{CO to $\mathbf{H_{2}}$ ratio} 
The results shown in Figure~\ref{N-xy} indicate that the amount of CO in the gas depends 
in part upon the column density of the material shielding the gas, but being two-dimensional 
projections, they do not easily allow us to quantify this dependence, or to separate the effects 
of increased shielding from the effects of increased gas number density. To do this, we need 
to make use of the full three-dimensional spatial information contained in the datacube. 

\begin{figure}
\centering
\epsfig{figure=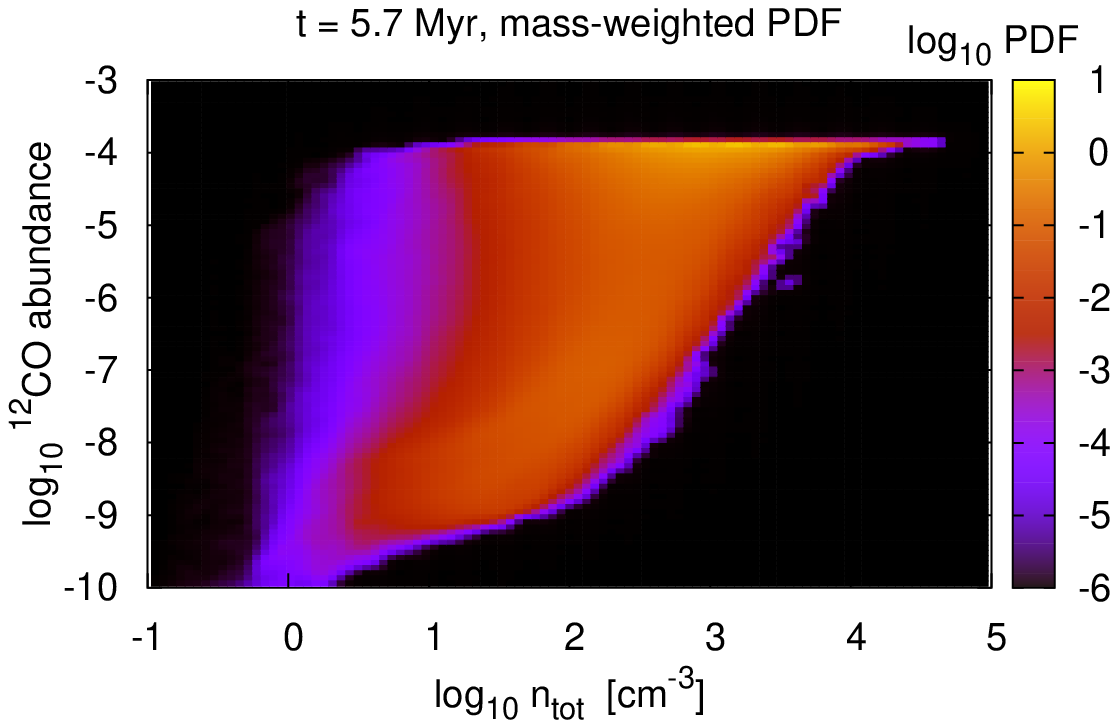,width=20pc,angle=0,clip=}
\epsfig{figure=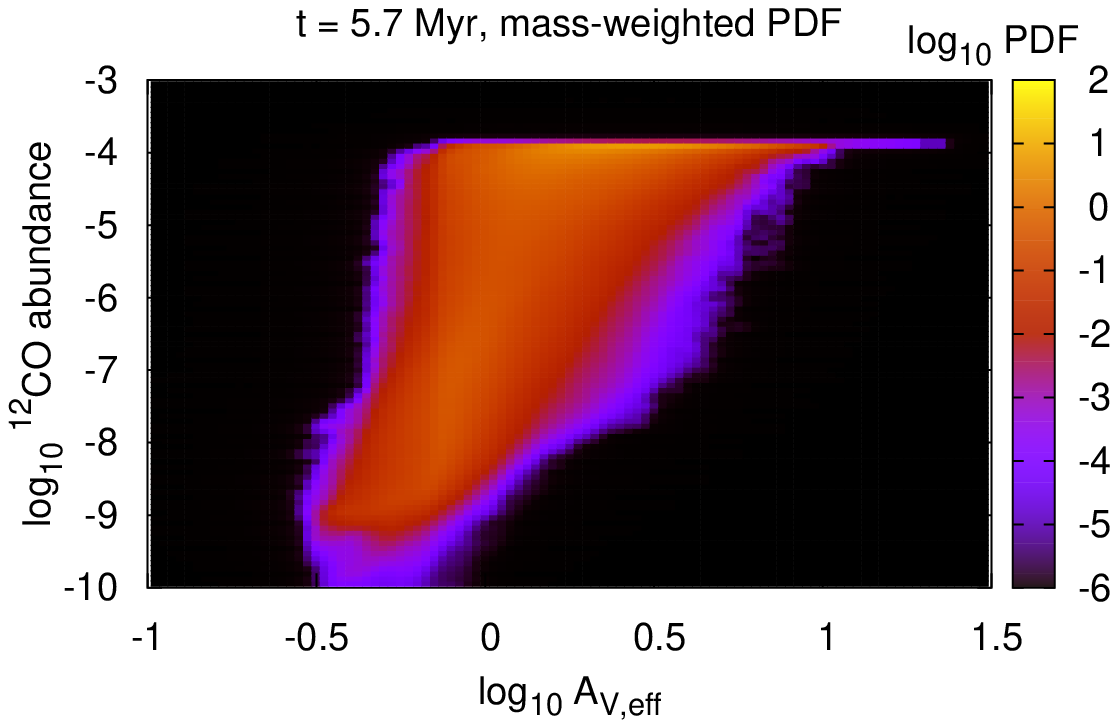,width=20pc,angle=0,clip=}
\epsfig{figure=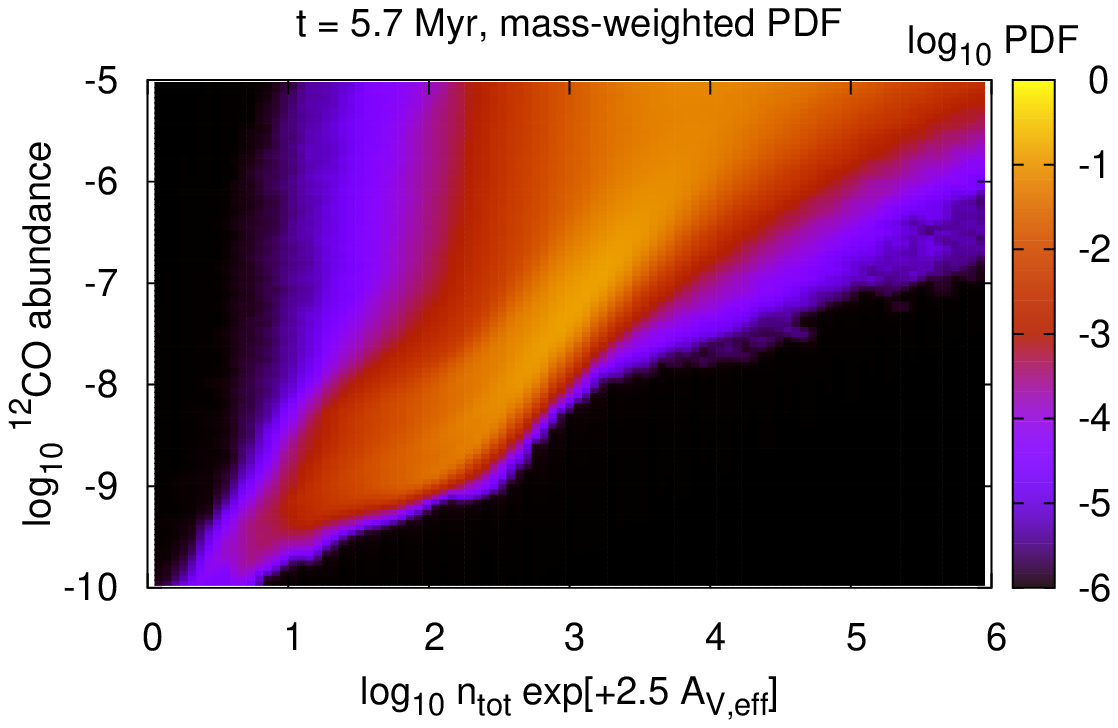,width=20pc,angle=0,clip=}
\epsfig{figure=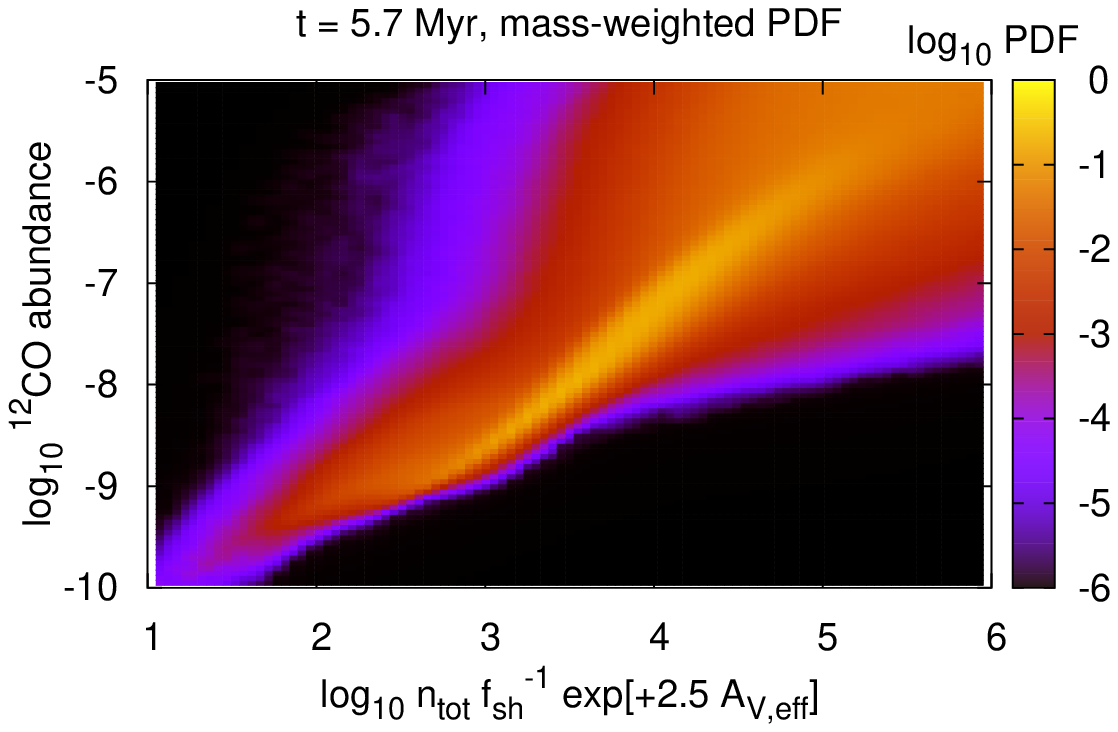,width=20pc,angle=0,clip=}
\caption{(a) Mass-weighted two-dimensional PDF of CO fractional abundance
$x_{\rm CO}$ versus gas number density $n_{\rm tot}$.
(b) Mass-weighted two-dimensional PDF of CO fractional abundance
$x_{\rm CO}$ versus effective visual extinction $A_{\rm V, eff}$ (defined in
Equation~\ref{aveff}).
(c) Mass-weighted two-dimensional PDF of $x_{\rm CO}$ versus 
$\exp(+2.5 A_{\rm V, eff}) n_{\rm tot}$.
(d) Mass-weighted two-dimensional PDF of $x_{\rm CO}$
versus $\xi$, defined in Equation~\ref{def-xi}.
\label{xco-ntot}}
\end{figure}

In Figure~\ref{xco-ntot}a, we show how the fractional abundance of CO, $x_{\rm CO}$,
depends on the number density $n_{\rm tot}$ by plotting the mass-weighted two-dimensional
PDF of these two quantities. In Figure~\ref{xco-ntot}b, we show a similar plot of
$x_{\rm CO}$ versus the effective visual extinction $A_{\rm V, eff}$. 
For a grid cell at position $\vec{x}$, 
we define the effective visual extinction $A_{\rm V, eff}$ as
\begin{equation}
A_{\rm V, eff} = - \frac{1}{2.5} \ln \left[ \frac{1}{6}
\left(\sum_{p=1}^3 e^{-2.5A_{\rm V}(x_{p+})} + e^{-2.5A_{\rm V}(x_{p-})} \right) \right] 
\label{aveff}
\end{equation}
where $A_{\rm V}(x_{p+})$ is the visual extinction of material between the
cell and the edge of the volume in the positive direction along the
$x_p$ axis, and $A_{\rm V}(x_{p-})$ is the same in the negative direction.
The choice of the factor of 2.5 occurs because
the CO photodissociation rate scales with the visual extinction $A_{\rm V}$ as 
$\exp(-2.5 A_{\rm V})$. The value of  $A_{\rm V, eff}$ corresponds to the visual extinction
used in our code, in the context of our six-ray approximation, for computing the CO 
photodissociation rate (see Section~\ref{photochem}).

Figure~\ref{xco-ntot}a demonstrates that there is only a weak correlation between the CO
abundance and the number density. At densities $n_{\rm tot} < 10 \: {\rm cm^{-3}}$, most
of the gas has a very low CO abundance, $x_{\rm CO} < 10^{-8}$. On the other hand,
at densities $n_{\rm tot} > 10^{4} \: {\rm cm^{-3}}$, almost all of the carbon in the gas is
found in the form of CO. At densities between these two extremes, however, there is a
very wide spread in $x_{\rm CO}$ at any given $n_{\rm tot}$. For instance, at 
$n_{\rm tot} = 100 \: {\rm cm^{-3}}$, values of $x_{\rm CO}$ between $10^{-9}$ and
$1.4 \times 10^{-4}$ are almost equally probable. This large scatter in $x_{\rm CO}$
is unexpected and demands an explanation. 

A large part of the puzzle is explained by Figure~\ref{xco-ntot}b, which shows that 
there is a correlation between $x_{\rm CO}$ and the effective visual extinction,
$A_{\rm V, eff}$, that is stronger than the correlation between CO fraction and gas 
number density. This is in line with our expectations based on our discussion of the
H$_{2}$:CO column density ratio above, and is also readily explained by considering
the microphysics of CO formation and destruction. Consider the following simplified 
model for the CO abundance. CO is formed from gas phase C or C$^{+}$ by a variety
of reactions, but the most important ones involve either a hydrocarbon radical (e.g.\ CH) or
its ion (e.g.\ CH$^{+}$), or the OH radical or its ion, OH$^{+}$. In cold gas, the formation
of one of these intermediate species is the rate-limiting step in forming CO, as the various
routes by which these species can be formed typically involve either a radiative association
reaction with H$_{2}$, or the cosmic ray ionization of H$_{2}$, both of which are slow
processes. Let us suppose, for simplicity, that reactions involving hydrocarbon radicals and
ions dominate.\footnote{We could construct a very similar model in the case that reactions
with OH and OH$^{+}$ dominate, only with the number density of atomic oxygen, 
$n_{\rm O}$, playing the role of $n_{\rm C, tot}$ above.} In that case, we can write the CO
formation rate as $R_{\rm form} n_{\rm C, tot} n_{\rm H_{2}}$, where $n_{\rm C, tot} = 
n_{\rm C} + n_{\rm C^{+}}$, and where $R_{\rm form}$ is the formation rate of our intermediate 
species, multiplied by a factor that accounts for the fact that some of the intermediate radicals 
and ions will be photodissociated, rather than reacting to form CO (or a further intermediate, 
such as CO$^{+}$, that reacts rapidly to form CO). If CO is primarily destroyed by 
photodissociation, at a rate $R_{\rm pd}  n_{\rm CO}$, then in chemical equilibrium, the 
CO fractional abundance is given by
\begin{equation}
x_{\rm CO} = \left(\frac{R_{\rm form}}{R_{\rm pd}} \right) x_{\rm C, tot} n_{\rm H_{2}}.
\label{xco-toy-1}
\end{equation}
The photodissociation rate $R_{\rm pd}$ can be written in terms of $A_{\rm V, eff}$
as $R_{\rm pd} = 2 \times 10^{-10} f_{\rm sh} \exp(-2.5 A_{\rm V, eff}) $, 
where $f_{\rm sh} = f_{\rm CO} f_{\rm H_{2}}$ is the product of the shielding factors
due to CO self-shielding ($f_{\rm CO}$) and due to the shielding of CO by H$_{2}$
($f_{\rm H_{2}}$) that we introduced in \S\ref{photochem}. We can therefore
rewrite Equation~\ref{xco-toy-1} as
\begin{equation}
x_{\rm CO} = \left(\frac{R_{\rm form}}{2 \times 10^{-10}} \right) f_{\rm sh}^{-1}
e^{2.5A_{\rm V, eff}} x_{\rm C, tot} n_{\rm H_{2}}.
\label{xco-toy-2}
\end{equation}
Consideration of the different processes involving C or C$^{+}$ that lead to the
formation of CH or CH$^{+}$ suggests that $R_{\rm form}$ should have a value
of roughly $10^{-17} \: {\rm cm^{3}} \: {\rm s^{-1}}$, give or take an order of magnitude.
If we assume, again for simplicity, that in fact $R_{\rm form} = 2 \times 10^{-17} \: 
{\rm cm^{3} \: s^{-1}}$, then Equation~\ref{xco-toy-2} becomes
\begin{equation}
x_{\rm CO} = 10^{-7} e^{2.5A_{\rm V, eff}} f_{\rm sh}^{-1} 
x_{\rm C, tot} n_{\rm H_{2}}.
\label{xco-toy-3}
\end{equation}
If we further simplify matters by assuming that  the H$_{2}$ fraction is of order one, and 
that most of the carbon in the gas is still in the form of C or C$^{+}$, so that $x_{\rm C, tot} 
\sim 10^{-4}$, then we obtain the following expression for  $x_{\rm CO}$
\begin{equation}
x_{\rm CO} \sim 10^{-11} e^{2.5A_{\rm V, eff}} f_{\rm sh}^{-1} n.
\label{xco-toy-4}
\end{equation}

Given the large number of assumptions and simplifications that we have made above, this
expression should clearly be treated only as a rough order-of-magnitude estimate of the
CO abundance. Nevertheless, this simplified model does highlight some of the behaviour
that we see for the actual CO abundance in our simulations. Our simplified model predicts 
that the abundance of CO should vary only {\em linearly} with the gas number density, and with 
$f_{\rm sh}^{-1}$ (which itself is a complicated function of $N_{\rm H_{2}}$ and $N_{\rm CO}$), 
but should vary {\em exponentially} with $A_{\rm V, eff}$. Thus, small changes in $A_{\rm V, eff}$ 
will produce a much larger change in $x_{\rm CO}$ than even relatively large changes in $n$, 
and so we would expect to find a much stronger correlation between $x_{\rm CO}$ and 
$A_{\rm V, eff}$ than between $x_{\rm CO}$  and $n_{\rm tot}$, as indeed is the case in our 
simulations (see Figure~\ref{xco-ntot}).

\begin{figure}
\centering
\epsfig{figure=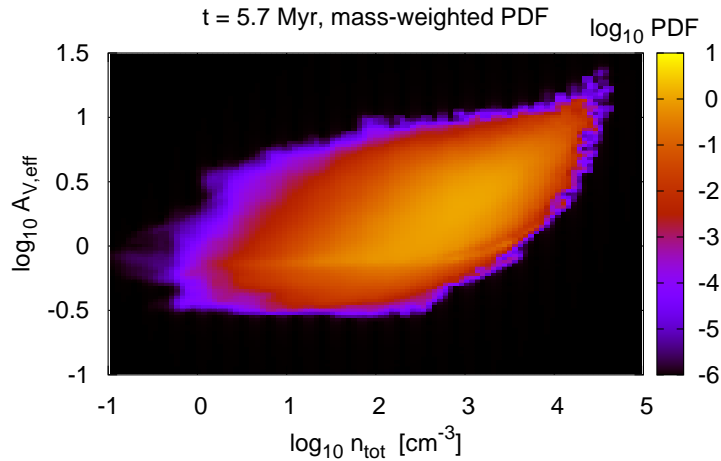,width=25pc,angle=0,clip=}
\caption{Mass-weighted two-dimensional PDF of $A_{\rm V, eff}$ versus
$n_{\rm tot}$.
\label{AV-ntot}}
\end{figure}

If the behaviour of $x_{\rm CO}$ is determined primarily by the amount of shielding by
dust and molecules, rather than by the density, then we would expect to obtain a much
tighter correlation if, instead of plotting $x_{\rm CO}$ against $n_{\rm tot}$, we 
plot $x_{\rm CO}$ against  $n_{\rm tot} \exp(2.5 A_{\rm V, eff})$ or, even better, against
\begin{equation} \xi \equiv
n_{\rm tot} \exp(2.5 A_{\rm V, eff}) f_{\rm sh}^{-1}.
\label{def-xi}
\end{equation}
This we have done in 
Figures~\ref{xco-ntot}c and \ref{xco-ntot}d.  These Figures show that when the CO fraction 
is small, then we obtain a much tighter correlation between $x_{\rm CO}$ and 
   $\xi$ than between $x_{\rm CO}$ and $n_{\rm tot}$,
while accounting for the effects of shielding from CO and H$_{2}$ improves the
correlation even further. This confirms that most of the scatter we see in the 
$x_{\rm CO}$--$n_{\rm tot}$ plot at low $x_{\rm CO}$ is due to the fact that the CO
fraction is far more sensitive to changes in the amount of shielding than to changes 
in the density, and that the amount of shielding is not well correlated with the gas
density. This point is further emphasised in Figure~\ref{AV-ntot}, which shows that
there is indeed a large scatter in  the effective visual extinction at most densities.

If the CO abundance is large, then the relationship between $x_{\rm CO}$ and the
amount of shielding becomes quite different, and the clear correlation found at 
low $x_{\rm CO}$ rapidly vanishes. Our simplified model for $x_{\rm CO}$ again can 
give us insight into why this happens. Consider that in order to produce a CO
abundance $x_{\rm CO} \sim 10^{-5}$, our model predicts that  we must have
         $\xi \sim 10^{6}$. For a gas density $n \sim 
300 \: {\rm cm^{-3}}$, this means that 
the CO photodissociation rate is reduced by a factor of roughly 3000 compared
to the optically thin value, i.e.\ $R_{\rm pd}  \sim 10^{-13} \: {\rm s^{-1}}$. The 
corresponding photodissociation timescale is 
then $0.3 \: {\rm Myr}$, which is not negligible in comparison to the dynamical
timescale of the gas, and so it is probably no longer safe to assume either that 
the gas is in chemical equilibrium or that photodissociation dominates the
destruction of CO. As our model was based on both of these assumptions, it
is unsurprising that it breaks down at high $x_{\rm CO}$.

To summarise, what these results are showing us is that in low extinction gas,
the CO abundance is determined by a combination of the extinction and the
density, with the extinction playing the primary role, while at high extinction, 
other physics, such as the dynamical history of the gas, or the impact of other
destruction mechanisms, such as charge transfer with $\Hep$, becomes far
more important.

The relationship between $A_{\rm V, eff}$ and $n_{\rm tot}$, shown in
Figure~\ref{AV-ntot}, is also worth discussing. As noted above, there is no strong 
correlation between these two quantities. However, they are not completely
uncorrelated either. There is a clear deficit of zones with high $A_{\rm V, eff}$ at low 
densities, and there is also a deficit of zones with low $A_{\rm V, eff}$ at high 
densities. The latter is  a consequence of the finite resolution of our simulation, 
since the value of $A_{\rm V, eff}$ in a given zone has a lower bound
\begin{equation}
A_{\rm V, eff} =  \frac{0.5 n \Delta x}{1.87 \times 10^{21} \: {\rm cm^{-2}}},
\end{equation}
where $\Delta x$ is the size of a grid zone, corresponding to absorption within the zone 
itself. However, this does not explain the deficit of points with high extinction and low
density. Instead, this is due to the fact  that there are only very few voids in the density distribution 
that are entirely surrounded by high extinction material. In addition, little of
the mass in the simulation volume is to be found at these very low densities.

Also of interest is the fact that the smallest values of $A_{\rm V, eff}$ found in our
simulation are roughly 0.1--0.3, rather than zero. The reason for this is quite
simple. The regions in the simulation with the lowest values of $A_{\rm V, eff}$ are
those at the edges and corners. These regions are highly exposed to radiation
propagating inwards from the nearest edges of the box. However, at the same time, 
they are very well shielded from radiation propagating inwards from the opposite 
edges of the box, since they are protected by the full width of gas in the simulation.
Therefore, these zones see a mean intensity that is smaller than the mean intensity
that they would see if the opacity in every direction were zero, and hence they have
a non-zero $A_{\rm V, eff}$, even though the visual extinction between them and the
closest edge of the box may be very close to zero. It is also clear that our ``six-ray''
approximation tends to overestimate the degree to which the zones at the edges of
the box are shielded. For instance, consider a zone located at the edge of the box,
in the center of one of the six faces. In one direction it has a very low $A_{\rm V}$;
however, in the other five directions, it may have
a very high $A_{\rm V}$. Therefore, using our approximation we would predict that
it would see a mean intensity of radiation that was roughly one-sixth of the value in
the absence of shielding, while in reality it should see a mean intensity that is only
one-half of the optically thin value. Fortunately, this is only really a problem for zones
right at the edge of our simulation, and most of the gas should not be significantly
affected. To test this, we verified that the removal of 32 zones of material from each
of the edges of our $256^{3}$ simulation did not significantly affect the PDFs reported
in this paper.

\begin{figure}
\centering
\epsfig{figure=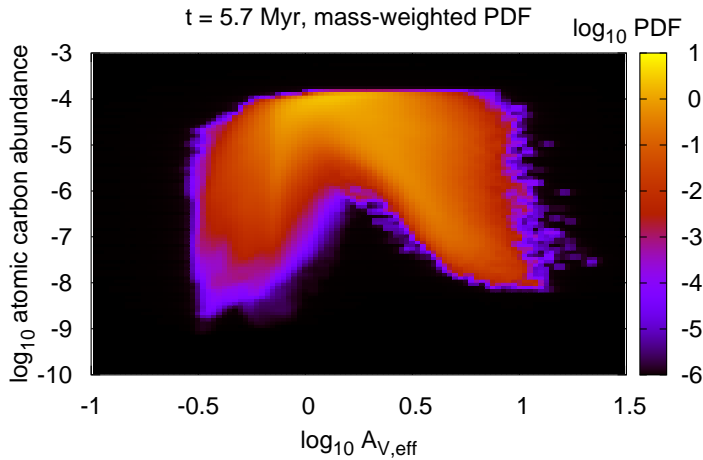,width=25pc,angle=0,clip=}
\caption{Mass-weighted two-dimensional PDF of the fractional abundance
of atomic carbon, $x_{\rm C}$ versus effective visual extinction $A_{\rm V, eff}$.
\label{xcI-AV}}
\end{figure}

\subsection{Comparison to photodissociation region models}
To conclude this section, it is interesting to compare our results for the CO 
distribution with what we would expect based on classical models of uniform
density photodissociation regions \citep[see e.g.][and references
therein]{ht99}.
These predict that we should find an outer shell of C$^{+}$ surrounding a shell 
of neutral carbon, that in
turn surrounds a central region dominated by CO. If we examine how the C
and CO abundances vary with visual extinction (Figures~\ref{xco-ntot}b 
and \ref{xcI-AV}), then they are to some degree consistent with this picture. 
We see a peak in the typical atomic carbon abundance at $A_{\rm V, eff} \sim 1$,
followed by a decline at higher $A_{\rm V, eff}$, as CO begins to dominate.
However, what is apparent in both figures is the large degree of scatter 
around this underlying trend. There are many regions of high extinction in
which the C abundance is very small, as the PDR models would predict,
but at the same time there are also regions in which $x_{\rm C}$ remains
large at high $A_{\rm V}$. Similarly, although most regions with $A_{\rm V} > 2$
are CO dominated, there is also a significant amount of mass with high
extinction that still has a very low CO abundance. 

Some of this scatter is surely due to the highly inhomogeneous nature of
the density field. As we have already seen, the density of gas with a 
given $A_{\rm V, eff}$ can vary over several orders of magnitude. In 
addition, it is likely that turbulent mixing of material between high extinction
and low extinction sightlines also plays an important role in determining
the spatial distribution of C$^{+}$, C and CO \citep[see also][]{gm07b,fed08}.
To properly disentangle these effects lies beyond the scope of this introductory paper, 
but is something we plan to revisit in future work.

\section{Summary}
\label{summ}
In this paper, we have outlined the methods that we have developed to model
the coupled thermal, chemical and dynamical evolution of the turbulent gas
making up giant molecular clouds. We have shown that it is now possible to
perform high-resolution simulations in three dimensions that track the thermal
and chemical evolution of the gas in a realistic fashion. The chemical network
that we use in our simulations is significantly simplified compared to the most
detailed models available, but nevertheless performs with acceptable 
accuracy for our purposes, and the largest errors in our models come not from
our simplified chemistry, but from our approximate treatment of the effects of 
the external radiation field.

We have performed simulations with numerical resolutions of $64^{3}$, 
$128^{3}$ and $256^{3}$, and have demonstrated that most of our results
are well-converged in our $256^3$ run. The main exception is the high
density tail of the density PDF, which is not fully converged
at a resolution of $256^{3}$ zones (although the peak of the density PDF
is well converged). This results from the improvement of our ability to 
resolve dense post-shock gas with increasing numerical resolution, and
is also responsible for the small dependence on resolution that remains
in our determination of the O$_{2}$ and H$_{2}$O abundances, and in 
the gas temperature distribution at the lowest temperatures. 

We have also quantified the timescales on which various quantities of 
interest reach a statistical steady state. The density PDF, which has the
same log-normal form as found in many previous simulations of 
interstellar turbulence, reaches a steady state after roughly one and a half 
turbulent crossing times, as does the H$_{2}$ number density distribution in all but
the lowest density gas. On the other hand, the CO number density 
distribution takes closer to two crossing times to settle into a steady state,
and the temperature distribution still has 
not reached a final steady state after three crossing times.

We find that CO formation occurs rapidly in the
dense, turbulent gas studied here. Most of the CO is produced within
the first 2--3~Myr of the simulation, i.e. within 1--2 turbulent 
crossing times. For comparison, most of the H$_{2}$ produced in the 
simulation forms within a single crossing time. These short chemical
timescales suggest that the limiting timescale in the formation of a
molecular cloud is not the time required to convert the hydrogen to
H$_{2}$ and the carbon to CO, but rather the time required to 
assemble the cloud material from the low density ISM. Once large
enough spatial and column densities are reached, conversion of the
gas to molecular form should follow very quickly.

Our simulations also demonstrate that the gas in molecular clouds
is not isothermal. Most of the 
molecular gas has a low temperature, in the range of 10--20~K, but
there is also a clear power-law tail in the temperature PDF at higher 
temperatures, made up of low extinction gas heated by photoelectric
emission from dust grains.

We have found that the CO abundances produced in our 
simulations are not well correlated with the gas density. This 
means that the use of a density cut to identify molecular regions
in turbulence simulations that do not self-consistently follow the
chemical evolution of the gas is dangerous,
and may give misleading results. Similarly, the assumption that
all observed CO is at roughly the same gas density may be
similarly misleading, as is the assumption that all of the CO in
a molecular cloud is located at densities that are high enough
for its rotational lines to be thermalized.

The poor correlation that we find between CO abundance and
gas density is explained by the sensitivity of the CO to the amount
of UV shielding provided by the dust, H$_{2}$ and CO. Variations
in the amount of shielding produce a far greater variation in the
CO abundance than do variations in the gas density. In addition, the amount 
of shielding provided by the gas is not well correlated with the 
density, owing to the highly inhomogeneous nature of the density
field. The important role that density inhomogeneities play in 
determining the chemical structure of molecular clouds has long
been recognized \citep[see e.g.][]{sg90,gsw92,wbs95}, but the most sophisticated current
PDR models continue to rely on a somewhat artificial picture of
their structure, visualising them as an ensemble of clumps with
well-defined densities surrounded by a much lower density
interclump medium \citep[e.g.][]{sun08,kra08}.  The picture suggested by our
simulations is rather different. We find a continuous  density
distribution, with no clear separation between `clump' and
`interclump' material, and with continual turbulent mixing of
gas between low-density and high-density regions. We look
forward to exploring the observational consequences of this
picture in greater detail in future work.


\section*{Acknowledgments}
The authors would like to thank R.~Banerjee,  P.~Brand,  P.~Clark, J.~Duval,
A.~Goodman, F.~Heitsch, P.~Hennebelle, C.~McKee, J.~Pineda, 
N.~Schneider-Bontemps,  and E.~V\'azquez-Semadeni 
for useful discussions on the physics and 
chemistry of molecular clouds. The simulations presented in this paper were 
performed on the IBM Blue Gene/P computer Babel at  the Institute for Development 
and Resources in Intensive Scientific computing (IDRIS) as part of the DEISA 
Extreme Computing Initiative project CHPTMC. In addition, we also made use of
computing facilities provided by the Parallel Computing Facility of the American 
Museum of Natural History, the Texas Advanced Computing Center and the
Forschungzentrum J\"ulich. SCOG acknowledges financial support from
DFG grant no.\ KL1358/4. 
CF acknowledges financial support by the International
Max Planck Research School for Astronomy and Cosmic Physics (IMPRS-HD),
and the Heidelberg Graduate School of Fundamental Physics (HGSFP). 
The HGSFP is funded by the Excellence Initiative of the DFG under grant 
number GSC 129/1. RSK acknowledges support from a Frontier grant
of Heidelberg University, also sponsored by the German Excellence 
Initiative, from the DFG SFB {\em Galaxies in the Early Universe}, 
and from the Landesstiftung Baden-W{\"u}rttemberg via their program 
{\em International Collaboration II}. SCOG, CF and RSK are also grateful 
for support from the ASTRONET project STAR FORMAT (05A09VHA).

\appendix

\section{Ensuring consistent advection of chemical species}
\label{cma}
\subsection{Consistent Multi-fluid Advection (CMA)}
Let $x_{m}(i,j,k)$ be the fractional abundance relative to hydrogen
of chemical species $m$ in grid zone $(i,j,k)$, let $X_{n}(i,j,k)$
be the elemental abundance, relative to hydrogen, of element
$n = {\rm H}, {\rm He}, {\rm C}$ or O in grid zone $(i,j,k)$, and let
$N_{n, m}$ be the number of nuclei of element $n$ in species $m$.
Then for each element $n$, we can construct a constraint equation
of the form
\begin{equation}
\sum_{m} N_{n, m} x_{m}(i,j,k) = X_{n}(i,j,k). 
\end{equation}
Now, if we assume that the elemental composition of the gas remains
the same throughout our simulation volume, i.e.\ that the elemental 
abundances $X_{n}$ are independent of position,\footnote{This is a reasonable
assumption in small-scale simulations of the local interstellar medium, where
we can treat all of the gas as having the same metallicity \citep[see e.g.][]{meyer98}, 
but will not hold on scales large enough that Galactic metallicity gradients become
important. Note also that we are only assuming a homogeneous metallicity,
not homogeneous chemical abundances.} then these constraint equations become
\begin{equation}
\sum_{m} N_{n, m} x_{m}(i,j,k) = X_{n}. \label{basic_cons}
\end{equation}
Although it is trivial to ensure that these constraints are satisfied at the
start of our simulations, ensuring that they remain satisfied as the gas
is evolved forward in time is not so simple. 

The root cause of the problem is to be found in the advection scheme. 
Most Eulerian treatments of multi-species flows model the species
abundances $x_{m}$ as passive scalars that are advected with the
flow. If we consider a one-dimensional scheme, for simplicity, then
we can write the advection equation for a given species $m$ in 
grid zone $i$ in finite volume form as
\begin{equation}
\rho_{i} x_{m, i}(t + \Delta t) \Delta V_{i} = \rho_{i} x_{m, i}(t) \Delta V_{i}
 - \Delta t \left[A_{i+1/2} F_{i+1/2} - A_{i-1/2} F_{i-1/2} \right], \label{fv_advect}
\end{equation}
where $F_{i+1/2} = \rho_{i+1/2} v_{i+1/2} x_{m, i+1/2}$ is the flux of $x_{m}$ 
from zone $i$ into zone $i+1$, $F_{i-1/2} = \rho_{i-1/2} v_{i-1/2} x_{m, i-1/2}$
is the flux of  $x_{m}$ from zone $i-1$ into zone $i$, $A_{i-1/2}$ and $A_{i+1/2}$
are the areas of the interfaces between zones $i-1$ and $i$, and between
zones $i$ and $i+1$, respectively, and $\Delta V_{i}$ is the zone volume.

If the elemental abundances $X_{n}$ are independent of position, then
at both interfaces of zone $i$, the species fluxes should satisfy the 
constraint equations
\begin{equation}
\left[ \sum_{m} N_{n, m} x_{m, i\pm1/2} \right] \rho_{i\pm1/2} v_{i\pm1/2}
= X_{n} \rho_{i\pm1/2} v_{i\pm1/2}. \label{flux_cons}
\end{equation}
In order to ensure stability, fluxes are often computed
from the distribution of the flow variables upstream of the interface
(``upwinding'') using some interpolation scheme. For instance, in
ZEUS-MP, which uses \citet{vl77} advection, the upwinded interpolated 
value $q_{i}^{*}$ of a zone-centered scalar $q$ at the negative interface
of zone $i$ is given by 
\citep{sn92a}
\begin{equation}
q_{i}^{*} = \left \{ \begin{array}{lr}
q_{i-1} + (\Delta x_{i-1} - v_{i-1/2} \Delta t)(dq_{i-1} / 2) & \hspace{.5in} v_{i-1/2} > 0 \\
q_{i} + (\Delta x_{i} + v_{i-1/2} \Delta t)(dq_{i} / 2) & \hspace{.5in} v_{i-1/2} < 0
\end{array} \right.
\label{vla}
\end{equation}
where $q_{i}$ is the zone-centered value of $q$ in zone $i$, $\Delta x_{i}$ is the
size of zone $i$, and $dq_{i}$ is the monotonized van Leer slope,
given by
\begin{equation}
dq_{i} = \left \{ \begin{array}{lr}
\frac{2(\Delta q_{i-1/2} \Delta q_{i+1/2})}{\Delta q_{i-1/2} + \Delta q_{i+1/2}} & \hspace{.5in}
\Delta q_{i+1/2} \Delta q_{i-1/2} > 0 \\
0 & \hspace{.5in} \Delta q_{i+1/2} \Delta q_{i-1/2} \leq 0
\end{array} \right.
\end{equation}
where $\Delta q_{i+1/2} = (q_{i+1} - q_{i}) / \Delta x_{i}$. (Note that in Equation~\ref{vla}, 
we have neglected any motion of the underlying coordinate grid, for simplicity).
Given $q_{i}^{*}$, the flux then follows from $F_{i-1/2} = q_{i}^{*} v_{i-1/2}$.
Also in common usage is the Piecewise Parabolic Method (PPM)
of \citet{cw84}, which uses higher-order interpolation, and so produces
a better quality solution (at the cost of increased computational effort).

Unfortunately, as several authors have noted \citep{fma89,larr91,pm99}, 
the fact that the interpolation profiles in these schemes are constructed independently 
for each chemical species means that there is no guarantee that the resulting
fluxes actually satisfy Eq.~\ref{flux_cons}. In general, they do not do so. Moreover, 
this violation occurs even when the underlying advection scheme is 
conservative.  Thus, even if our initial chemical abundances satisfy the constraints
represented by Equation~\ref{basic_cons}, as soon as we begin advecting them with a standard 
higher-order scheme, these constraints will be violated.

This problem is often dealt with by a renormalization of species abundances following
the advection step to ensure that Eq.~\ref{basic_cons} is satisfied.
However, as \citet{pm99} note, this procedure lacks any formal justification and can lead 
to large systematic errors in the abundances of the least abundant species. It can also destroy
the conservative nature of the scheme. \citet{pm99} suggest an improved way of
dealing with this problem, which they term the Consistent Multi-fluid Advection (CMA)
method. They consider the case in which one has only a single constraint equation
\begin{equation}
\sum_{m} N_{1,m} x_{m}(i,j,k) = 1, \label{cma_const}
\end{equation}
where we have chosen to set $X_{n} = 1$ for simplicity. The example they consider
is a fluid consisting of many different elements (but no composite molecules), whose 
mass fractions must sum to unity. An alternative example, more in keeping with
our discussion here, is a gas consisting of only one element (e.g.\ hydrogen) that
can form several different stable chemical species (e.g.\ $\Hp$, $\Hm$, $\mH$, $\mHt$).
They show that in this case, one can satisfy Equation~\ref{cma_const} 
while preserving the conservative nature of the advection scheme by replacing the 
interpolated abundances  $x_{i\pm1/2,m}$ used to construct the fluxes in 
Equation~\ref{fv_advect} with the modified abundances
\begin{equation}
\chi_{i\pm1/2,m} = \frac{x_{i\pm1/2,m}}{\sum_{m} N_{1,m} x_{i\pm1/2,m}}.
\end{equation}
In other words, instead of ensuring that the constraint equation is satisfied by
normalizing the abundances after the advection step, we ensure that it is
satisfied by normalizing the fluxes {\em during} the advection step.

When dealing with the PPM code, additional modifications are necessary to 
avoid problems related to PPM's contact discontinuity detection algorithm,
as discussed in Section~2.3 of \citet{pm99}. However, these problems do not
occur with the simpler van Leer advection scheme used in ZEUS-MP, and 
so we do not discuss them further here. 

\subsection{Extending the CMA algorithm}
Although \citeauthor{pm99}'s CMA algorithm is simple to implement and is highly
effective in practice, its applicability to chemically reacting flow is limited. The
problem comes from the assumption that the chemical abundances must satisfy
only a single constraint equation. In a flow containing compounds of multiple
elements, such as CO, this is not the case. In such a flow, our modified 
interface abundances for carbon and oxygen must satisfy both
\begin{equation}
\sum_{m} N_{{\rm C}, m} x_{m} = X_{\rm C}
\end{equation}
and
\begin{equation}
\sum_{m} N_{{\rm O}, m} x_{m} = X_{\rm O}
\end{equation}
If we rescale the abundances of those species containing carbon according to
\begin{equation}
\chi_{i\pm1/2,m} = \frac{X_{\rm C}}{\sum_{m} N_{{\rm C},m} x_{i\pm1/2,m}}  x_{i\pm1/2,m}
\end{equation}
and those species containing oxygen according to
\begin{equation}
\chi_{i\pm1/2,m}^{\prime} = \frac{X_{\rm O}}{\sum_{m} N_{{\rm O},m} x_{i\pm1/2,m}}  x_{i\pm1/2,m}
\end{equation}
then we will find, in general, that for species containing both carbon and oxygen,
$\chi_{i\pm1/2,m}^{\prime} \neq \chi_{i\pm1/2,m}$. We therefore cannot use the simple
CMA prescription to rescale the interface abundances (and the fluxes) in this kind of flow.

To avoid this problem, we propose an extension of the CMA algorithm, which we term
modified CMA (or MCMA, for short). The basic idea behind this algorithm is the same
as that motivating the CMA algorithm: we aim to modify the chemical abundances used
to construct the species fluxes, such that the fluxes satisfy the appropriate chemical 
abundance constraint equations, while the advection scheme itself remains conservative.
We accomplish this by writing the rescaled abundances as
\begin{equation}
\chi_{i\pm1/2,m} = \left( \sum_{k} \eta_{k} \frac{N_{k,m}}{N_{{\rm tot, m}}} \right) x_{i\pm1/2,m}
\end{equation}
where $N_{k, m}$ is the number of nuclei of element $k$ in species $m$, $N_{{\rm tot}, m}$
is the total number of nuclei in species $m$, and the $\eta_{k}$ are correction factors chosen
so that the set of constraint equations 
\begin{equation}
\sum_{m} N_{k,m} \chi_{i\pm1/2,m} = X_{k}
\end{equation}
are simultaneously satisfied. The required correction factors $\eta_{k}$ can be found
by solving the matrix equation
\begin{equation}
\mathbf{M \eta} = \mathbf{X}
\end{equation}
where $\mathbf{\eta} = (\eta_{1}, ..., \eta_{k})$, $\mathbf{X} = (X_{1}, ..., X_{k})$, and
where the elements of the matrix $\mathbf{M}$ are given by
\begin{equation}
M_{kl}  = \sum_{m} \frac{N_{k,m}}{N_{{\rm tot}, m}} N_{l,m} x_{i\pm1/2,m},
\end{equation}
where we sum over all species $m$. In a flow with a mix of species containing only 
a single element, $N_{1,m} / N_{\rm tot, m} = 1$ for all $m$, and this prescription
reduces to \citeauthor{pm99}'s CMA scheme.

The main drawback of this scheme compared to the original \citet{pm99} scheme
is that we must now perform a matrix inversion for every flux on every advection step.
However, the rank of this matrix scales only as the number of elements, {\em not}
the number of species, and so this additional cost will generally be dwarfed by the
cost of solving the chemical rate equations themselves. 

\section{List of chemical reactions}

\begin{table*}
\begin{minipage}{126mm}
\caption{List of collisional gas-phase reactions included in our chemical model \label{chem_gas}}
\begin{tabular}{llllc}
\hline
No.\  & Reaction & Rate coefficient $({\rm cm}^{3} \: {\rm s}^{-1})$ & Notes & Ref.\ \\
\hline
1 & $\mH  + \me  \rightarrow \Hm + \gamma$ & 
$k_{1} = {\rm dex}[-17.845 + 0.762 \log{T} + 0.1523 (\log{T})^{2}$ & & 1 \\
& & $\phantom{k_{1}= {\rm dex}[} \mbox{} - 0.03274 (\log{T})^{3}] $ &
$T \le 6000 \: {\rm K}$ & \\
& & $ \phantom{k_{1}} = {\rm dex}[-16.420 + 0.1998 (\log{T})^{2}$ & & \\ 
& & $ \phantom{k_{1} = {\rm dex}} \mbox{}-5.447  \times 10^{-3}  (\log{T})^{4}$ & & \\ 
& & $ \phantom{k_{1} = {\rm dex}} \mbox{}+ 4.0415 \times 10^{-5} (\log{T})^{6}]$ 
& $T > 6000 \: {\rm K}$ & \\
2 & $\Hm  + \mH  \rightarrow \mHt + \me$ &
$k_{2} = 1.5 \times 10^{-9}$ & $T \le 300 \: {\rm K}$& 2 \\
& & $\phantom{k_{2}} = 4.0 \times 10^{-9} T^{-0.17}$ & $T > 300 \: {\rm K}$ & \\
3 & $\mH  + \Hp  \rightarrow \mHtp + \gamma$ &
$k_{3} = {\rm dex}[-19.38 - 1.523 \log{T} $ & & 3 \\
& & $\phantom{k_{3}} \mbox{}+1.118 (\log{T})^{2}  - 0.1269 (\log{T})^{3}]$ & & \\
4 & $\mH + \mHtp \rightarrow \mHt + \Hp$ & $k_{4} = 6.4 \times 10^{-10}$ & & 4 \\
5 & $\Hm  + \Hp  \rightarrow \mH + \mH$ & $k_{5} = 2.4 \times 10^{-6} T^{-1/2} (1.0 + T / 20000)$
& & 5 \\
 6 & $\mHtp + \me \rightarrow \mH + \mH$ & $k_{6} = 
 1.0\times 10^{-8}$ &  $T \le 617 \: {\rm K}$ & 6 \\
 & & $\phantom{k_{6}} = 1.32 \times 10^{-6} T^{-0.76}$ & $T > 617 \: {\rm K}$ & \\
7 & $\mHt + \Hp  \rightarrow \mHtp + \mH$ &
$k_{7} = [- 3.3232183 \times 10^{-7}$ & & 7 \\
 & & $\phantom{k_{7}=}  \mbox{} + 3.3735382 \times 10^{-7}  \ln{T}$ & & \\
 & & $\phantom{k_{7}=}  \mbox{} - 1.4491368 \times 10^{-7}  (\ln{T})^2$ & & \\
 & & $\phantom{k_{7}=}  \mbox{} + 3.4172805 \times 10^{-8}  (\ln{T})^3$ & & \\
 & & $\phantom{k_{7}=}  \mbox{} - 4.7813720 \times 10^{-9}  (\ln{T})^4$ & & \\
 & & $\phantom{k_{7}=}  \mbox{} + 3.9731542 \times 10^{-10} (\ln{T})^5$ & & \\
 & & $\phantom{k_{7}=}  \mbox{}  - 1.8171411 \times 10^{-11}  (\ln{T})^6$ & & \\
 & & $\phantom{k_{7}=}  \mbox{}  + 3.5311932 \times 10^{-13} (\ln{T})^7 ]$ & & \\
 & & $\phantom{k_{7}=} \mbox{} \times \exp \left(\frac{-21237.15}{T} \right)$ & & \\
8 & $\mHt + \me  \rightarrow  \mH + \mH +  \me$ &
$ k_{8} = 3.73 \times 10^{-9} T^{0.1121} \exp\left(\frac{-99430}{T}\right) $
& & 8 \\
9 & $\mHt + \mH  \rightarrow  \mH + \mH + \mH$  & 
$ k_{\rm 9, l} = 6.67 \times 10^{-12} T^{1/2} \exp \left[-(1+ \frac{63590}{T}) \right]$ & & 9 \\
& & $k_{\rm 9, h} = 3.52 \times 10^{-9} \expf{-}{43900}{T}$ & & 10 \\
& & $n_{\rm cr, H} = {\rm dex}\left[3.0 - 0.416 \log \left(\frac{T}{10000}\right) - 0.327  
\left\{\log \left(\frac{T}{10000}\right)\right\}^{2} \right]$ & & 10 \\
10 & $\mHt + \mHt \rightarrow  \mHt + \mH + \mH$ & 
$k_{\rm 10, l} = \frac{5.996 \times 10^{-30} T^{4.1881}}{(1.0 + 6.761 \times 10^{-6} T)^{5.6881}}
 \exp \left(-\frac{54657.4}{T} \right)$ & & 11 \\
& & $k_{\rm 10, h} = 1.3 \times 10^{-9} \expf{-}{53300}{T}$ & & 12 \\
& & $n_{\rm cr, H_{2}} = {\rm dex}\left[4.845 - 1.3 \log \left(\frac{T}{10000}\right) + 1.62  
\left\{\log \left(\frac{T}{10000}\right)\right\}^{2} \right]$ & & 12 \\
11 & $\mH  + \me  \rightarrow \Hp + \me + \me$ & 
$k_{11} =  \exp[-3.271396786 \times 10^{1}$ & & 13 \\
& & $\phantom{k_{11}=}  \mbox{}  + 1.35365560 \times 10^{1} \ln T_{\rm e}$ & & \\
& & $\phantom{k_{11}=}  \mbox{}  - 5.73932875 \times 10^{0} (\ln T_{\rm e})^{2}$ & & \\
& & $\phantom{k_{11}=}  \mbox{}  + 1.56315498 \times 10^{0} (\ln T_{\rm e})^{3}$ & & \\
& & $\phantom{k_{11}=}  \mbox{}  -  2.87705600 \times 10^{-1} (\ln T_{\rm e})^{4}$ & & \\
& & $\phantom{k_{11}=}  \mbox{}  + 3.48255977 \times 10^{-2} (\ln T_{\rm e})^{5}$ & & \\
& & $\phantom{k_{11}=}  \mbox{}   - 2.63197617 \times 10^{-3} (\ln T_{\rm e})^{6}$ & & \\ 
& & $\phantom{k_{11}=}  \mbox{}  + 1.11954395\times 10^{-4} (\ln T_{\rm e})^{7}$ & & \\
& & $\phantom{k_{11}=}  \mbox{}   -  2.03914985 \times 10^{-6} (\ln T_{\rm e})^{8}]$ & & \\
12 & $\Hp  + \me  \rightarrow \mH +  \gamma$ & 
 $k_{12, {\rm A}} = 1.269 \times 10^{-13} \left(\frac{315614}{T}\right)^{1.503}$ & Case A & 14 \\
 & & $\phantom{k_{13}=} \mbox{} \times
 [1.0+ \left(\frac{604625}{T}\right)^{0.470}]^{-1.923} $ & & \\
& & $k_{12, {\rm B}} = 2.753 \times 10^{-14} \left(\frac{315614}{T}\right)^{1.500}$ & Case B & 14 \\
 & & $\phantom{k_{13}=} \mbox{} \times
 [1.0+ \left(\frac{115188}{T}\right)^{0.407}]^{-2.242} $ & & \\
13 & $\Hm  + \me  \rightarrow \mH + \me + \me$ &
$ k_{13}  = \exp [-1.801849334 \times 10^{1}$ & & 13 \\
& & $\phantom{k_{15}=}  \mbox{} + 2.36085220 \times 10^{0} \ln T_{\rm e}$ & & \\
& & $\phantom{k_{15}=} \mbox{} - 2.82744300 \times 10^{-1} (\ln T_{\rm e})^{2}$ & & \\
& & $\phantom{k_{15}=} \mbox{}  +1.62331664\times 10^{-2} (\ln T_{\rm e})^{3}$ & & \\
& & $\phantom{k_{15}=} \mbox{} -3.36501203 \times 10^{-2} (\ln T_{\rm e})^{4}$ & & \\
& & $\phantom{k_{15}=} \mbox{}   +1.17832978\times 10^{-2} (\ln T_{\rm e})^{5}$ & & \\ 
& & $\phantom{k_{15}=} \mbox{}  -1.65619470\times 10^{-3} (\ln T_{\rm e})^{6}$ & & \\ 
& & $\phantom{k_{15}=} \mbox{}   +1.06827520\times 10^{-4} (\ln T_{\rm e})^{7}$ & & \\
& & $\phantom{k_{15}=} \mbox{}  -2.63128581\times 10^{-6} (\ln T_{\rm e})^{8} ]$ & & \\
\hline
\end{tabular}
\end{minipage}
\end{table*}
 
\begin{table*}
\begin{minipage}{126mm}
\contcaption{}
\begin{tabular}{llllc}
\hline 
14 & $\Hm  + \mH  \rightarrow  \mH + \mH +  \me$ &
$k_{14} = 2.5634 \times 10^{-9} T_{\rm e}^{1.78186}$ & $ T_{\rm e} \le 0.1 \: \rm{eV}$ & 13 \\
& & $\phantom{k_{14}} = \exp[-2.0372609 \times 10^{1}$ & & \\
& & $\phantom{k_{14}=}  \mbox{}+1.13944933 \times 10^{0} \ln T_{\rm e}$ & & \\
& & $\phantom{k_{14}=} \mbox{}-1.4210135 \times 10^{-1} (\ln T_{\rm e})^{2}$ & & \\
& & $\phantom{k_{14}=} \mbox{}+8.4644554 \times 10^{-3} (\ln T_{\rm e})^{3}$ & & \\
& & $\phantom{k_{14}=}  \mbox{}-1.4327641 \times 10^{-3} (\ln T_{\rm e})^{4}$  & & \\
& & $\phantom{k_{14}=}  \mbox{}+2.0122503 \times 10^{-4} (\ln T_{\rm e})^{5}$ & & \\
& & $\phantom{k_{14}=}  \mbox{}+8.6639632 \times 10^{-5} (\ln T_{\rm e})^{6}$ & & \\
& & $\phantom{k_{14}=}  \mbox{}-2.5850097 \times 10^{-5} (\ln T_{\rm e})^{7}$ & & \\
& & $\phantom{k_{14}=} \mbox{}+ 2.4555012\times 10^{-6} (\ln T_{\rm e})^{8}$ & & \\
& & $\phantom{k_{14}=} \mbox{} -8.0683825\times 10^{-8} (\ln T_{\rm e})^{9}]$ & 
$T_{\rm e} > 0.1 \: \rm{eV}$ & \\ 
15 & $\Hm + \Hp   \rightarrow \mHtp + \me$ & 
$k_{15}= 6.9\times 10^{-9}  T^{-0.35}$  & $T \le 8000 \: {\rm K}$ & 15 \\
 & & $\phantom{k_{15}} = 9.6 \times 10^{-7} T^{-0.90}$ & $T > 8000 \: {\rm K}$ & \\
16 & $\He + \me \rightarrow \Hep + \me + \me$ &
$k_{16} =  \exp[-4.409864886 \times 10^{1}$ & & 13 \\
& & $\phantom{k_{16} = } \mbox{}  + 2.391596563 \times 10^{1} \ln T_{\rm e}$ & & \\
& & $\phantom{k_{16} = } \mbox{}  - 1.07532302 \times 10^{1} (\ln T_{\rm e})^{2}$ & & \\
& & $\phantom{k_{16} = } \mbox{}  + 3.05803875 \times 10^{0} (\ln T_{\rm e})^{3}$ & & \\
& & $\phantom{k_{16} = } \mbox{}  -  5.6851189 \times 10^{-1} (\ln T_{\rm e})^{4}$ & & \\
& & $\phantom{k_{16} = } \mbox{}  + 6.79539123 \times 10^{-2} (\ln T_{\rm e})^{5}$ & & \\
& & $\phantom{k_{16} = } \mbox{}   - 5.0090561 \times 10^{-3} (\ln T_{\rm e})^{6}$ & & \\ 
& & $\phantom{k_{16} = } \mbox{}  + 2.06723616\times 10^{-4} (\ln T_{\rm e})^{7}$ & & \\
& & $\phantom{k_{16} = } \mbox{}   -  3.64916141 \times 10^{-6} (\ln T_{\rm e})^{8}]$ & & \\
17 & $\Hep + \me \rightarrow \He + \gamma$ & 
$k_{17, {\rm rr, A}} = 10^{-11} T^{-0.5} \left[12.72 - 1.615 \log{T} \right. $ & Case A & 16 \\
& & $\left. \phantom{k_{17, {\rm rr, A}} = } \mbox{} - 0.3162 (\log{T})^{2} + 0.0493 (\log{T})^{3}\right]$ & & \\
& & $k_{17, {\rm rr, B}} = 10^{-11} T^{-0.5} \left[11.19 - 1.676 \log{T} \right. $ & Case B & 16 \\
& & $\left. \phantom{k_{17, {\rm rr, A}} = } \mbox{} - 0.2852 (\log{T})^{2} + 0.04433 (\log{T})^{3} \right]$ & & \\
& & $k_{17, {\rm di}} = 1.9 \times 10^{-3} T^{-1.5} \expf{-}{473421}{T}$ & & \\
& & $\phantom{k_{17, {\rm di}} = } \mbox{} \times \left[1.0 + 0.3 \expf{-}{94684}{T} \right] $ & & 17 \\
18 & $\Hep + \mH \rightarrow \He + \Hp$ & 
$k_{18} = 1.25 \times 10^{-15} \left(\frac{T}{300}\right)^{0.25}$ & & 18 \\
19 & $\He + \Hp \rightarrow \Hep + \mH$ &
$k_{19} = 1.26 \times 10^{-9} T^{-0.75} \expf{-}{127500}{T}$ & $ T \leq 10000 \: {\rm K}$ & 19 \\
& & $\phantom{k_{19}} = 4.0 \times 10^{-37} T^{4.74}$ & $T > 10000 \: {\rm K}$ &  \\
20 & $\Cp + \me \rightarrow \mC  + \gamma$ &
$k_{20} = 4.67 \times 10^{-12}  \left(\frac{T}{300}\right)^{-0.6}$ & $T \le 7950 \: {\rm K}$ & 20 \\
& & $\phantom{k_{20} } =1.23 \times 10^{-17}  \left(\frac{T}{300}\right)^{2.49} 
\exp \left(\frac{21845.6}{T} \right)$ & $ 7950 \: {\rm K} < T \le 21140 \: {\rm K}$ & \\
& & $\phantom{k_{20}} = 9.62 \times 10^{-8} \left(\frac{T}{300}\right)^{-1.37} 
 \exp \left(\frac{-115786.2}{T} \right)$ & $T > 21140 \: {\rm K}$ & \\
 21 & $\Op + \me  \rightarrow \mO + \gamma$ &
$k_{21} = 1.30 \times 10^{-10} T^{-0.64}$ &  $T \le 400 \: {\rm K}$ & 21 \\
& & $\phantom{k_{21}} = 1.41 \times 10^{-10} T^{-0.66} + 7.4 \times 10^{-4}  T^{-1.5}$ & & \\
& & $\phantom{k_{21}=} \mbox{} \times \exp \left(-\frac{175000}{T}\right) [1.0 + 0.062 \times 
\exp \left(-\frac{145000}{T}\right) ]$ & $T > 400 \: {\rm K}$ & \\
22 & $\mC  + \me  \rightarrow \Cp  + \me + \me$ & 
$k_{22} = 6.85 \times 10^{-8} (0.193 + u)^{-1} u^{0.25} e^{-u}$ & $u = 11.26 / T_{\rm e}$ & 22 \\
23 & $\mO  + \me  \rightarrow \Op  + \me + \me$ &
$k_{23} = 3.59 \times 10^{-8} (0.073 + u)^{-1} u^{0.34} e^{-u}$ & $u = 13.6 / T_{\rm e}$ & 22 \\
24 & $\Op  + \mH  \rightarrow \mO  + \Hp$ &
$ k_{24} = 4.99 \times 10^{-11} T^{0.405} +
7.54 \times 10^{-10} T^{-0.458} $ & & 23 \\
25 & $\mO  + \Hp  \rightarrow \Op  + \mH$ &
$k_{25} = [1.08 \times 10^{-11} T^{0.517} $ & & 24 \\
& & $\phantom{k_{25} = } \mbox{} + 4.00 \times 10^{-10} T^{0.00669}] \exp 
\left(-\frac{227}{T}\right)$ & & \\
 26 & $\mO + \Hep \rightarrow \Op + \He$ & 
 $k_{26} = 4.991 \times 10^{-15} \left(\frac{T}{10000}\right)^{0.3794} 
 \expf{-}{T}{1121000}$ & & 25 \\
 & & $\phantom{k_{26} = } \mbox{} + 2.780 \times 10^{-15} 
 \left(\frac{T}{10000}\right)^{-0.2163} \expf{}{T}{815800}$ & & \\
27 & $\mC  + \Hp  \rightarrow \Cp  + \mH$ & $k_{27} = 3.9 \times 10^{-16} T^{0.213}$ & & 24 \\
28 & $\Cp  + \mH  \rightarrow \mC  + \Hp$ & $k_{28} = 6.08 \times 10^{-14} 
\left(\frac{T}{10000}\right)^{1.96} \expf{-}{170000}{T}$ & & 24 \\
 29 & $\mC + \Hep \rightarrow \Cp + \He$ & 
 $k_{29} = 8.58 \times 10^{-17}  T^{0.757}$ & $T \leq 200 \: {\rm K}$ & 26 \\
 & & $\phantom{k_{29}} = 3.25 \times 10^{-17} T^{0.968}$ & $200 < T \leq 2000 \: {\rm K}$ & \\
 & & $\phantom{k_{29}} = 2.77 \times 10^{-19} T^{1.597}$ & $T > 2000 \: {\rm K}$ & \\
30 & $\mHt + \He \rightarrow \mH + \mH + \He$ &
$k_{\rm 30, l} = {\rm dex}\left[\mbox{}-27.029 + 3.801 \log{(T)} - 29487/T \right]$ & & 27 \\ 
& & $k_{\rm 30, h} = {\rm dex}\left[-2.729-1.75 \log{(T)} - 23474/T\right]$ & & \\
& & $n_{\rm cr, He} = {\rm dex} \left[5.0792 (1.0 - 1.23 \times 10^{-5} (T - 2000) \right]$ & & 27 \\
31 & $\oh + \mH \rightarrow \mO + \mH + \mH $ & $k_{31} = 6.0 \times 10^{-9} \expf{-}{50900}{T}$ & & 28 \\
32 & ${\rm HOC^{+}} + \mHt \rightarrow {\rm HCO^{+}} + \mHt $ & $k_{32} = 3.8 \times 10^{-10}$ & & 29 \\
33 & ${\rm HOC^{+}} + \co \rightarrow {\rm HCO^{+}} + \co $ & $k_{33} = 4.0 \times 10^{-10}$ & & 30 \\
34 & $\mC + \mHt \rightarrow \ch + \mH $ &
$k_{34} = 6.64 \times 10^{-10} \expf{-}{11700}{T}$ & & 31 \\
35 & $\ch + \mH \rightarrow \mC + \mHt$ & $k_{35} = 1.31 \times 10^{-10} \expf{-}{80}{T}$ & & 32 \\
\hline
\end{tabular}
\end{minipage}
\end{table*}
 
\begin{table*}
\begin{minipage}{126mm}
\contcaption{}
\begin{tabular}{llllc}
\hline 
36 & $\ch + \mHt \rightarrow  \ch_{2} + \mH$ & 
$k_{36} = 5.46 \times 10^{-10} \expf{-}{1943}{T}$ & & 33 \\
37 & $\ch + \mC \rightarrow \mC_{2} + \mH $ &
$k_{37} = 6.59 \times 10^{-11}$ & & 34 \\
38 & $\ch + \mO \rightarrow \co + \mH $ & $k_{38} = 6.6 \times 10^{-11}$ & $T \le 2000 \: {\rm K}$  & 35 \\
& & $\phantom{k_{38}} = 1.02 \times 10^{-10} \expf{-}{914}{T}$ & $T > 2000 \: {\rm K}$ & 36 \\
39 & $\ch_{2} + \mH \rightarrow \ch + \mHt$ & $k_{39} = 6.64 \times 10^{-11}$ & & 37 \\
40 & $\ch_{2} + \mO \rightarrow \co + \mH + \mH$ & $k_{40} =1.33 \times 10^{-10}$ & & 38 \\
41 & $\ch_{2} + \mO \rightarrow \co + \mHt$ & $k_{41} = 8.0 \times 10^{-11}$ & & 39 \\
42 & $\mC_{2} + \mO \rightarrow \co + \mC $ & $k_{42} = 5.0 \times 10^{-11} \tmpt{0.5}$ & 
$T \le 300 \: {\rm K}$ & 40 \\
& & $\phantom{k_{42}} = 5.0 \times 10^{-11} \tmpt{0.757}$ & $T > 300 \: {\rm K}$ & 41 \\
43 & $\mO + \mHt \rightarrow \oh + \mH$ & 
$k_{43} = 3.14 \times 10^{-13} \tmpt{2.7} \expf{-}{3150}{T}$ & & 42 \\
44 & $\oh + \mH \rightarrow \mO + \mHt $ & 
$k_{44} = 6.99 \times 10^{-14} \tmpt{2.8} \expf{-}{1950}{T}$ & & 43 \\
45 & $\oh + \mHt \rightarrow \hto + \mH$ & 
$k_{45} = 2.05 \times 10^{-12} \tmpt{1.52} \expf{-}{1736}{T}$ & & 44 \\
46 & $\oh + \mC \rightarrow \co + \mH$ & $k_{46} = 1.0 \times 10^{-10}$ & & 34 \\
47 & $\oh + \mO \rightarrow \mO_{2} + \mH $ &
$k_{47} = 3.50 \times 10^{-11}$  & $T \le 261 \: {\rm K}$ & 45 \\ 
& & $\phantom{k_{47}} = 1.77 \times 10^{-11} \expf{}{178}{T}$ & $T > 261 \: {\rm K}$ & 33 \\
48 & $\oh + \oh \rightarrow \hto + \mH$ & 
$k_{48} = 1.65 \times 10^{-12} \tmpt{1.14} \expf{-}{50}{T}$ &  & 34 \\ 
49 & $\hto + \mH \rightarrow \mHt + \oh $ &
$k_{49} =1.59 \times 10^{-11} \tmpt{1.2} \expf{-}{9610}{T}$ & & 46 \\
50 & $\mO_{2} + \mH \rightarrow \oh + \mO $ & $k_{50} = 2.61 \times 10^{-10} \expf{-}{8156}{T}$ & & 33 \\
51 & $\mO_{2} + \mHt \rightarrow \oh + \oh $ & $k_{51} = 3.16 \times 10^{-10} 
\expf{-}{21890}{T}$ & & 47 \\
52 & $\mO_{2} + \mC \rightarrow \co + \mO $ &
$k_{52} = 4.7 \times 10^{-11} \tmpt{-0.34}$ & $T \le 295 \: {\rm K}$ & 34 \\
& & $ \phantom{k_{52}} = 2.48 \times 10^{-12} \tmpt{1.54} \expf{}{613}{T}$ & $T > 295 \: {\rm K}$ & 33 \\
53 & $\co + \mH \rightarrow \mC + \oh $ & 
$k_{53} = 1.1 \times 10^{-10} \tmpt{0.5} \expf{-}{77700}{T}$ & & 28 \\
54 & $\mHtp + \mHt \rightarrow \htp + \mH$ &
$k_{54} = 2.24 \times 10^{-9} \tmpt{0.042} \expf{-}{T}{46600}$ & & 48 \\
55 & $\htp + \mH \rightarrow \mHtp + \mHt$ & $k_{55} = 7.7 \times 10^{-9} \expf{-}{17560}{T}$ & & 49 \\
56 & $\mC + \mHtp \rightarrow \ch^{+} + \mH$ & $k_{56} = 2.4 \times 10^{-9}$ & & 28 \\
57 & $\mC + \htp \rightarrow \ch^{+} + \mHt$ & $k_{57} = 2.0 \times 10^{-9}$ & & 28 \\
58 & $\Cp + \mHt \rightarrow \ch^{+}  + \mH $ & 
$k_{58} = 1.0 \times 10^{-10} \expf{-}{4640}{T}$ & & 50 \\
59 & $\ch^{+} + \mH \rightarrow \Cp + \mHt $ & $k_{59} = 7.5 \times 10^{-10}$ & & 51 \\
60 & $\ch^{+} + \mHt \rightarrow \ch_{2}^{+} + \mH $ & $k_{60} = 1.2 \times 10^{-9}$ & & 51 \\
61 & $\ch^{+} + \mO \rightarrow \co^{+} + \mH $ & $k_{61} = 3.5 \times 10^{-10}$ & & 52 \\
62 & $\ch_{2} + \Hp \rightarrow \ch^{+} + \mHt $ & $k_{62} = 1.4 \times 10^{-9}$ & & 28 \\
63 & $\ch_{2}^{+} + \mH \rightarrow \ch^{+} + \mHt $ & 
$k_{63} = 1.0 \times 10^{-9} \expf{-}{7080}{T}$ & & 28 \\
64 & $\ch_{2}^{+} + \mHt \rightarrow \ch_{3}^{+} + \mH $ & $k_{64} = 1.6 \times 10^{-9}$ & & 53 \\
65 & $\ch_{2}^{+} + \mO \rightarrow {\rm HCO}^{+} + \mH $ & $k_{65} = 7.5 \times 10^{-10}$ & & 28 \\
66 & $\ch_{3}^{+} + \mH \rightarrow \ch_{2}^{+} + \mHt$ & 
$k_{66} = 7.0 \times 10^{-10} \expf{-}{10560}{T}$ & & 28 \\
67 & $\ch_{3}^{+} + \mO \rightarrow {\rm HCO}^{+} + \mHt $ & $ k_{67} = 4.0 \times 10^{-10}$ & & 54 \\
68 & $\mC_{2} + \Op \rightarrow \co^{+} + \mC $ & $k_{68} = 4.8 \times 10^{-10}$ & & 28 \\
69 & $\Op + \mHt   \rightarrow  \oh^{+} + \mH $  & $k_{69} = 1.7 \times 10^{-9}$ & & 55 \\
70 & $\mO + \mHtp  \rightarrow  \oh^{+} + \mH $ & $k_{70} = 1.5 \times 10^{-9}$ & & 28 \\
71 & $\mO + \htp  \rightarrow  \oh^{+} + \mHt $  & $k_{71} = 8.4 \times 10^{-10}$ & & 56 \\
72 & $\oh + \htp \rightarrow \hto^{+} + \mHt$ & $k_{72} = 1.3 \times 10^{-9}$ & & 28 \\
73 & $\oh + \Cp \rightarrow \co^{+} + \mH $ & $k_{73} = 7.7 \times 10^{-10}$ & & 28 \\
74 & $\oh^{+} + \mHt  \rightarrow  \hto^{+} + \mH $  & $k_{74} = 1.01 \times 10^{-9}$ & & 57 \\
75 & $\hto^{+} + \mHt  \rightarrow  {\rm H_{3}O}^{+} + \mH $  & 
$k_{75} = 6.4 \times 10^{-10}$ & & 58 \\
76 & $\hto + \htp \rightarrow {\rm H_{3}O^{+}} + \mHt$ &$k_{76} = 5.9 \times 10^{-9}$ & & 59 \\
77 & $\hto + \Cp \rightarrow {\rm HCO^{+}} + \mH$ & $k_{77} = 9.0 \times 10^{-10}$ & & 60 \\
78 & $\hto +\Cp \rightarrow {\rm HOC^{+}} + \mH$ & $k_{78} = 1.8 \times 10^{-9}$ & & 60 \\
79 & ${\rm H_{3}O}^{+} + \mC  \rightarrow {\rm HCO^{+}} + \mHt$ & $k_{79} = 1.0 \times 10^{-11}$  
&  & 28 \\
80 & $\mO_{2} + \Cp \rightarrow \co^{+} + \mO $ & $k_{80} = 3.8 \times 10^{-10}$ & & 53 \\
81 & $\mO_{2} + \Cp \rightarrow \co + \Op $ & $k_{81} = 6.2 \times 10^{-10}$ & & 53 \\
82 & $\mO_{2} + \ch_{2}^{+} \rightarrow {\rm HCO}^{+} + \oh $ & 
$k_{82} = 9.1 \times 10^{-10}$ & & 53 \\
83 & $\mO_{2}^{+} + \mC \rightarrow \co^{+} + \mO $ & $k_{83} = 5.2 \times 10^{-11}$ & & 28 \\
84 & $\co + \htp \rightarrow {\rm HOC^{+}} + \mHt$ & $k_{84} = 2.7 \times 10^{-11}$ & & 61 \\
85 & $\co + \htp \rightarrow {\rm HCO^{+}} + \mHt$ & $k_{85} = 1.7 \times 10^{-9}$ & & 61 \\
86 & ${\rm HCO^{+}} + \mC \rightarrow \co + \ch^{+}$ & $k_{86} = 1.1 \times 10^{-9}$ & & 28 \\
87 & ${\rm HCO^{+}} + \hto \rightarrow \co + {\rm H_{3}O^{+}} $ & 
$k_{87} = 2.5 \times 10^{-9}$ & & 62 \\
\hline
\end{tabular}
\end{minipage}
\end{table*}
 
\begin{table*}
\begin{minipage}{126mm}
\contcaption{}
\begin{tabular}{llllc}
\hline 
88 & $\mHt + \Hep \rightarrow \He + \mHtp$ & $k_{88} = 7.2 \times 10^{-15}$ & & 63 \\
89 & $\mHt  + \Hep \rightarrow \He + \mH + \Hp$ & $k_{89} = 3.7 \times 10^{-14} \expf{}{-35}{T}$ 
& & 63 \\  
90 & $\ch + \Hp \rightarrow \ch^{+} + \mH$ & $k_{90} = 1.9 \times 10^{-9}$ & & 28 \\
91 & $\ch_{2} + \Hp \rightarrow \ch_{2}^{+} + \mH$ & $k_{91} = 1.4 \times 10^{-9}$ & & 28 \\
92 & $\ch_{2} + \Hep \rightarrow \Cp + \He + \mHt$ & $k_{92} = 7.5 \times 10^{-10}$ & & 28 \\
93 & $\mC_{2} + \Hep \rightarrow \Cp + \mC + \He $ & $k_{93} =1.6 \times 10^{-9}$  & & 28 \\
94 & $\oh + \Hp \rightarrow \oh^{+} + \mH$ & $k_{94} = 2.1 \times 10^{-9}$ & & 28 \\
95 & $\oh + \Hep \rightarrow \Op + \He + \mH$ & $k_{95} = 1.1 \times 10^{-9}$ & & 28 \\
96 & $\hto + \Hp \rightarrow \hto^{+} + \mH $ & $k_{96} = 6.9 \times 10^{-9}$ & & 64 \\
97 & $\hto + \Hep \rightarrow \oh + \He + \Hp $ & $k_{97} = 2.04 \times 10^{-10}$ & & 65 \\
98 & $\hto + \Hep \rightarrow \oh^{+} + \He + \mH $ & $k_{98} = 2.86 \times 10^{-10}$ & & 65 \\
99 & $\hto + \Hep \rightarrow \hto^{+} + \He $ & $k_{99} = 6.05 \times 10^{-11}$ & & 65 \\
100 & $\mO_{2} + \Hp  \rightarrow \mO_{2}^{+} + \mH $ & $k_{100} = 2.0 \times 10^{-9}$ & & 64 \\
101 & $\mO_{2} + \Hep \rightarrow \mO_{2}^{+} + \He $ & $k_{101} = 3.3 \times 10^{-11}$ & & 66 \\
102 & $\mO_{2} + \Hep \rightarrow \Op + \mO + \He $ & $k_{102} = 1.1 \times 10^{-9}$ & & 66 \\
103 & $\mO_{2}^{+} + \mC \rightarrow \mO_{2} + \Cp $ & $k_{103} = 5.2 \times 10^{-11}$ & & 28 \\
104 & $\co + \Hep \rightarrow \Cp + \mO + \He $ & $k_{104} = 1.4 \times 10^{-9} \tmpt{-0.5}$ & & 67 \\
105 & $\co + \Hep \rightarrow \mC + \Op + \He $ & $k_{105} = 1.4 \times 10^{-16} \tmpt{-0.5}$ & & 67 \\
106 & $\co^{+} + \mH \rightarrow \co + \Hp $ & $k_{106} = 7.5 \times 10^{-10}$ & & 68 \\
107 & $\Cm + \Hp \rightarrow \mC + \mH $ & $k_{107} = 2.3 \times 10^{-7} \tmpt{-0.5}$ & & 28 \\
108 & $\Om + \Hp \rightarrow \mO + \mH $ & 
$k_{108} = 2.3 \times 10^{-7} \tmpt{-0.5}$ & & 28 \\ 
109 & $\Hep + \Hm \rightarrow \He + \mH$ & $k_{109} = 2.32 \times 10^{-7} \tmpt{-0.52} 
\expf{}{T}{22400}$ & & 69 \\
110 & $\htp + \me \rightarrow \mHt + \mH$ & $k_{110} = 2.34 \times 10^{-8} \tmpt{-0.52}$ & & 70 \\
111 & $\htp + \me \rightarrow \mH + \mH + \mH$ & $k_{111} = 4.36 \times 10^{-8} \tmpt{-0.52}$ & & 70 \\
112 & $\ch^{+} + \me \rightarrow \mC + \mH $ & $k_{112} = 7.0 \times 10^{-8} \tmpt{-0.5}$ & & 71 \\
113 & $\ch_{2}^{+} + \me \rightarrow \ch + \mH$ & $k_{113} = 1.6 \times 10^{-7} \tmpt{-0.6}$  & & 72 \\
114 & $\ch_{2}^{+} + \me \rightarrow \mC + \mH + \mH$ & 
$k_{114} = 4.03 \times 10^{-7} \tmpt{-0.6}$ & & 72 \\
115 & $\ch_{2}^{+} + \me \rightarrow \mC + \mHt$ & $k_{115} = 7.68 \times 10^{-8} \tmpt{-0.6}$  & & 72 \\
116 & $\ch_{3}^{+} + \me \rightarrow  \ch_{2} + \mH$ & 
$k_{116} = 7.75 \times 10^{-8} \tmpt{-0.5}$ & & 73 \\
117 & $\ch_{3}^{+} + \me \rightarrow \ch + \mHt$ & $k_{117} = 1.95 \times 10^{-7} \tmpt{-0.5}$ & & 73 \\
118 & $\ch_{3}^{+} + \me \rightarrow \ch + \mH + \mH$ & 
$k_{118} = 2.0 \times 10^{-7} \tmpt{-0.4}$ & & 28 \\
119 & $\oh^{+} + \me \rightarrow \mO + \mH$ & $k_{119} = 6.3 \times 10^{-9} \tmpt{-0.48}$ & & 74 \\
120 & $\hto^{+} + \me \rightarrow \mO + \mH + \mH$ & 
$k_{120} = 3.05 \times 10^{-7} \tmpt{-0.5}$ & & 75 \\
121 & $\hto^{+} + \me \rightarrow \mO + \mHt $ & $k_{121} = 3.9 \times 10^{-8} \tmpt{-0.5}$  & & 75 \\
122 & $\hto^{+} + \me \rightarrow \oh + \mH $ & $k_{122} = 8.6 \times 10^{-8} \tmpt{-0.5}$ & & 75 \\
123 & ${\rm H_{3}O}^{+} + \me  \rightarrow \mH + \hto $  & 
$k_{123} = 1.08 \times 10^{-7} \tmpt{-0.5}$ & & 76 \\
124 & ${\rm H_{3}O}^{+} + \me  \rightarrow \oh + \mHt $  & 
$k_{124} = 6.02 \times 10^{-8} \tmpt{-0.5}$ & & 76 \\
125 & ${\rm H_{3}O}^{+} + \me  \rightarrow \oh + \mH + \mH $  & 
$k_{125} = 2.58 \times 10^{-7} \tmpt{-0.5}$ & & 76 \\
126 & ${\rm H_{3}O}^{+} + \me  \rightarrow \mO + \mH + \mHt $  & 
$k_{126} = 5.6 \times 10^{-9}  \tmpt{-0.5}$ & & 76 \\
127 & $\mO_{2}^{+} + \me \rightarrow \mO + \mO $ & $k_{127} = 1.95 \times 10^{-7} \tmpt{-0.7}$ & & 77 \\
128 & $\co^{+} + \me \rightarrow \mC + \mO$ & $k_{128} = 2.75 \times 10^{-7} \tmpt{-0.55}$ & & 78 \\
129 & ${\rm HCO^{+}} + \me \rightarrow \co + \mH $ & 
$k_{129} = 2.76 \times 10^{-7} \tmpt{-0.64}$ & & 79 \\
130 & ${\rm HCO^{+}} + \me \rightarrow \oh + \mC $ & 
$k_{130} = 2.4 \times 10^{-8} \tmpt{-0.64}$  & & 79 \\
131 & ${\rm HOC^{+}} + \me \rightarrow \co + \mH $ & $k_{131} = 1.1 \times 10^{-7} \tmpt{-1.0}$ & & 28 \\ 
132 & $\Hm + \mC \rightarrow \ch + \me$ & $k_{132} = 1.0 \times 10^{-9}$ & & 28 \\
133 & $\Hm + \mO \rightarrow \oh + \me$ & $k_{133} = 1.0 \times 10^{-9}$ & & 28 \\
134 & $\Hm + \oh \rightarrow \hto + \me$ & $k_{134} = 1.0 \times 10^{-10}$ & & 28 \\
135 & $\Cm + \mH \rightarrow \ch + \me $ & $k_{135} = 5.0 \times 10^{-10}$ & & 28 \\
136 & $\Cm + \mHt \rightarrow \ch_{2} + \me $ & $k_{136} = 1.0 \times 10^{-13}$ & & 28 \\
137 & $\Cm + \mO \rightarrow \co + \me $ & $k_{137} = 5.0 \times 10^{-10}$ & & 28 \\
138 & $\Om + \mH \rightarrow \oh + \me$ & $k_{138} = 5.0 \times 10^{-10}$ & & 28\\
139 & $\Om + \mHt \rightarrow \hto + \me $ & $k_{139} = 7.0 \times 10^{-10}$ & & 28 \\
140 & $\Om + \mC \rightarrow \co + \me $ & $k_{140} = 5.0 \times 10^{-10}$ & & 28 \\
141 & $\mHt + \Hp \rightarrow \htp + \gamma$ & 
$k_{141} = 1.0 \times 10^{-16}$ & & 80 \\
\hline
\end{tabular}
\end{minipage}
\end{table*}
 
\begin{table*}
\begin{minipage}{166mm}
\contcaption{}
\begin{tabular}{llllc}
\hline 
142 & $\mC + \me \rightarrow \Cm + \gamma $ & $k_{142} = 2.25 \times 10^{-15}$ & & 81 \\
143 & $\mC + \mH \rightarrow \ch + \gamma$ & $k_{143} = 1.0 \times 10^{-17}$ & & 82 \\
144 & $\mC + \mHt \rightarrow \ch_{2} + \gamma$ & $k_{144} = 1.0 \times 10^{-17}$ & & 82 \\
145 & $\mC + \mC \rightarrow \mC_{2} + \gamma $ & 
$k_{145} = 4.36 \times 10^{-18} \tmpt{0.35} \expf{-}{161.3}{T}$ & & 83 \\
146 & $\mC + \mO \rightarrow \co + \gamma$ & 
$k_{146} = 2.1 \times 10^{-19}$ & $T \le 300 \: {\rm K}$ & 84 \\
& & $\phantom{k_{146}} = 3.09 \times 10^{-17} \tmpt{0.33} \expf{-}{1629}{T}$ & 
$T > 300 \: {\rm K}$ & 85 \\
147 & $\Cp + \mH  \rightarrow \ch^{+}  + \gamma $ & 
$k_{147} = 4.46 \times 10^{-16} T^{-0.5} \expf{-}{4.93}{T^{2/3}}$ & & 86 \\
148 & $\Cp + \mHt \rightarrow \ch_{2}^{+} + \gamma $ & 
$k_{148} = 4.0 \times 10^{-16} \tmpt{-0.2}$ & & 87 \\
149 & $\Cp + \mO \rightarrow \co^{+} + \gamma$ & $k_{149} = 2.5 \times 10^{-18}$ & 
$T  \le 300 \: {\rm K}$ & 84 \\
& & $\phantom{k_{149}} = 3.14 \times 10^{-18} \tmpt{-0.15} \expf{}{68}{T}$ & $T > 300 \: {\rm K}$ & \\
150 & $\mO + \me \rightarrow \mO^{-} + \gamma$ & $k_{150} = 1.5 \times 10^{-15}$ & & 28 \\
151 & $\mO + \mH \rightarrow \oh + \gamma$ & $k_{151} = 9.9 \times 10^{-19} \tmpt{-0.38}$ & & 28 \\
152 & $\mO + \mO \rightarrow \mO_{2} + \gamma $ & 
$k_{152} = 4.9 \times 10^{-20} \tmpt{1.58}$ & & 82 \\
153 & $\oh + \mH \rightarrow \hto + \gamma$ & 
$k_{153} = 5.26 \times 10^{-18} \tmpt{-5.22} \expf{-}{90}{T}$ & & 88 \\
154 & $\mH + \mH + \mH \rightarrow \mHt + \mH$ & 
$k_{154} = 1.32 \times 10^{-32} \tmpt{-0.38}$ & $T \le 300 \: {\rm K}$ & 89 \\ 
& & $\phantom{k_{154}} = 1.32 \times 10^{-32} \tmpt{-1.0}$ & $T > 300 \: {\rm K}$ & 90 \\ 
155 & $\mH + \mH + \mHt \rightarrow \mHt + \mHt$ &
$k_{155} = 2.8 \times 10^{-31} T^{-0.6}$ & & 91 \\ 
156 & $\mH + \mH + \He \rightarrow \mHt + \He$ & $k_{156} = 6.9 \times 10^{-32} T^{-0.4}$ & & 92 \\
157 & $\mC + \mC + {\rm M} \rightarrow \mC_{2} + {\rm M}$ &
$k_{157} = 5.99 \times 10^{-33} \tmptscl{5000}{-1.6}$ & $T \le 5000 \: {\rm K}$ & 93 \\
& & $\phantom{k_{157}} = 5.99 \times 10^{-33} \tmptscl{5000}{-0.64} \expf{}{5255}{T}$ &
$T > 5000 \: {\rm K}$ & 94 \\ 
158 & $\mC + \mO + {\rm M} \rightarrow \co + {\rm M}$ &
$k_{158} = 6.16 \times 10^{-29} \tmpt{-3.08}$& $T \le 2000 \: {\rm K}$ & 35 \\ 
& & $\phantom{k_{158}} = 2.14 \times 10^{-29} \tmpt{-3.08} \expf{}{2114}{T}$ & 
$T > 2000 \: {\rm K}$ & 67 \\
159 & $\Cp + \mO + {\rm M} \rightarrow \co^{+} + {\rm M}$ & $k_{159} = 100 \times k_{210}$ & & 67 \\
160 & $\mC + \Op + {\rm M} \rightarrow \co^{+} + {\rm M}$ & $k_{160} = 100 \times k_{210}$ & & 67 \\
161 & $\mO + \mH + {\rm M} \rightarrow \oh+ {\rm M}$ & 
$k_{161} = 4.33 \times 10^{-32} \tmpt{-1.0}$ & & 43 \\
162 & $\oh + \mH + {\rm M} \rightarrow \hto + {\rm M}$ & 
$k_{162} = 2.56 \times 10^{-31} \tmpt{-2.0}$ & & 35 \\
163 & $\mO + \mO + {\rm M} \rightarrow \mO_{2} + {\rm M}$ &
$k_{163} = 9.2 \times 10^{-34} \tmpt{-1.0}$ & & 37 \\
164 & $\mO + \ch \rightarrow {\rm HCO^{+}} + \me$ & 
$k_{164} = 2.0 \times 10^{-11} \tmpt{0.44}$ & & 95 \\
165 & $\mH + \mH({\rm s}) \rightarrow \mHt$ & $k_{165} = 3.0 \times 10^{-18} T^{0.5} f_{\rm A}
[1.0 + 0.04(T + T_{\rm d})^{0.5}$ & $f_{\rm A} = \left[1.0 + 10^{4} \exp \left(-\frac{600}{T_{\rm d}}\right) \right]^{-1}$ & 96 \\
& & $ \phantom{k_{165} =} \mbox{} + 0.002 \, T + 8 \times 10^{-6} T^{2}]^{-1}$ &  & \\
\hline
\end{tabular}
\medskip
\\
{\bf Notes:}
$T$ and $T_{\rm e}$ are the gas temperature in units of Kelvin and eV respectively, 
while $T_{\rm d}$ is the dust temperature in Kelvin. $\mH({\rm s})$ denotes a hydrogen
atom adsorbed on a grain surface. References are to the primary source of data for each 
reaction. \\
{\bf References}:
1: \citet{WIS79}, 2: \citet{LAU91}, 3: \citet{RAM76}, 4: \citet{KAR79},
5: \citet{cdg99}, 6: \citet{SCH94}, 7: \citet{SAV04}, 8: \citet{STI99}, 9: \citet{MAC86},
10: \citet{ls83}, 11: \citet{mkm98}, 12: \citet{sk87},
13: \citet{JAN87}, 14: \citet{FER92}, 15: \citet{POU78}, 
16: \citet{hs98}, 17: \citet{ap73}, 18: \citet{z89}, 19: \citet{kldd93}, 
20: \citet{NAH97}, 21: \citet{NAH99}, 22: \citet{VOR97},
23: \citet{STA99}, 24: \citet{STA98}, 25: \citet{z04}, 26: \citet{kd93},
27: \citet{drcm87}, 28: \citet{TEU00},  29: \citet{ssrg02}, 30: \citet{wr85}, 
31: \citet{ddh91}, 32: \citet{hgs93}, 33: Fit by \citet{TEU00} to data from the NIST 
chemical kinetics database; original source or sources unclear,
34: \citet{shc04},  35: \citet{bau92}, 36: \citet{mr86},
37: \citet{war84}, 38: \citet{fra86}, 39: Fit by \citet{TEU00} to 
data from \citet{fj84} and \citet{fbj88}, 40: \citet{md83}, 
41: \citet{fa69}, Glover (2009, in prep.), 42: \citet{nr87}, 43: \citet{th86},
44: \citet{old92}, 45: \citet{cgk06}, Glover (2009, in prep.) 
46: \citet{cw79}, 47: \citet{aat75}, 48: \citet{ljb95}, 49: \citet{smt92},
50: \citet{asm84}, 51: \citet{ms99}, 52: \citet{vig80},
53: \citet{sa77a,sa77b}, 54: \citet{feh76}, 55: \citet{sam78,asp80},
56: \citet{mm00}, 57: \citet{jbt81}, 58: \citet{rw80}, 59: \citet{kth74,afhk75},
60: \citet{ahf76,wah76}, 61: \citet{kth75}, 62: \citet{asg78}, 
63: \citet{ba84}, 64: \citet{ssm92}, 
65: \citet{mdm78a,mdm78b}, 66: \citet{as76a, as76b}, 67: \citet{pd89}, 
68: \citet{fed84a,fed84b}, 69: \citet{ph94}
70: Fit by \citet{umist07} to data from \citet{mac04}, 71: \citet{tkl91},
72: \citet{lls98}, 73: \citet{mi90},  74: \citet{gub95}, 75: \citet{rds00}, 
76: \citet{jbs00}, 77: \citet{aas83}, 78: \citet{rpl98},  79: \citet{gep05},
80: \citet{gh92}, 81: \citet{sd98}, 82: \citet{ph80}, 83: \citet{as97}, 
84: \citet{dd90}, 85: \citet{ssbo99}, 86: \citet{bvh06}, 87: \citet{he85}, 
88: \citet{fa80}, 89: \citet{or87}, 90: \citet{abn02}
91: \citet{cw83},  92: \citet{wk75},  93: Glover (2009, in prep.), 
94: Fit by \citet{TEU00} to data from \citet{fa69} and \citet{sla76},
95: \citet{mb73}, 96: \citet{hm79}
\end{minipage}
\end{table*}

\begin{table}
\caption{List of photochemical reactions included in our chemical model \label{chem_photo}}
\begin{tabular}{llllc}
\hline
No.\  & Reaction & Optically thin rate $({\rm s}^{-1})$ & $\gamma$ & Ref.\ \\
\hline
166 & $\Hm + \gamma \rightarrow \mH + \me$ & $R_{166} = 7.1 \times 10^{-7}$ & 0.5 & 1 \\
167 & $\mHtp + \gamma \rightarrow \mH + \Hp$ & $R_{167} = 1.1 \times 10^{-9}$ & 1.9 & 2 \\
168 & $\mHt + \gamma \rightarrow \mH + \mH$ & $R_{168} = 5.6 \times 10^{-11}$ & 
See \S\ref{photochem} & 3 \\
169 & $\htp + \gamma \rightarrow \mHt + \Hp$ & $R_{169} = 4.9 \times 10^{-13}$ & 1.8 & 4 \\
170 & $\htp + \gamma \rightarrow \mHtp + \mH$ & $R_{170} = 4.9 \times 10^{-13}$ & 2.3 & 4 \\
171 & $\mC + \gamma \rightarrow  \Cp  + \me$ & $R_{171} = 3.1 \times 10^{-10}$ & 3.0 & 5 \\ 
172 & $\Cm + \gamma \rightarrow \mC + \me $ & $R_{172} = 2.4 \times 10^{-7}$ & 0.9 & 6 \\
173 & $\ch + \gamma \rightarrow \mC + \mH$&  $R_{173} = 8.7 \times 10^{-10}$ & 1.2 & 7 \\
174 & $\ch + \gamma \rightarrow \ch^{+} + \me$& $R_{174} = 7.7 \times 10^{-10}$ & 2.8 & 8 \\
175 & $\ch^{+} + \gamma \rightarrow \mC + \Hp$ & $R_{175} = 2.6 \times 10^{-10}$ & 2.5 & 7 \\
176 & $\ch_{2} + \gamma \rightarrow \ch + \mH$&  $R_{176} = 7.1 \times 10^{-10}$ & 1.7 & 7 \\
177 & $\ch_{2} + \gamma \rightarrow \ch_{2}^{+} + \me$& $R_{177} = 5.9 \times 10^{-10}$ & 2.3 & 6 \\
178 & $\ch_{2}^{+} + \gamma \rightarrow \ch^{+} + \mH$ & $R_{178} = 4.6 \times 10^{-10}$ & 1.7 & 9 \\ 
179 & $\ch_{3}^{+} + \gamma \rightarrow \ch_{2}^{+} + \mH$ &  $R_{179} = 1.0 \times 10^{-9}$ 
& 1.7 & 6 \\
180 & $\ch_{3}^{+} + \gamma \rightarrow \ch^{+} + \mHt$ &  $R_{180} = 1.0 \times 10^{-9}$ & 1.7 & 6 \\
181 & $\mC_{2} + \gamma \rightarrow \mC + \mC $ & $R_{181} = 1.5 \times 10^{-10}$ & 2.1 & 7 \\
182 & $\Om + \gamma \rightarrow \mO + \me $ & $R_{182} = 2.4 \times 10^{-7}$ & 0.5 & 6 \\
183 & $\oh + \gamma \rightarrow \mO + \mH$ & $R_{183} = 3.7 \times 10^{-10}$ & 1.7 & 10 \\
184 & $\oh + \gamma \rightarrow \oh^{+} + \me$ & $R_{184} = 1.6 \times 10^{-12}$ & 3.1 & 6 \\
185 & $\oh^{+} + \gamma \rightarrow \mO + \Hp$ & $R_{185} = 1.0 \times 10^{-12}$ & 1.8 & 4 \\
186 & $\hto + \gamma \rightarrow \oh + \mH$ & $R_{186} = 6.0 \times 10^{-10}$ & 1.7 & 11 \\
187 & $\hto + \gamma \rightarrow \hto^{+} + \me$ & $R_{187} = 3.2 \times 10^{-11}$ & 3.9 & 8 \\
188 & ${\rm H_{2}O^{+}} + \gamma \rightarrow \mHtp + \mO$ & $R_{188} = 5.0 \times 10^{-11}$ &
See \S\ref{photochem} & 12 \\
189 & ${\rm H_{2}O^{+}} + \gamma \rightarrow \Hp + \oh$ & $R_{189} = 5.0 \times 10^{-11}$ & See \S\ref{photochem} & 12 \\
190 & ${\rm H_{2}O^{+}} + \gamma \rightarrow \Op + \mHt$ & $R_{190} = 5.0 \times 10^{-11}$ & See \S\ref{photochem} & 12 \\
191 & ${\rm H_{2}O^{+}} + \gamma \rightarrow \oh^{+} + \mH$ & 
$R_{191} = 1.5 \times 10^{-10}$ &See \S\ref{photochem} & 12 \\
192 & ${\rm H_{3}O^{+}} + \gamma \rightarrow \Hp + \hto$ & $R_{192} = 2.5 \times 10^{-11}$ & See \S\ref{photochem} & 12 \\
193 & ${\rm H_{3}O^{+}} + \gamma \rightarrow \mHtp + \oh$ & $R_{193} = 2.5 \times 10^{-11}$ & See \S\ref{photochem} & 12 \\
194 & ${\rm H_{3}O^{+}} + \gamma \rightarrow \hto^{+} + \mH$ & 
$R_{194} = 7.5 \times 10^{-12}$&See \S\ref{photochem} & 12 \\
195 & ${\rm H_{3}O^{+}} + \gamma \rightarrow \oh^{+} + \mHt$ & 
$R_{195} = 2.5 \times 10^{-11}$& See \S\ref{photochem} & 12 \\
196 & $\mO_{2} + \gamma \rightarrow \mO_{2}^{+} + \me $ & $R_{196}= 5.6 \times 10^{-11} $ & 3.7 & 7 \\
197 & $\mO_{2} + \gamma \rightarrow \mO + \mO $ & $R_{197} = 7.0 \times 10^{-10}$ & 1.8 & 7 \\
198 & $\co + \gamma \rightarrow \mC + \mO $ & $R_{198} = 2.0 \times 10^{-10}$ & 
See  \S\ref{photochem} & 13 \\
\hline
\end{tabular}
\medskip
\\
{\bf Note:}
Rates are computed assuming the standard interstellar radiation field from \citet{d78}, with
field strength $G_{0} = 1.7$ in \citet{habing68} units. $\gamma$ quantifies the dependence of
the rate on the visual extinction $A_{\rm V}$ in optically thick gas: $R_{\rm thick} = R_{\rm thin}
\exp(-\gamma A_{\rm V})$ (see Eq.~\ref{thick}).\\
{\bf References:}
1: \citet{dj72}, 2: \citet{DUN68}, 3: \citet{db96}, 4: \citet{vd87},
5: \citet{VER96}, 6: \citet{TEU00}, 7: \citet{rjld91}, 8: \citet{vd88},
9: \citet{vd06}, 10: \citet{vdd84}, 11: \citet{lee84}, 12: \citet{sd95},
13: \citet{vdb88}
\end{table}

\begin{table}
\caption{List of reactions included in our chemical model that
involve cosmic rays or cosmic-ray induced UV emission \label{tab:cosmic}}
\begin{tabular}{lllc}
\hline
No.\ & Reaction & Rate (${\rm s^{-1}} \zeta_{\mH}^{-1}$) & Ref.\  \\
\hline
199 & $\mH + {\rm c.r.} \rightarrow \Hp + \me$ & $R_{199} = 1.0$ & --- \\
200 & $\He  + {\rm c.r.} \rightarrow \Hep + \me$ & $R_{200} = 1.1$ & 1 \\
201 & $\mHt + {\rm c.r.} \rightarrow \Hp + \mH + \me$ & $R_{201} = 0.037$ & 1 \\
202 & $\mHt + {\rm c.r.} \rightarrow \mH + \mH$ & $R_{202} = 0.22$ & 1 \\
203 & $\mHt + {\rm c.r.} \rightarrow \Hp + \Hm$ & $R_{203} = 6.5 \times 10^{-4}$ & 1 \\
204 & $\mHt + {\rm c.r.} \rightarrow \mHtp + \me$ & $R_{204} = 2.0$ & 1 \\
205 & $\mC  + {\rm c.r.} \rightarrow \Cp + \me$ & $R_{205} = 3.8$ & 1 \\
206 & $\mO  + {\rm c.r.} \rightarrow \Op + \me$ & $R_{206} = 5.7$ & 1 \\
207 & $\co + {\rm c.r.} \rightarrow \co^{+} + \me$ & $R_{207} = 6.5$ & 1 \\
208 & $\mC + \gamma_{\rm c.r.} \rightarrow \Cp + \me$ & $R_{208} = 2800$ & 2 \\
209 & $\ch + \gamma_{\rm c.r.} \rightarrow \mC + \mH$  & $R_{209} = 4000$  & 3 \\
210 & $\ch^{+} + \gamma_{\rm c.r.} \rightarrow \Cp + \mH$  & $R_{210} = 960$ & 3 \\
211 & $\ch_{2} + \gamma_{\rm c.r.} \rightarrow \ch_{2}^{+} + \me$ & $R_{211} = 2700 $ & 1 \\
212 & $\ch_{2} + \gamma_{\rm c.r.} \rightarrow \ch + \mH$ & $R_{212} = 2700 $ & 1 \\
213 & $\mC_{2} + \gamma_{\rm c.r.} \rightarrow \mC + \mC$ & $R_{213} = 1300$  & 3 \\
214 & $\oh + \gamma_{\rm c.r.} \rightarrow \mO + \mH$ & $R_{214} = 2800 $  & 3 \\
215 & $\hto + \gamma_{\rm c.r.} \rightarrow \oh + \mH$ & $R_{215} = 5300 $  & 3 \\
216 & $\mO_{2} + \gamma_{\rm c.r.} \rightarrow \mO + \mO$ & $R_{216} = 4100 $ & 3 \\
217 & $\mO_{2} + \gamma_{\rm c.r.} \rightarrow \mO_{2}^{+} + \me$ & $R_{217} = 640 $ & 3 \\
218 & $\co + \gamma_{\rm c.r.} \rightarrow \mC + \mO$ & $R_{218} = 0.21 T^{1/2} x_{\mHt} 
x_{\co}^{-1/2} $  & 4 \\
\hline
\end{tabular}
\medskip
\\
{\bf Note:}
Rates are quoted relative to the cosmic ray ionization rate of atomic
hydrogen, $\zeta_{\mH}$, which is an adjustable parameter in our 
models. Rates for cosmic-ray induced photoionizations and photodissociations
(reactions 208--218) are quoted assuming a grain albedo $\omega = 0.6$,
following reference 1. In the expression for $R_{218}$, $T$ is the gas temperature
in Kelvin and $x_{\mHt}$ and $x_{\co}$ are the fractional abundances of $\mHt$ and 
CO, respectively.
 \\
{\bf References:}
1: \citet{TEU00}, 2: \citet{gld87}, 3: \citet{gldh89}, 4: \citet{mht96}

\end{table}

\end{document}